\documentclass[12pt]{article}
\newif\ifpdf\ifx\pdfoutput\undefined\pdffalse\else\pdfoutput=1\pdftrue\fi
\newcommand{\pdfgraphics}{\ifpdf\DeclareGraphicsExtensions{.pdf,.jpg}
\else\fi} 
\usepackage{graphicx,epstopdf,amssymb,amsfonts,amsmath,amsthm,array,mathrsfs}
\newcommand{\pbs}[1]{\let\temp=\\#1\let\\=\temp}
\numberwithin{equation}{section}
\def\be{\begin{equation}}\def\ee{\end{equation}}
\def\cvp{\raise 2pt\hbox{,}} 
 
  \def\Tr{\mathop{\rm Tr}\nolimits}

 \def\d{{\rm d}}\def\nn{{\cal
N}}

 \def\Nf{N_{\mathrm f}}
\def\suN{{\rm SU}(N)} \def\uN{{\rm U}(N)} 

\def\u{\text{U}(1)}

\def\wl{W_{\rm low}}\def\wt{W_{\rm tree}}\def\weff{W_{\rm eff}}
\def\La{\Lambda}
\def\g{\boldsymbol{g}}\def\q{\boldsymbol{q}} \def\ring{\mathsf A}\def\hring{\mathbb A}\def\pfield{\mathsf k} \def\pring{\mathsf a}\def\ideal{\mathscr I}
\def\ringp{\ring_{|\varphi)}}\def\hringp{\hring_{|\varphi)}}

\theoremstyle{plain}
\newtheorem{thm}{Theorem}
\newtheorem{lem}[thm]{Lemma}
\newtheorem{prop}[thm]{Proposition}

\theoremstyle{definition}
\newtheorem{defn}{Definition}
\newtheorem{exmp}{Example}
\theoremstyle{remark}
\newtheorem*{rem}{Remark}

\def\plb#1#2#3{{\it Phys.\ Lett.\ }{\bf B #1} (#2) #3}
\def\npb#1#2#3{{\it Nucl.\ Phys.\ }{\bf B #1} (#2) #3}
\def\npps#1#2#3{{\it Nucl.\ Phys.\ Proc.\ Suppl.\ }{\bf #1} (#2) #3}

\def\jhep#1#2#3{{\it J. High Energy Phys.\ }{\bf #1} (#2) #3}
\def\prd#1#2#3{{\it Phys.\ Rev.\ }{\bf D #1} (#2) #3}

\def\atmp#1#2#3{{\it Adv.\ Theor.\ Math.\ Phys.\ }{\bf #1} (#2) #3}

\def\pr#1#2#3{{\it Phys.\ Rep.\ }{\bf #1} (#2) #3}

\def\cras#1#2#3{{\it C.\ R.\ Physique }{\bf #1} (#2) #3}
\begin{document}
\pdfgraphics 
%
\pagestyle{empty}
{\parskip 0in

\hfill LPTENS-08/24

\vfill
\begin{center}
{\LARGE On the Geometry of Super Yang-Mills Theories:}

\bigskip

{\LARGE Phases and Irreducible Polynomials}

\vspace{0.4in}

Frank F{\scshape errari}
\\
\medskip
{\it Service de Physique Th\'eorique et Math\'ematique\\
Universit\'e Libre de Bruxelles and International Solvay Institutes\\
Campus de la Plaine, CP 231, B-1050 Bruxelles, Belgique}\\
\smallskip
{\tt frank.ferrari@ulb.ac.be}
\end{center}
\vfill\noindent

We study the algebraic and geometric structures that underly the space
of vacua of $\nn=1$ super Yang-Mills theories at the non-perturbative
level. Chiral operators are shown to satisfy polynomial equations over
appropriate rings, and the phase structure of the theory can be
elegantly described by the factorization of these polynomials into
irreducible pieces. In particular, this idea yields a powerful method
to analyse the possible smooth interpolations between different
classical limits in the gauge theory. As an application in $\uN$
theories, we provide a simple and completely general proof of the fact
that confining and Higgs vacua are in the same phase when fundamental
flavors are present, by finding an irreducible polynomial equation
satisfied by the glueball operator. We also derive the full phase
diagram for the theory with one adjoint when $N\leq 7$ using
computational algebraic geometry programs.

\vfill

\medskip
%
\begin{flushleft}
\today
\end{flushleft}
\newpage\pagestyle{plain}
\baselineskip 16pt
\setcounter{footnote}{0}

%
\section{Introduction}
\label{IntroSec}
\subsection{General presentation}
\label{gpSec}

The study of the non-perturbative aspects of $\nn=1$ supersymmetric
gauge theories has revealed over the years many remarkable physical
phenomena that can be described in a rich mathematical framework. The
fundamental tool is the existence of special \emph{chiral} operators
that preserve half of the supercharges. The expectation values of
these operators are space-time independent and depend
\emph{holomorphically} on the various parameters of the theory. This
allows to use the many tools of complex analysis, in particular it is
possible to make analytic continuations to derive results at strong
coupling from semi-classical instanton calculations. Recently, a
completely general and first-principle approach has been developed to
compute any chiral operator expectation values along these lines
\cite{mic1,mic2,mic3,micrev}. The results have an interpretation in
the context of the open/closed string duality (geometric transitions,
matrix models), brane engineering, mirror symmetry, integrable systems
etc\ldots They lie at the heart of many developments in Quantum Field
Theory and String Theory over the last 15 years.

In the present work we are going to formulate the results in an
algebraic and geometric language that turns out to be extremely
natural and efficient to understand the general structure of the
theory and to derive the physical consequences of the solutions. In
particular, we revisit some fundamental notions like the chiral ring,
whose full significance has not been fully understood and exploited in
previous works. We also correct some confusions that have appeared in
the literature.

This research is motivated by the fact that a satisfactory
understanding of the global properties of the space of vacua of
supersymmetric theories, including the phase structure and the
possible interpolations between different classical vacua, requires
new powerful computational tools. The framework that we are going to
develop allows to reduce many interesting physical questions to simple
arithmetic properties of polynomials. Moreover, when necessary, our
approach lends itself very well to calculations on the computer.

An important conceptual issue is to understand the nature of the
various phases in which the gauge theories can be realized. For
example, is it possible to distinguish the phases using some symmetry
principle? This is an outstanding open problem. The standard
't~Hooft's and Wilson's order parameters provide a partial answer, but
it is known that they fail to provide a complete classification
\cite{csw1}. The results of our work can be used to shed a new
interesting light on these questions, as will be explained in a
separate publication \cite{Galois}.

\subsection{Vacua versus phases}
\label{vvpsec}

One of the most interesting aspect of supersymmetric gauge theories is
to have a very rich and complex landscape of vacua. The number of
vacua can be very large, growing exponentially with the number of
colours. We shall be able to study examples with several thousands of
vacua in the following. The vacua realized in a given theory can have
very different physics, with various particle spectra and gauge
groups. The structure has actually many similarities with the M/string
theory landscape.

The notion of vacuum is very central in the usual approaches to
quantum field theory (and actually to any quantum theory). One of the
basic reason is that quantum mechanics is usually formulated by
starting from a classical solution (a classical vacuum) and then
quantizing around this solution. Typically, one expands the
observables in powers of a parameter measuring the strength of the
quantum corrections around the classical solution under consideration.
In essence, this is an \emph{analytic} approach. In the favorable
cases where the expansion converges (this is what happens in the
chiral sector of the theory), one can then have access to the genuine
quantum regime. The resulting analytic formulas can be very cumbersome
and the underlying strongly quantum physics can be hard, if not
impossible, to describe.

On the other hand, from a purely quantum point of view, independently
of any semi-classical approximation, the notion of a classical (or
quantum) vacuum is peripheral. This fundamental fact will become
clearer and clearer the further we advance in the paper. The central
invariant concept is the one of \emph{phase}. A precise definition
will be given later, but we can already describe the most relevant
features. A given gauge theory may be realized in various phases, but
the main property of individual phases is that by varying the
parameters in arbitrary ways the theory always remains in the same
phase. In this sense, a phase can be considered to be by itself a
consistent quantum theory. Many vacua can belong to the same phase,
which means equivalently that a given phase can have many different
classical limits. Any classical limit in a given phase can be obtained
from any other classical limit in the same phase by a suitable
analytic continuation. These analytic continuations can be strongly
quantum mechanical, involving highly non-trivial effects like the
exchange between D-brane like objects and solitonic branes and the
changing of the unbroken gauge groups
\cite{fer1,fer2,ferstr,csw1,csw2}.

It is when one wishes to study the phases in a fully quantum way, in
particular taking into account all the possible classical limits at
the same time, that the algebro-geometric approach that we shall use
is very powerful. 

\subsection{Algebraic geometry}

The geometric picture is actually very simple. It is known that
supersymmetry implies that the space of vacua $\mathscr M$ must be a
complex manifold. This is particularly clear at the classical level,
where the classical space of vacua $\mathscr M_{\text{cl}}$ is
described by the $F$-term constraints on the set of gauge invariant
chiral operators. The variety $\mathscr M_{\text{cl}}$, even though it
doesn't know about the strongly coupled gauge dynamics, can be quite
non-trivial and interesting as recent works have shown
\cite{Rclpapers}. In Section \ref{FoundSec}, we are going to explain
in details how to define the quantum algebraic variety $\mathscr M$,
describing explicitly its defining equations. The ring of chiral
observables of the theory coincides with the ring of functions defined
on the variety. A crucial aspect is that the variety $\mathscr M$ is
not in general irreducible. The existence of distinct phases
$|\varphi)$ in the gauge theory precisely corresponds to the
decomposition of $\mathscr M$ into irreducible components,
\be\label{Mdecompose}\mathscr M = \bigcup_{|\varphi)}\mathscr
M_{|\varphi)}\, .\ee
Algebraically, a given irreducible factor $\mathscr M_{|\varphi)}$ is
characterized by a set of special relations satisfied by the chiral
operators in the phase $|\varphi)$, making the ideal of operator
relations prime. In practice, this can be described by the
factorization of certain polynomials defined over appropriate rings
into irreducible pieces. An extremely simple description of the
operator algebra in a given phase in terms of ``primitive operators''
can then be given. All these aspects are explained in Section
\ref{CRPSec}.

In the above picture, the vacua simply correspond to the intersection
points between $\mathscr M$ and a set of hyperplanes that corresponds
to fixing the parameters of the gauge theory to some special values.
If $v$ is the total number of vacua and $p$ the total number of
parameters, $\mathscr M$ can then be seen as a $v$-fold cover of
$\mathbb C^{p}$. However this description is quite arbitrary. For
example, one could slice $\mathscr M$ with generic hyperplanes. The
number of intersection points, which is the degree of the variety, is
then in general larger than $v$. On the other hand, the decomposition
\eqref{Mdecompose} expresses an intrinsic property of the space
$\mathscr M$ and of the quantum gauge theory.

One advantage of the algebraic description of the space of vacua that
we shall set up is that methods from computational algebraic geometry
become available. This field has been developing rapidly over the last
few years, with a profound impact on research in algebraic geometry
and commutative algebra. A list of available softwares can be found in
\cite{soft}. We have used both \textsc{Singular} (for symbolic
computations) and PHC (for numerical computations)
\cite{singular,PHC}. These programs implement powerful algorithms that
are able to compute the decomposition \eqref{Mdecompose} into
irreducible components, see Section \ref{CoulSec}.

\subsection{Applications}
\label{introappSec}

One outstanding application that we are going to study is the
following. Consider a $\uN$ gauge theory with fields in the
fundamental representation. In this case, a test charge in any
representation of the gauge group can be screened by the dynamical
fundamental fields, and the usual criteria used to distinguish the
confining and the Higgs regimes do not work. In fact, it has been
known for almost 30 years that the confining and Higgs regimes can be
smoothly connected and are thus in the same phase when the theory is
formulated on the lattice \cite{FF}.

In the continuum, the problem is much more difficult to study because
the interpolation cannot be described perturbatively or
semi-classically. In \cite{csw2}, it was convincingly argued that the
solutions to $\nn=1$ supersymmetric gauge theories with fundamentals
seemed to have the required features for describing a single
Higgs/confining phase. A proof could not be given, however, because of
the apparent complexity of the explicit solution of the model. One
uses auxiliary algebraic curves and meromorphic functions with a
complicated pole structure defined on these curves. To understand the
phase structure one then has to study in great details how the
algebraic curves and the poles are deformed when the parameters are
varied. This is made extremely difficult and cumbersome by the fact
that the curves and the positions of the poles must obey complicated
non-linear constraints. This problem was further studied in
\cite{phaseref} using very detailed calculations and numerical
analysis in special cases.

In our framework, the equivalence between the ``Higgs'' and
``confining'' phases follow from the fact that the corresponding vacua
belong to the same irreducible component of the space of vacua. We
shall be able to provide a completely general and simple proof of this
fact in Section \ref{HiggsConfSec}, by finding an irreducible
polynomial equation satisfied by the gluino condensate.

Another interesting model, that has been much studied in the
literature, is the $\uN$ theory with only one adjoint matter chiral
superfield. The landscape of vacua for this model is very interesting,
with a highly non-trivial phase structure. We shall give a complete
description of the space of vacua for all $N\leq 7$ in Section
\ref{CoulSec}, providing in particular many explicit and non-trivial
examples of irreducible polynomial equations. For example, the
$\text{U}(7)$ theory can be realized in 10 distinct phases and a model
that realizes all these phases must have at least 11075 vacua. The
decomposition into phases is worked out by proving the irreducibility
of several complicated polynomials of degrees up to 126.

\subsection{Remarks and terminology}

The aim of this paper is to develop a general framework in which
the solutions of the theories can be naturally expressed and
exploited. However, we do not explain how the explicit solutions are
obtained. Let us simply stress that direct derivations from first
principles are now available \cite{mic1,mic2,mic3}.

All the necessary algebraic notions are introduced in a pedagogical
way and are motivated by physical questions. The tools we need are
fairly elementary and do not go beyond the beginning graduate level.
Excellent references that we have used are listed in \cite{books}.

A \emph{field} in the following always refers to the notion of an
algebraic field. A field is thus a commutative ring in which every
non-zero element has an inverse. A basic result explained in Section 2
is that in a given phase the ring of chiral operators of the theory is
actually a field, i.e.\ every non-zero operator has an inverse.

If $k$ is a field, we denote by $k[X_{1},\ldots,X_{n}]$ the ring of
polynomials with $n$ indeterminates $X_{1},\ldots,X_{n}$ and
coefficients in $k$. Thus the $X_{1},\ldots,X_{n}$ are always
unconstrained variables. On the other hand, we denote by $k[\mathcal
O_{1},\ldots,\mathcal O_{n}]$ the ring generated by arbitrary variables
$\mathcal O_{1},\ldots,\mathcal O_{n}$ over $k$. These variables may
satisfy polynomial relations over $k$. If $I$ is the ideal generated
by these relations, then
\be\label{canonicalh} k[\mathcal O_{1},\ldots,\mathcal O_{n}]=
k[X_{1},\ldots,X_{n}]/I\, .\ee

An ideal $I$ is said to be \emph{prime} if $ab\in I$ implies that
either $a$ or $b$ is in $I$. The quotient ring \eqref{canonicalh} is
then an integral domain and one can build a field of fractions from it
in the same way as one builds the field of rational numbers $\mathbb
Q$ from the ring of integers $\mathbb Z$.

\section{Foundations}
\label{FoundSec}
\subsection{Generalities}
\label{GeneSec}

We consider a general $\nn=1$ supersymmetric gauge theory in four
dimensions. The lowest components of gauge invariant chiral
superfields are called \emph{chiral operators}. Equivalently, chiral
operators are local gauge invariant operators that commute (in the
case of bosonic operators) or anticommute (in the case of fermionic
operators) with the left-handed supersymmetry charges.

The Lie algebra $\mathfrak g$ of the gauge group decomposes into a
direct sum of $\mathfrak{u}(1)$ factors and simple non-abelian
factors,
\be\label{Lie}\mathfrak g =\mathfrak u(1)\oplus\cdots\oplus\mathfrak
u(1) \oplus_{\alpha}\mathfrak g_{\alpha}\, .\ee
To each non-abelian factor $\mathfrak g_{\alpha}$ is associated a
complex gauge coupling constant
\be\label{gcoupling} \tau_{\alpha} = \frac{\theta_{\alpha}}{2\pi} +
i\frac{4\pi}{g_{\alpha}^{2}}\,\cdotp\ee
In the quantum theory, the gauge couplings run,
\be\label{taumu}\tau_{\alpha}(\mu) =
\frac{i\beta_{\alpha}}{2\pi}\ln\frac{\mu}{\La_{\alpha}}\,\cdotp\ee
The coefficients $\beta_{\alpha}$ can be computed at one loop and the 
higher loop effects are included in the complex scales $\La_{\alpha}$.
These scales, or more conveniently the instanton factors
\be\label{defq} q_{\alpha} = \La_{\alpha}^{\beta_{\alpha}} =
\mu^{\beta_{\alpha}}e^{2i\pi\tau_{\alpha}}\, ,\ee
can be interpreted as being the lowest components of background chiral
superfields \cite{hol1,hol2}. The $q_{\alpha}$ will be denoted
collectively by $\q$.

On top of the $\q$, the theory has parameters $\g=(g_{k})$ that couple
to chiral operators $\mathcal O_{k}$ in the tree-level superpotential,
\be\label{treeW} \wt = \sum_{k}g_{k}\mathcal O_{k}\, .\ee
As for the $\q$, the parameters $\g$ are best viewed as background
chiral operators. 

A fundamental property of the expectation values of chiral operators
is that they depend \emph{holomorphically} on $\g$ and $\q$.
\emph{Solving} the theory means computing the analytic functions
$\langle\mathcal O\rangle (\g,\q)$ for all the chiral operators
$\mathcal O$. We are going to describe some general properties of
these analytic functions below.

\subsection{On the number of vacua}
\label{numbervacsSec}
\subsubsection{With or without a moduli space}

We are interested in models that do not break supersymmetry. For a
generic superpotential \eqref{treeW}, one typically finds a finite
number of supersymmetric vacua. In some special cases, when
\eqref{treeW} has flat directions that are not lifted in the quantum
theory, there is a moduli space of vacua.

A theory with a moduli space can often we obtained from the more
generic case without a moduli space by turning off certain parameters
in \eqref{treeW}. In this situation, the solution with a moduli space
is a special case of the solution with a finite number of vacua.
Independently of this observation, it turns out that the cases with
and without a moduli space can be formally studied along the same
lines. This can be easily understood as follows. A moduli space of
dimension $d$ can be parametrized by $d$ coordinates that correspond
to the expectation values of $d$ massless chiral operators $\mathcal
O_{1},\ldots,\mathcal O_{d}$.\footnote{As will become clear in the
following, the moduli space may have various irreducible components
corresponding to different phases of the theory. The dimension can
vary from one component to the other and thus, strictly speaking, the
discussion in this paragraph applies for each irreducible component
independently.} Once the parameters of the theory and the
$\langle\mathcal O_{i}\rangle$ are fixed, all the other expectation
values are unambiguously determined, up to a possible finite
degeneracy. If we treat the $\langle\mathcal O_{i}\rangle$ for $1\leq
i\leq d$ as the other parameters $\g$ and $\q$, the solution can then
be described as in the case of the theories with a finite number of
vacua.

For the above reasons and if not explicitly stated otherwise, we shall
focus in the following on theories that have a finite number $v$ of
vacua.

\subsubsection{Counting the vacua}
\label{numbervacSec}

Let $|i\rangle$, $1\leq i\leq v$, be the supersymmetric vacua of the
theory. Mathematically, the existence of multiple vacua is equivalent
to the multi-valuedness of the analytic functions $\langle\mathcal
O\rangle (\g,\q)$. Each possible value $\mathcal O_{i}(\g,\q)$
corresponds to the expectation in a vacuum $|i\rangle$,
\be\label{vevi}\langle i|\mathcal O|i\rangle = \mathcal
O_{i}(\g,\q)\, .\ee
The number of vacua is thus equal to the degree of the analytic
functions $\langle\mathcal O\rangle (\g,\q)$. This number cannot
change when the parameters are varied, except at special points where
the expectation values may go to infinity and the associated vacuum
disappears from the spectrum.

From the above remarks it is easy to compute $v$ explicitly in any
given model by looking at the small $q_{\alpha}$ expansion of the
expectation values, which can be straightforwardly obtained from the
explicit solutions. At the classical level, $q_{\alpha}=0$, the vacua
are found by extremizing the tree-level superpotential \eqref{treeW}.
To each classical solution $|a\rangle_{\text{cl}}$ is associated a
certain pattern of gauge symmetry breaking. The Lie algebra $\mathfrak
h^{|a\rangle_{\text{cl}}}$ of the unbroken gauge group in
$|a\rangle_{\text{cl}}$ decomposes as
\be\label{Liebroken}\mathfrak h^{|a\rangle_{\text{cl}}} =\mathfrak
u(1)\oplus\cdots \oplus\mathfrak u(1) \oplus_{\beta}\mathfrak
h^{|a\rangle_{\text{cl}}}_{\beta}\, .\ee
In each simple non-abelian factor $\smash{\mathfrak
h^{|a\rangle_{\text{cl}}}_{\beta}}$, with associated dual Coxeter
number $\smash{h^{\mathsf{V}}(\mathfrak
h^{|a\rangle_{\text{cl}}}_{\beta})}$, chiral symmetry breaking implies
a $\smash{h^{\mathsf{V}}(\mathfrak
h^{|a\rangle_{\text{cl}}}_{\beta})}$-fold degeneracy. The number of
quantum vacua associated to the classical solution \eqref{Liebroken}
is thus given by $\smash{\prod_{\beta}h^{\mathsf{V}}(\mathfrak
h^{|a\rangle_{\text{cl}}}_{\beta})}$. The total number of vacua is
then obtained by summing over all the classical solutions,
\be\label{vform} v =
\sum_{|a\rangle_{\text{cl}}}\prod_{\beta}h^{\mathsf{V}}(\mathfrak
h^{|a\rangle_{\text{cl}}}_{\beta})\, .\ee
We see in particular that $v$ changes precisely when the number of
classical solutions changes. This happens when some of the $g_{k}$ in
\eqref{treeW} vanish and the asymptotic behaviour of the tree-level
superpotential is changed.

\begin{exmp}\label{Ex1}
In the case of the pure gauge theory based on a simple gauge group
$G$, $v=h^{\mathsf{V}}(\mathfrak g)$. For example, for $G=\suN$,
$v=N$. If $G=\uN$, one also has $v=N$, because the $\mathfrak u(1)$
factor in \eqref{Liebroken} does not change $v$.
\end{exmp}

\begin{exmp}\label{Ex2}
Let us consider the $\uN$ gauge theory, with $\Nf$ flavours of quarks
corresponding to chiral superfields $Q^{\mathsf a}_{f}$ and $\tilde
Q_{\mathsf a}^{f}$ in the fundamental and anti-fundamental
representations respectively ($\mathsf a$ and $\mathsf a'$ are gauge
indices and $f$ and $f'$ are flavour indices). Let us choose the
tree-level superpotential to be
\be\label{wtex2} \wt =\tilde Q m Q = \sum_{1\leq f,f'\leq
\Nf}\sum_{1\leq\mathsf a\leq N}\tilde Q_{\mathsf a}^{f}
m_{f}^{\ f'} Q^{\mathsf a}_{f'}\, , \ee
where $m=(m_{f}^{\ f'})$ is an invertible mass matrix. The classical
solutions correspond to $Q=\tilde Q = 0$ and thus to an unbroken gauge
group. The number of vacua is thus $v=N$. Physically, one can
integrate out the quarks and find at low energy a pure $\uN$ gauge
theory.
\end{exmp}

\begin{exmp}\label{Ex3}
Let us now consider the paradigmatic example of the $\uN$ gauge theory
with one adjoint chiral superfield $\phi$ and tree-level
superpotential
\be\label{wtex3} \wt = \Tr W(\phi)\, ,\ee
where $W$ is a polynomial such that
\be\label{Wder} W'(z) = \sum_{k=0}^{d}g_{k}z^{k} =
g_{d}\prod_{i=1}^{d}(z-w_{i})\, .\ee
The classical solutions $|N_{1},\ldots,N_{d}\rangle_{\text{cl}}$ are
labeled by non-negative integers $(N_{1},\ldots,N_{d})$ satisfying
$\sum_{i=1}^{d}N_{i} = N$. The integer $N_{i}$ corresponds to the
number of eigenvalues of the matrix $\phi$ that are equal to $w_{i}$.
The number $v_{\text{cl}}$ of classical vacua is thus equal to the
number of partitions of $N$ by $d$ non-negative integers,
\be\label{vclex} v_{\text{cl}} = \binom{N+d-1}{d-1} =
\frac{(N+d-1)!}{(d-1)!N!}\,\cdotp\ee
To a given classical solution
$|N_{1},\ldots,N_{d}\rangle_{\text{cl}}$, we associate an integer $r$
that counts the number of non-zero $N_{i}$. We call $r$ the
\emph{rank} of the solution (this terminology comes from the fact that
the low energy gauge group in the quantum theory is $\u^{r}$ in this
case). Taking into account a trivial combinatorial factor
$\binom{d}{r}$ corresponding to the choice of the non-zero positive
integers $N_{i}$, there are
\be\label{vclr} v_{\text{cl},\, r} = \binom{d}{r}\binom{N-1}{r-1}\ee
classical solutions of rank $r$, and obviously
$v_{\text{cl}}=\sum_{r=1}^{\min(d,N)}v_{\text{cl},\, r}$.

The Lie algebra of the unbroken gauge group in the classical vacuum
$|N_{1},\ldots,N_{d}\rangle_{\text{cl}}$ is given by
\be\label{algbex3}\mathfrak h^{|N_{1},\ldots,N_{d}\rangle_{\text{cl}}}
= \mathfrak u(1)^{r}\oplus\mathfrak{su}(N_{i_{1}})\oplus\cdots\oplus
\mathfrak{su}(N_{i_{r}})\, , \ee
where the $r$ distinct indices $i_{k}$ correspond to the
$N_{i_{k}}>0$. Equation \eqref{vform} shows that the quantum vacua can
be labeled as $|N_{1},k_{1};\ldots;N_{d},k_{d}\rangle$ where the
integers $k_{i}$ are defined modulo $N_{i}$. The total number of
quantum vacua at rank $r$ is thus given by
\be\label{vrquan} v_{r} = \binom{d}{r}\hat v_{r}(N)\ee
with
\be\label{vhr} \hat v_{r}(N) = \sum_{\sum_{i=1}^{r}N_{i}=N}N_{1}\cdots
N_{r}\, .\ee
It is not difficult to find a generating function for $\hat v_{r}(N)$.
If
\be\label{genfun}
f(x_{1},\ldots,x_{r})=\prod_{i=1}^{r}\frac{x_{i}}{1-x_{i}} =
\sum_{N_{1}\geq 1,\ldots,N_{r}\geq 1}x_{1}^{N_{1}}\cdots
x_{r}^{N_{r}}\, ,\ee
then
\be\label{genfun2} g(x) = \frac{\partial^{r}f}{\partial
x_{1}\cdots\partial x_{r}}\bigl( x_{1}=x,\ldots,x_{r}=x\bigr)
=\frac{1}{(1-x)^{2r}}= \sum_{N\geq r}\hat v_{r}(N) x^{N-r}\ee
and this yields
\be\label{vrhform} \hat v_{r}(N) = \binom{N+r-1}{2r-1}\, .\ee
We list in Table \ref{tablevrhat} the numbers $\hat v_{r}(N)$ for low
values of $N$. These numbers are typically very large, which gives a
first indication of the high level of complexity of the model. The
case $r=1$ corresponds to an unbroken gauge group. The $N$-fold
degeneracy, $\hat v_{1}(N)=N$, is similar to what is found in the pure
gauge theory. The case $r=N$ corresponds to the Coulomb branch with
unbroken gauge group $\u^{N}$. This branch can be made arbitrarily
weakly coupled and there is no chiral symmetry breaking, which
explains why $\hat v_{N}(N) =1$. 

Finally, let us note that the number of vacua at rank $r$
\eqref{vrquan}, or the total number of vacua
$v=\sum_{r=1}^{\min(d,N)}v_{r}$, changes only when the degree of the
tree-level superpotential changes, which occurs when $g_{d}=0$.
\begin{table}\label{tablevrhat}
\be\nonumber
\begin{matrix}
\hphantom{N=} &\hphantom{1}\ \vline & r=1 & 2 & 3 & 4 & 5 & 6 & 7\\
\hline N=&1\ \vline & 1 & & & & & &\\
& 2\ \vline & 2& 1& & & & &\\
& 3\ \vline & 3& 4& 1& & & &\\
& 4\ \vline & 4& 10& 6& 1& & &\\
& 5\ \vline & 5& 20& 21& 8& 1& &\\
& 6\ \vline & 6& 35& 56& 36& 10& 1&\\
& 7\ \vline & 7& 56& 126&120 & 55& 12&1\\
\end{matrix}
\ee
\caption{Values of $\hat v_{r}(N)$ for $1\leq N\leq 7$ and $1\leq
r\leq N$.}   
\end{table}
\end{exmp}

\begin{exmp}\label{Ex4}
Our last example is the $\uN$ gauge theory with one adjoint chiral
superfield $\phi = (\phi^{\mathsf a}_{\ \mathsf b})$, $\Nf$ flavours
of quarks $Q^{\mathsf a}_{f}$ and $\tilde Q_{\mathsf a}^{f}$ and
tree-level superpotential
\be\label{wtex4} \wt = \frac{1}{2}\mu\Tr \phi^{2} + \tilde Q_{\mathsf
a}^{f} m_{f}^{\ f'}(\phi)^{\mathsf a}_{\ \mathsf b} Q^{\mathsf b}_{f'}\,
.\ee
The matrix-valued polynomial $m_{f}^{\ f'}(\phi)$ is chosen to be
\be\label{defmff}m_{f}^{\ f'}(\phi) = \delta^{f'}_{f}\bigl(\phi -
m_{f})\, .\ee
There is no difficulty in considering more general possibilities, with
arbitrary polynomial $m_{f}^{\ f'}(\phi)$ and a general term $\Tr
W(\phi)$ instead of $\frac{1}{2}\mu\Tr\phi^{2}$ in $\wt$, but the
cases \eqref{wtex4} and \eqref{defmff} are enough to illustrate all
the relevant physics of the models (we shall come back on this point
in Section \ref{HiggsConfSec}). The classical solutions can be easily
obtained by extremizing \eqref{wtex4}. It is found that the
eigenvalues of the matrix $\phi$ can be either equal to zero (which
extremizes $W(z)=\frac{1}{2}\mu z^{2}$) or equal to the $m_{f}$.
Moreover, at most one eigenvalue of $\phi$ can be equal to any given
$m_{f}$. The solutions are thus labeled as
$|n;\nu_{1},\ldots,\nu_{\Nf}\rangle_{\text{cl}}$, with $n$ denoting
the number of zero eigenvalues and $\nu_{f}=0$ or $1$ according to
whether there is an eigenvalue equal to $m_{f}$ or not. Taking into
account the constraint $n+\sum_{f}\nu_{f}=N$, we find that the total
number of classical vacua is given by
\be\label{vclex4} v_{\text{cl}} =
\sum_{k=0}^{\min(\Nf,N)}\binom{\Nf}{k}\, . \ee
In particular,
\be\label{vclex4p} v_{\text{cl}} = 2^{\Nf}\quad\text{for}\ \Nf\leq N\,
.\ee

In $|n;\nu_{1},\ldots,\nu_{\Nf}\rangle_{\text{cl}}$, the quarks have
non-zero expectation values when some of the $\nu_{f}$ are non-zero
and the gauge group is Higgsed down to $\text U(n)$. In the quantum
theory, there are thus
\be\label{vex4} v_{1} = \sum_{k=0}^{\min(\Nf,N)}(N-k)\binom{\Nf}{k}\,
\ee
rank one vacua, corresponding to $n\geq 1$ and a low energy gauge
group $\u$. In particular,
\be\label{vex4p} v_{1} =(2N-\Nf) 2^{\Nf-1}\quad\text{for}\ \Nf\leq N\,
.\ee
If $\Nf\geq N$, there are also $v_{0}=\binom{\Nf}{N}$ rank zero vacua
in which the gauge group is completely broken.
\end{exmp}

The model \eqref{wtex4} is ideal to study the relation between the
confining and Higgs regimes as described in Section \ref{introappSec}.
Consider for example the case $\Nf = N-1$ (all the other cases display
similar phenomena). This model has $n\binom{N-1}{N-n}$ quantum vacua
corresponding to classical solutions with unbroken gauge group $\text
U(n)$, for any $1\leq n\leq N$. When $n=N$, the gauge group is
unbroken and we find the usual $N$ strongly coupled ``confining''
vacua, similar to the vacua of the pure gauge theory. In particular,
classically, the quark fields have zero expectation values in these
vacua. On the other hand, when $n=1$, the gauge group is completely
broken (except for the trivial global $\u$ factor in $\uN$) by the
quarks expectation values and we find the weakly coupled ``Higgs''
vacuum. Intermediate values of $n$ correspond to partially Higgsed
vacua. At the classical or semi-classical levels, vacua with different
values of $n$ look completely different, and in particular it is
impossible to interpolate smoothly between them by varying the
parameters. However we shall prove in Section \ref{HiggsConfSec} that
in the full quantum theory the $(N+1)2^{N-2}$ vacua of this model,
with all the possible patterns of gauge symmetry breaking
$\uN\rightarrow\text U(n)$ for $1\leq n\leq N$, are actually in the
same phase!

\subsection{The theory space and monodromies}
\subsubsection{Global coordinates on theory space}

The parameters $\g$ and $\q$ play a distinguished r\^ole. For example,
the definition of the $q_{\alpha}$ in \eqref{defq} is motivated by the
$2\pi$ periodicity in the angles $\theta_{\alpha}$ given in
\eqref{gcoupling}. The precise statement is as follows.
\begin{prop}\label{propParameters}
The parameters $(\g,\q)$ are good global coordinates in theory space.
In other words, the theory is uniquely defined once we choose $\g$ and
$\q$ and conversely, to a given theory corresponds a unique choice of
$\g$ and $\q$.
\end{prop}
\noindent For example, the theories corresponding to the angles
$\theta_{\alpha}$ and $\theta_{\alpha}+2\pi n_{\alpha}$, for any
integers $n_{\alpha}$, must be the same and are associated with the
same values of $\q$. On the other hand, fractional powers of the
instanton factors are not good coordinates since for example
$q_{\alpha}^{1/2}$ and $-q_{\alpha}^{1/2}$ both correspond to the same
theory. Similarly, $q_{\alpha}^{2}$ is not a good coordinate, because
two distinct theories, corresponding to $q_{\alpha}$ and
$-q_{\alpha}$, both have the same $q_{\alpha}^{2}$.

How can we prove Prop.\ \ref{propParameters}? In perturbation theory,
it is a trivial statement. Beyond perturbation theory, the standard
argument is to invoke instantons. Instanton contributions are indeed
proportional to some powers of the $q_{\alpha}$. However, this
argument, in its simplest form, is not correct. The instanton calculus
is a semi-classical approximation and thus applies only at weak
coupling. On the other hand, Prop.\ \ref{propParameters} is supposed
to be valid in all cases, including in theories like the pure gauge
theories that have strongly coupled vacua.

Providing a full proof of Prop.\ \ref{propParameters} requires a
rigorous, axiomatic definition of the super Yang-Mills theories. This
definition doesn't exist for arbitrary correlators, but it does exist
in the case of the chiral sector we are interested in
\cite{mic1,mic2,mic3}. The validity of Prop.\ \ref{propParameters} is
then a direct consequence of the formalism. We cannot provide the full
details here, but the idea is as follows. It turns out that the full
information on the chiral sector can be encoded in a microscopic
quantum effective superpotential $W_{\text{mic}}$ that can always be
computed in the instanton approximation for reasons explained in
details in \cite{mic1}. The physics is described by the critical
points of the microscopic superpotential. The instanton series for
$W_{\text{mic}}$ has a finite radius of convergence. The critical
points that are located inside the radius of convergence correspond to
weakly coupled vacua and the other critical points correspond to
strongly coupled vacua. As discussed in the next subsection, in these
vacua the expectation values are not $2\pi$ periodic in the
$\theta_{\alpha}$, but this is still consistent with Prop.\
\ref{propParameters}.\footnote{The $2\pi$ periodicity in the $\theta$
angles is conjectured to be valid in non-supersymmetric theories as
well. A rigorous justification of this fact must await the rigorous
construction of the quantum gauge theories. A heuristic argument in
favour of $2\pi$ periodicity is that the definition of the theory is
essentially a UV problem. For asymptotically free gauge theories, the
UV is arbitrarily weakly coupled, and thus arguments based on
instantons are likely to be correct for this particular purpose (even
though they do not give a sensible approximation to the physical
correlators). The microscopic construction of the supersymmetric
models in \cite{mic1,mic2,mic3} is perfectly consistent with this
heuristic idea.}

\subsubsection{Monodromies amongst the vacua}
\label{monodSec}

Let us first consider a weakly coupled vacuum $|i\rangle$ in which the
instanton approximation is valid. The analytic function $\mathcal
O_{i}(\g,\q)$ is then given by a power series in the $q_{\alpha}$. In
particular, $\mathcal O_{i}(\g,\q)$ is $2\pi$ periodic in the $\theta$
angles,
\be\label{noperm} \mathcal O_{i}(\g, e^{2i\pi n_{\alpha}} q_{\alpha})
= \mathcal O_{i}(\g,\q)\ee
for any integers $n_{\alpha}$. 

Prop.\ \ref{propParameters} allows a more general behaviour than
\eqref{noperm} and actually the $2\pi$ periodicity of the correlators
can be violated \cite{thetafer}. To understand the most general
possibility, let us start for some values $(\g,\q)$ of the parameters
and perform an analytic continuation along a closed loop in theory
space. Prop.\ \ref{propParameters} implies that the theory and thus
the set of vacua $\{|i\rangle\}$ must be the same before and after the
analytic continuation. In other words, if $|i\rangle$ is transformed
into $|i\rangle'$ under the analytic continuation, then there must
exist a permutation $\sigma$ such that
\be\label{permvac} |i\rangle' = |\sigma(i)\rangle\, .\ee
Equivalently, the analytic functions $\mathcal O_{i}$ transform as
\be\label{analcont} \langle i|\mathcal O|i\rangle =\mathcal
O_{i}(\g,\q)\longrightarrow\mathcal O_{\sigma(i)}(\g,\q)=\langle
\sigma(i)|\mathcal O|\sigma(i)\rangle\, .\ee
When strongly coupled vacua are present the permutation $\sigma$ can
be non-trivial.

Performing $2\pi$ shifts in the $\theta$ angles correspond to
particular closed loops in theory space and thus \eqref{analcont}
implies that in general \eqref{noperm} is replaced by
\be\label{yesperm} \mathcal O_{i}(\g, e^{2i\pi n_{\alpha}} q_{\alpha})
= \mathcal O_{\sigma(i)}(\g,\q)\, ,\ee
for some permutation $\sigma$ that depends on the integers
$n_{\alpha}$. We see explicitly that vacuum expectation values are not
necessarily $2\pi$ periodic in the $\theta$ angles. In some simple
cases, as in the pure gauge theories, the vacua $|i\rangle$ and
$|\sigma(i)\rangle$ are related by broken symmetry generators and are
thus physically equivalent. However, this is not the case in general:
\emph{the physics (i.e.\ the physical measurements) of the theories is
not, in general, $2\pi$ periodic in the $\theta$ angles.} The meaning
of Prop.\ \ref{propParameters} is that the theory must be $2\pi$
periodic \emph{as a whole}, when all the vacua are taken into account
at the same time. Note that in the special cases where there is only
one vacuum, or when all the vacua are related by broken symmetry
generators, then the physics is automatically $2\pi$ periodic. This is
what is believed to happen in non-supersymmetric models.

\subsection{The polynomial equations}
\label{poleqSec}

Prop.\ \ref{propParameters} can be used to derive a very useful
property of the analytic functions $\langle\mathcal O\rangle (\g,\q)$.
\begin{thm}\label{thmPolynomial}
For any supersymmetric gauge theory with a finite number $v$ of vacua,
there exists a ring $\pring$, called the ring of parameters, which is
a subring of the ring of entire functions in the parameters $\g$ and
$\q$, such that the expectation value of any chiral operator $\mathcal
O$ satisfies a degree $v$ polynomial equation with coefficients in
$\pring$:
\be\label{fulls} P_{\mathcal O}(\langle\mathcal O\rangle) = 0\,
,\quad P_{\mathcal O}\in\pring[X]\, ,\ \deg P_{\mathcal O}=v\, .\ee
Moreover, if there exists a $\u$ symmetry for which the charges of the
fundamental chiral fields and of the parameters $\g$ and $\q$ are all
strictly positive, then $\pring=\mathbb C[\g,\q]$ is the polynomial
ring in the variables $\g$ and $\q$.
\end{thm}
\subsubsection{Discussion of the theorem}
\label{DiscussThmSec}
\begin{table}
\label{asign1}
\be\nonumber
\begin{matrix}
& \vartheta & \lambda & \phi & \psi &\mathcal O_{k} & g_{k} & q \\
\u_{\text R} & 3/2& 3/2 & 1 & -1/2& \delta_{k}
& 3-\delta_{k} & 3N - \frac{1}{2}\sum_{\chi}I_{\chi} \, .
\end{matrix}\ee
\caption{Charge asignments for the $\u_{\text R}$ symmetry of a
general gauge theory. The variables $\vartheta$ are the superspace
coordinates, $\lambda$ is the gluino, $\phi$ is the lowest component
of an arbitrary scalar chiral superfield, $\psi$ its supersymmetric
partner, $\mathcal O_{k}$ a chiral operator in the tree-level
superpotential \eqref{treeW}, $g_{k}$ the associated coupling, and $q$
an instanton factor. The charge of $q$ is given by the usual chiral
anomaly, the sum over $\chi$ corresponding to a sum over all the
spinor fields coupled to the simple factor of the gauge group
associated with $q$, $I_{\chi}$ being the index of the gauge group
representation in which $\chi$ transforms.}
\end{table}

The non-trivial content of Th.\ \ref{thmPolynomial} is not in the
existence of algebraic equations satisfied by the expectation values
(by itself this is an empty statement), but in the fact that \emph{the
coefficients of these algebraic equations are contrained to be
elements of a particular ring}. In this sense, the analytic functions
$\langle\mathcal O\rangle$ are similar with respect to the ring
$\pring$ to numbers like $\sqrt{2}$ with respect to the ring of
integers $\mathbb Z$.

For many purposes the ring $\pring$ can be replaced by it field of
fractions $\pfield$,\footnote{The field of fraction exists because
$\pring$, being a subring of the ring of entire functions, is an
integral domain.} that we shall call the \emph{field of parameters}.
One interest in using $\pfield$ instead of $\pring$ is that the
polynomials in Th.\ \ref{thmPolynomial} can be constrained to be
monic, i.e.\ of the form $P_{\mathcal O}(X) = X^{v}+\cdots$ For
example, if $\pring=\mathbb C[\g,\q]$, then $\pfield=\mathbb C(\g,\q)$
is the field of rational functions in the parameters $\g$ and $\q$. In
this case, an equation with coefficients in $\pfield$ actually
automatically yields an equation with coefficients in $\pring$, since
we can always clear the denominators of the coefficients by
multiplying by their least common multiple.

In the following, the reader may always assume that $\pring=\mathbb
C[\g,\q]$ is the polynomial ring. The assumption in Th.\
\ref{thmPolynomial} that ensures that this is the case is a relatively
minor technical requirement satisfied in a lot of models. For example,
all the super Yang-Mills theories have a $\u_{\text R}$ symmetry
defined by identifying the $\u_{\text R}$ charges with the canonical
dimensions of the chiral superfields, see Table \ref{asign1}. This
symmetry satisfies the conditions of the theorem provided the model is
asymptotically free (which yields a positive charge for $q$) and the
tree-level superpotential includes only super-renormalizable terms
(which corresponds to positive charges for the $g_{k}$). For instance,
Ex.\ \ref{Ex2} in Section \ref{numbervacSec} is of this type.
Renormalizable (but not super-renormalizable) terms, associated with
couplings of zero $\u_{\text R}$ charge, can also be included in many
cases, because the renormalizable couplings can often be absorbed in
suitable field redefinitions (in other words, the dependence in these
couplings can be straightforwardly derived by simple rescalings). For
instance, this is what we have done in Ex.\ \ref{Ex4} by choosing the
leading term in \eqref{defmff} to be $\delta_{f'}^{f}$ instead of
$g\delta_{f'}^{f}$ for an arbitrary coupling $g$. Even models
including non-renormalizable terms\footnote{These terms occur in a
string theory context where the field theory is viewed as a low energy
approximation and yield interesting physics. Even from a purely field
theoretic point of view it is perfectly consistent to include them
when one focuses on the chiral sector of the theory. This is so
because the necessary counterterms are governed by the UV cut-off
which is a real parameter and thus does not affect the chiral sector.}
often satisfy the assumption in the theorem. For instance, Ex.\
\ref{Ex3} does have non-renormalizable couplings in the tree-level
superpotential \eqref{wtex3} when $\deg W >3$. However, the model has
another R-symmetry $\u_{\text R}'$ with charge asignments
\be\label{asign2}
\begin{matrix}
& \vartheta & \lambda & \phi & \psi & g_{k} & q\\
\u_{\text R}' & 1 & 1 & 0 & -1 & 2 & 0 &.
\end{matrix}\ee
It is always possible to find a linear combination of $\u_{\text{R}}$
and $\u_{\text R}'$ that satisfies the conditions of the theorem. 

On the other hand, in theories with zero $\beta$ functions, the ring
$\pring$ can include arbitrary power series in the instanton factors.
For example, if the theory has a S-duality, the coefficients of the
polynomials of Th.\ \ref{thmPolynomial} typically involve modular
forms.

\subsubsection{Proof of the theorem}
\label{PSec}

Let $\mathcal O$ be a chiral operator, $\mathcal O_{i}=\langle
i|\mathcal O|i\rangle$ and consider the monic polynomial
\be\label{POdef} \hat P_{\mathcal O}(X) = \prod_{i=1}^{v}\bigl( X -
\mathcal O_{i}(\g,\q)\bigr) = X^{v} + \sum_{k=1}^{v} \hat a_{k}(\g,\q)
X^{v-k} \, .\ee
By construction, $\hat P_{\mathcal O}(\langle\mathcal
O\rangle) = 0$.

Let us perform an analytic continuation along an arbitrary closed loop
in the space of parameters $(\g,\q)$. From \eqref{analcont}, we find
that
\be\label{POacont} \hat P_{\mathcal O}(X)\longrightarrow
\prod_{i=1}^{v}\bigl( X - \mathcal O_{\sigma (i)}(\g,\q)\bigr) =
\prod_{i=1}^{v}\bigl( X - \mathcal O_{i}(\g,\q)\bigr)=\hat P_{\mathcal
O}(X)\, ,\ee
and thus the coefficients $\hat a_{k}(\g,\q)$ defined in \eqref{POdef}
are \emph{single}-valued analytic functions of $\g$ and $\q$.

Singularities of the functions $\hat a_{k}$ can only occur when some
of the vacua disappear from the spectrum. From the discussion of
Section \ref{numbervacSec}, we know that the positions of the
singularities can thus be derived by a purely classical analysis.
Moreover, near a singular point, the effective superpotential
evaluated in the vacuum that disappears at the singularity is
arbitrarily large. The leading singular behaviour of the expectation
values can then be obtained from a classical analysis as well. For
tree-level superpotentials of the form \eqref{treeW}, this is always
given by a power-law divergence. The conclusion is that the $\hat
a_{k}(\g,\q)$ are meromorphic functions with a finite number of poles.
Multiplying $\hat P_{\mathcal O}$ by a suitable polynomial in the
parameters $\g$ and $\q$ to clear up these poles, we obtain the
polynomial $P_{\mathcal O}$ of Th.\ \ref{thmPolynomial}.

Let us now assume that there exists a $\u$ symmetry for which the
charges of the fundamental chiral fields and of the parameters $\g$
and $\q$ are all strictly positive. An arbitrary chiral operator
$\mathcal O$ can be written as a sum of operators of strictly positive
$\u$ charges. Let $\delta>0$ be the greatest of these charges. If we
asign to the dummy variable $X$ in \eqref{POdef} the charge $\delta$,
then $\hat P_{\mathcal O}$ is a sum of terms whose charges are bounded
by $v\delta$. The polynomial $P_{\mathcal O}$, which is obtained from
$\hat P_{\mathcal O}$ by multiplying by a polynomial in $\g$ and $\q$,
is thus also a sum of terms of given $\u$ charges, these charges being
bounded by a certain strictly positive integer. Let us write
\be\label{POexp} P_{\mathcal O}(X) = \sum_{k=0}^{v}a_{k}(\g,\q)
X^{v-k}\, .\ee
The $a_{k}$ are entire functions and can thus be expanded as power
series in $\g$ and $\q$. From the above discussion, they have a
maximum $\u$ charge. Since the variables $\g$ and $\q$ have strictly
positive $\u$ charges, the power series must terminate after a finite
number of terms, and thus $a_{k}\in\mathbb C[\g,\q]$.

\subsubsection{The power of the polynomial equations}
\label{powerSec}
\begin{prop}\label{solPolProp} The full solution of the model, i.e.\
the full set of expectation values $\langle i|\mathcal O|i\rangle$ for
all chiral operators $\mathcal O$ and all the vacua $|i\rangle$,
$1\leq i\leq v$, can be derived from the knowledge of a finite number
of the polynomial equations of Theorem \ref{thmPolynomial}.
\end{prop}
\noindent Let us explain the significance of this result. If one picks
a given operator $\mathcal O$, then by construction there are $v$
solutions to the polynomial equation $P_{\mathcal O}=0$, corresponding
to the $v$ expectation values $\langle i|\mathcal O|i\rangle$. Which
solution corresponds to which vacuum is a matter of convention and we
can always choose to label the vacua according to a particular
labeling of the roots of $P_{\mathcal O}$. Let us assume that we have
chosen a particular labeling. Let us now consider another operator
$\mathcal O'$. One can find the \emph{unordered} set of $v$
expectation values of $\mathcal O'$ by solving $P_{\mathcal O'}=0$.
However, we do not know which root corresponds to which vacuum. This is
no longer a matter of arbitrary choice, since the vacua have already
been labeled. So we see that the knowledge of $P_{\mathcal O}$ and
$P_{\mathcal O'}$ is not enough to derive the expectation values of
$\mathcal O$ and $\mathcal O'$, there remains an ambiguity
corresponding to the permutation of the vacua. Of course, additional
constraints can be found by considering more polynomials, like
$P_{\mathcal O\mathcal O'}$ for example. Prop.\ \ref{solPolProp}
states that all the ambiguity, for all the expectation values, can be
cleared up by considering a finite set of equations of the form
\eqref{fulls}. The proof of this result will be given in
\ref{powerbisSec} after more technical tools are introduced.

\subsubsection{Simple examples}
\label{seSec}

\begin{exmp}\label{pureEx}
In the case of the pure gauge theory based on a simple gauge group
$G$, the most general chiral operator is a polynomial in the glueball
operator $S$, defined in terms of the super field strength
$W^{\alpha}$ as
\be\label{Sdef} S = -\frac{1}{16\pi^{2}}\Tr W^{\alpha}W_{\alpha}\,
.\ee
The expectation value of this operator satisfies the equation
\be\label{Seq} \langle S\rangle ^{h^{\mathsf{V}}} = q\, ,\ee
where $h^{\mathsf{V}}$ is the dual Coxeter number of $G$. Thus the
polynomial for $S$ is simply
\be\label{PS} P_{S}(X) = X^{h^{\mathsf{V}}} - q \in\mathbb C[q][X]\,
.\ee
\end{exmp}

\begin{exmp}\label{QCDEx}
In the model \eqref{wtex2}, the polynomial for the glueball operator
is simply $P_{S}(X) = X^{N} - q\det m$. The mesonic operator
$M_{f'}^{\ f}=\tilde Q^{f}Q_{f'}$ expectation values also satisfy
degree $N$ algebraic equations with coefficients in $\mathbb
C[q,m_{f}^{\ f'}]$ that straightforwardly follow from the relation
$\langle M_{f'}^{\ f}\rangle = (m^{-1})_{f'}^{\ f}\langle S\rangle$.
\end{exmp}

\subsubsection{A clarifying remark}

Let us here stress a point that has been at the origin of some
considerable confusion in the literature. The fact that the
coefficients of the polynomials $P_{\mathcal O}$ are polynomials in
the instanton factors (and not, for example, in arbitrary fractional
powers of these factors), might lead one to believe that the result
relies on some semi-classical instanton analysis. \emph{This is not
true}. The arguments that we have used to derive the result are valid
in the full strongly coupled quantum theory. The fact that only
integer powers of $q$ enter in the coefficients of $P_{\mathcal O}$
comes from an argument based on analyticity and \emph{not} from an
argument based on a weakly coupled approximation. In particular,
\emph{the coefficients of the polynomials $P_{\mathcal O}$ cannot be
computed in general from a straightforward instanton calculation}.
Another facet of this subtlety is that the expectation values, which
are the solutions of the polynomial equations $P_{\mathcal O}=0$,
usually do not have expansions in integer powers of $q$.

For example, the fact that only $q$ enters the equation \eqref{Seq}
suggested in the old literature that the relation could be derived by
a direct instanton calculation in the pure gauge theory and this
yielded some inconsistencies. This is not surprising. The pure gauge
theory is strongly coupled and \eqref{Seq} \emph{cannot} be derived by
a direct semi-classical calculation in this theory.

\subsection{The chiral ring}
\label{CRsSec}

We are now ready to define the fundamental notion of the quantum
chiral ring. This concept is well-known, but a precise definition in
the non-perturbative quantum theory does not seem to have appeared
previously in the literature. Much more importantly, the full
significance of this notion has not been fully appreciated and its
power was used only very recently \cite{CRcons}. Understanding the
structure of the chiral ring will give us the keys to understanding
the phase structure of the models.

Our definition in Section \ref{CRdefSec} of the quantum chiral ring is
motivated by the following two fundamental properties.
\begin{prop}\label{BasicProp}
(i) The full solution of the theory in the chiral sector is coded in
the chiral ring $\ring$, i.e.\ we can compute the analytic functions
$\langle\mathcal O\rangle$, for all the chiral operators $\mathcal O$,
from the knowledge of the ring $\ring$. (ii) The chiral ring contains
only physical information.
\end{prop}
\noindent Presenting the solution of a gauge theory via an algebraic
structure like a ring may be unfamiliar. The main interest in doing so
is that the ring $\ring$ does not contain any unphysical,
``scheme-dependent'' information. On the other hand, and as will
become clear in the examples below, the usual ways of presenting the
solutions, for example using effective superpotentials or generating
functions for expectation values, do contain a lot of unphysical
information that can obscure the physics.

\subsubsection{On the classical chiral ring}
\label{clCRSec}

Let us start by reviewing the simple notion of the classical chiral
ring. The construction starts by building all the chiral operators by
forming appropriate gauge invariant polynomials in the elementary
chiral fields. A very important property, that follows immediately
from the fact that a field theory has only a finite number of
elementary fields, is that the most general chiral operator $\mathcal
O$ can be written as a polynomial $\rho_{\mathcal O}$ in a
\emph{finite} number of generators,
\be\label{chiraldev1} \mathcal O = \rho_{\mathcal O}(\mathcal
O_{1},\ldots,\mathcal O_{m})\, .\ee
The generators satisfy algebraic identities that come from their
definitions in terms of the gauge-variant elementary fields (these
identities are called sygyzies). Moreover, there are relations that
follow from the extremization of the tree-level superpotential (the
so-called $F$-term conditions). A standard definition of the classical
chiral ring is then given by considering only the bosonic generators
$\mathcal O_{1},\ldots,\mathcal O_{n}$ and by taking the quotient of
the polynomial ring $\mathbb C[X_{1},\ldots,X_{n}]$ in $n$ variables
with the ideal $I$ generated by the set of all the above-mentioned
relations,
\be\label{Rclstandard} \ring_{\text{cl,\,standard}}  =
\mathbb C[X_{1},\ldots,X_{n}]/I\, .\ee
Sometimes, one considers only generators built from bosonic elementary
superfields, thus excluding fermion bilinears for
examples.\footnote{Discarding fermionic variables is justified at the
classical level since fields build from them will automatically have
zero classical expectation values.} In spite of the fact that
$\ring_{\text{cl,\,standard}}$ is a purely classical object, it can
have a rather complex and interesting structure. For example, in the
case of theories that are built in string theory by putting D-branes
at Calabi-Yau singularities, the classical chiral ring encodes in a
very interesting way the Calabi-Yau geometry and many additional
useful informations \cite{Rclpapers}.

In the standard approach, the polynomials in \eqref{chiraldev1} have
coefficients in the field of complex numbers and the parameters $\g$
of the classical theory are simply considered to be complex numbers as
well. This point of view is insufficient for our purposes, which is to
be able to reconstruct all the expectation values (for the moment in
the classical theory) as functions of $\g$ from the structure of the
chiral ring only. To do that, one must consider the parameters $\g$ to
be ``dummy variables,'' or in other words to include them as new
generators in the chiral ring. In this point of view, the polynomials
in \eqref{chiraldev1} will be elements of the polynomial ring $\mathbb
C[\g][X_{1},\ldots,X_{n}]$, and we define
\be\label{RclNstandard} \ring_{\text{cl}}  =
\mathbb C[\g][X_{1},\ldots,X_{n}]/I\, ,\ee
where $I$ is now the ideal generated by all the relations between the
generators that are polynomials with coefficients in $\mathbb C[\g]$.
This definition is sensible because it turns out that all the syzygies
and all the $F$-term constraints are equivalent to polynomial
constraints \emph{with coefficients in $\mathbb C[\g]$}. This is a
crucial point, that we are going to develop further in the general
case of the non-perturbative quantum theory.

\subsubsection{The definition of the quantum chiral ring}
\label{CRdefSec}

To define the chiral ring at the quantum level, with the properties
listed in Prop.\ \ref{BasicProp} in mind, we cannot, as we have just
done in the classical context, refer to the gauge-variant elementary
fields of the theory. Indeed, the physical content of the theory is
entirely coded in the gauge invariant variables. In particular, at the
quantum level, we want to be able to describe situations where
different classical theories, with different gauge-variant elementary
fields and/or gauge groups, can yield physically equivalent ``dual''
quantum theories.

The only data that we must borrow from the classical theory is a list
of chiral operators $\mathcal O_{1},\ldots,\mathcal O_{n}$ that form a
set of generators for all the chiral operators of the theory (note
that the identity operator is always present and in general we do not
include it explicitly in the list of generators). This can be seen as
a basic axiom of what we mean by quantizing a given classical theory.
If $\pring$ is the ring of parameters (see Th.\ \ref{thmPolynomial}),
we define the most general chiral operator of the theory to be any
finite sum of finite products of the generators $\mathcal O_{i}$ with
coefficients in $\pring$,
\be\label{COpgen} \mathcal O = \rho_{\mathcal O}(\mathcal
O_{1},\ldots,\mathcal O_{n})\, ,\quad \rho_{\mathcal
O}\in\pring[X_{1},\ldots,X_{n}]\, .\ee
\begin{defn}\label{oprelDef} 
Let $\mathcal O^{(1)},\ldots,\mathcal O^{(p)}$ be $p$ chiral
operators, i.e.\ operators of the form \eqref{COpgen}. An
\emph{operator relation} between the $\mathcal O^{(i)}$ is a
polynomial equation of the form
\be\label{oprelpol1} P(\mathcal O^{(1)},\ldots,\mathcal O^{(p)}) = 0\,
,\quad P\in\pring[X_{1},\ldots,X_{p}]\, ,\ee
such that $P(\langle i|\mathcal O^{(1)}|i\rangle,\ldots,\langle
i|\mathcal O^{(p)}|i\rangle)$ identically vanishes in all the vacua
$|i\rangle$ of the theory.
\end{defn}
\noindent Note that this definition is unambiguous because of the
well-known factorization of chiral operators expectation values,
\be\label{cluster} \langle\mathcal O\mathcal O'\rangle =
\langle\mathcal O\rangle\langle\mathcal O'\rangle\, ,\ee
which follows from the space-time independence of the chiral
correlators and from the cluster decomposition principle. In
particular,
\be\label{clustercons} P\bigl(\langle i|\mathcal
O^{(1)}|i\rangle,\ldots,\langle i|\mathcal O^{(p)}|i\rangle\bigr) =
\bigl\langle i\big|P(\mathcal O^{(1)},\ldots, \mathcal
O^{(p)})\big|i\bigr\rangle\, .\ee
An operator relation in the sense of Def.\ \ref{oprelDef} thus has two
basic properties: first it is a relation valid in all the vacua of the
theory; second it is a polynomial relation with coefficients in
$\pring$.
\begin{defn}\label{CRDef} 
The quantum chiral ring $\ring=\pring[\mathcal O_{1},\ldots,\mathcal
O_{n}]$ is the ring of all chiral operators of the form
\eqref{COpgen}, taking into account all the operator relations of the
form \eqref{oprelpol1}. In other words, there is a canonical
surjective ring homomorphism from the polynomial ring
$\pring[X_{1},\ldots,X_{n}]$ onto $\ring$ obtained by mapping $X_{i}$
to $\mathcal O_{i}$. The kernel of this mapping is the ideal $\ideal$
generated by all the operator relations and $\ring$ is isomorphic to
the ring quotient
\be\label{CRform} \ring = \pring[X_{1},\ldots,X_{n}]/\ideal\, .\ee
\end{defn}
\subsubsection{The perturbative chiral ring}
\label{pertCRSec}

It is useful to define the notion of a perturbative chiral ring
$\ring_{\text{pert}}$. The motivation behind this concept is to make
precise the notion of quantum corrections: the quantum corrections are
non-trivial is $\ring$ and $\ring_{\text{pert}}$ are not isomorphic
and are trivial otherwise.

In perturbation theory, the standard non-renormalization theorem
ensures that the chiral operators expectation values are not quantum
corrected. This motivates the following definition.
\begin{defn}\label{CRpertDef}
The perturbative chiral ring $\ring_{\text{pert}}$ is defined as the
quantum chiral ring in Def.\ \ref{CRDef}, except that we set to zero
all the instanton factors in the quantum operator relations,
\be\label{Rpertdef}\ring_{\text{pert}} = \ring/(\q)[\q]\, .\ee
\end{defn}
\noindent In many (but not necessarily all) cases, the perturbative
chiral ring simply coincides with the classical chiral ring defined in
\eqref{RclNstandard}, except that the variables $\q$ are added,
\be\label{RclRpert} \ring_{\text{pert}} = \ring_{\text{cl}}[\q]\, .\ee
\subsubsection{Simple algebraic properties of the quantum chiral ring}
\label{AlgpropSec}

The rings $\ring$ in Def.\ \ref{CRDef} are not generic rings but have
some special properties that we now discuss.

\smallskip

\noindent\emph{The ring $\ring$ is commutative}.\\
In general, gauge invariant chiral operators can include both bosonic
and fermionic operators. However, fermionic operators automatically
have zero expectation values in a Lorentz-invariant theory. Our
definition of the chiral ring then implies that only the bosonic
operators need to be taken into account and thus $\ring$ is always
commutative. Let us note that this requirement could be lifted by
introducing Lorentz-violating couplings to the fermionic chiral
operators in the tree-level superpotential. It is straightforward to
develop a generalized theory that includes these terms but, since we
are not aware of any useful physical application of such a
construction, we shall restrict ourselves to Lorentz invariant
theories.

\smallskip

\noindent\emph{The ring $\ring$ has no nilpotent element}.\\
A nilpotent element $x$ is a non-zero element such that $x^{r} = 0$
for some integer $r>1$. However, $x^{r}=0$ in $\ring$ means that
$\langle x^{r}\rangle = \langle x\rangle^{r}=0$ in all the vacua of
the theory. This in turn implies that $\langle x\rangle = 0$ in all
the vacua and thus that $x=0$ in $\ring$.

The fact that $\ring$ has no nilpotent element can be expressed in
terms of the ideal $\ideal$ of operator relations. In general, for any
ideal $I$ of a commutative ring $A$, one defines the radical $r(I)$ of
$I$ to be the set of elements $x$ of $A$ such that $x^{r}\in I$ for
some $r\geq 1$,
\be\label{radicaldef} r(I) = \{ x\in A \mid \exists\, r>0\, ,\ x^{r}\in
I\}\, .\ee
It is straightforward to check that $r(I)$ is itself an ideal, that
$r(r(I)) = r(I)$ and that $A/I$ has no nilpotent element if and only
if $r(I)=I$ in which case we say that $I$ is a radical ideal. Thus the
ideal $\ideal $ of operator relations is radical.

Let us note that the classical or perturbative rings as defined in
\ref{clCRSec} and \ref{pertCRSec} can have nilpotent elements. Thus
\emph{perturbative (or classical) chiral rings are not special cases
of quantum chiral rings.} This exhibits the singular nature of the
classical limit and will be illustrated in Ex.\ \ref{algpropEx} below.

\smallskip

\noindent\emph{A finite dimensional vector space}.\\
As we have already briefly discussed in Section \ref{DiscussThmSec},
it is natural for many purposes to enlarge the set of chiral operators
by allowing the coefficients of the polynomials in \eqref{COpgen} to
be elements of the field of fractions $\pfield = \text{Frac}(\pring)$
instead of $\pring$. The enlarged chiral ring $\hring$ will be simply
defined by
\be\label{CRform2}\hring =\pfield[\mathcal O_{1},\ldots,\mathcal
O_{n}]= \pfield[X_{1},\ldots,X_{n}]/\ideal\, ,\ee
to be compared with \eqref{CRform}. Considering the ring $\hring$
instead of $\ring$ doesn't change the physics but can help to simplify
the mathematics. For example, it is clear that the ring $\hring$ is a
$\pfield$ vector space. Interestingly, it is a finite dimensional
vector space. In particular, if $(\mathcal
B_{\alpha})_{1\leq\alpha\leq\dim_{\pfield}\hring}$ is a basis, then
any chiral operator $\mathcal O\in\hring$ can be expanded as
\be\label{basisexp} \mathcal O = \sum_{\alpha=1}^{\dim_{\pfield}\hring}
c_{\alpha}\mathcal B_{\alpha}\, ,\ee
where $c_{\alpha}\in\pfield$. 

The relation \eqref{basisexp} is interesting because it is
\emph{linear}, unlike the non-linear relations of the form
\eqref{COpgen}. The proof of the existence of a finite basis
$(\mathcal B_{\alpha})$ relies on Theorem \ref{thmPolynomial}. For
example, assume that the ring is generated by only one operator
$\mathcal O_{1}$. By using the polynomial equation satisfied by
$\mathcal O_{1}$ one can express $\mathcal O_{1}^{p}$, for any $p\geq
v$, as a linear combination of $(\mathbb I,\mathcal
O_{1},\ldots,\mathcal O_{1}^{v-1})$. This implies that
$\dim_{\pfield}\hring\leq v$. In the general case, the proof can be
easily done by induction on the number $n$ of generators.

\smallskip

\noindent\emph{The ring $\ring$ is graded}.\\
Each $\u$ global symmetry of the gauge theory induces a grading
\be\label{grading}
\ring = \bigoplus_{n}\ring_{n}\ee
where $\ring_{n}$ is the set of ring elements having charge $n$. The
important property is that $\ring_{n}\ring_{m}\subset\ring_{n+m}$.
Note that only $\ring_{0}$ is a subring. The grading implies that the
ideal $\ideal$ is generated by a set of homogeneous polynomials, i.e.\
by polynomials of fixed $\u$ charges.

\smallskip

\noindent\emph{The ideal $\ideal$ is finitely generated}.\\
The ideals of the ring $\pfield[X_{1},\ldots,X_{n}]$ are always
finitely generated (one says that the polynomial ring is noetherian).
This result applied to the ideal $\ideal$ implies that there always
exists a \emph{finite} number of polynomials
$R_{i}\in\pfield[X_{1},\ldots,X_{n}]$, $1\leq i\leq r$, such that any
operator relation can be written in the form
\be\label{oprelnoether} \sum_{i=1}^{r}a_{i}R_{i}=0\ee
with $a_{i}\in\pfield[X_{1},\ldots,X_{n}]$. In other words, all the
information about the ring $\hring$ is encoded in a \emph{finite}
number of relations $R_{1}=\cdots=R_{r}=0$. The same is true for the
ring $\ring$.\footnote{When $\pring$ is not noetherian this is a
consequence of Prop.\ \ref{solPolProp} in \ref{powerSec}.}

\begin{exmp}\label{algpropEx}
To illustrate the above properties, consider the simplest case of the
pure $\text{SU}(N)$ gauge theory. The ring of parameters is simply
$\pring = \mathbb C[q]$. The chiral ring is generated by the single
operator $S$ defined in \eqref{Sdef}. From \eqref{Seq}, we deduce that
it satisfies the operator relation
\be\label{oprelpure} S^{N}-q = 0\, .\ee
Hence $\ideal = (S^{N}-q)$ and
\be\label{Rpure}\ring = \mathbb C[q,S]/(S^{N}-q)\, .\ee
Taking into account \eqref{oprelpure}, we see that the most general
chiral operator can be written in the form
\be\label{basepure} \mathcal O = \sum_{k=0}^{N-1}a_{k}(q)S^{k}\, ,\ee
where $a_{k}\in\mathbb C[q]$ (if $\mathcal O\in\ring$) or
$a_{k}\in\mathbb C(q)$ (if $\mathcal O\in\hring$). Clearly,
$(1,S,\ldots,S^{N-1})$ is a base of $\hring$ over $\mathbb C(q)$ and
in particular $\dim_{\mathbb C(q)}\hring = N$. The ring $\ring$ is
graded with respect to a $\u$ symmetry under which $S$ has charge 1
and $q$ has charge $N$ (up to a rescaling of charges, this is the
$\u_{\text R}$ symmetry described in Section \ref{DiscussThmSec}).

The perturbative chiral ring is obtained by setting $q=0$ in
\eqref{oprelpure},
\be\label{Rclpure} \ring_{\text{pert}} = \mathbb C[q,S]/(S^{N})\, .\ee
We see that $S$ is nilpotent in $\ring_{\text{pert}}$.
\end{exmp}
\subsubsection{Physical properties of the chiral ring}
\label{PhysCRSec}

In this subsection, we are going to discuss the fundamental
Proposition \ref{BasicProp}. 

Property (ii) in the Proposition is trivially satisfied, because our
definition of the chiral ring relies exclusively on the knowledge of
the expectation values $\langle\mathcal O\rangle$.

Property (i) means that one can reconstruct in principle all the
chiral operators expectation values from the ring $\ring$. The
procedure to do so is as follows. One first considers the canonical
surjection $\pring[X_{1},\ldots,X_{n}]\rightarrow\ring =
\pring[\mathcal O_{1},\ldots,\mathcal O_{n}]$ that maps the dummy
unconstrained variables $X_{i}$ to the operator $\mathcal O_{i}$. The
kernel of this mapping is the radical ideal $\ideal$. We then find a
set of generators for the ideal, $\ideal = (R_{1},\ldots,R_{r})$,
which yields a set of algebraic equations for the expectation values,
\be\label{eqgenerators} R_{i}\bigl(\langle\mathcal
O_{1}\rangle,\ldots,\langle\mathcal O_{n}\rangle\bigr) = 0\, , \quad
R_{i}\in\pring[X_{1},\ldots,X_{n}]\, ,\ 1\leq i\leq r\, .\ee
The question is: do the algebraic equations \eqref{eqgenerators},
which are constrained to be with coefficients in $\pring$, determine
unambiguously the analytic functions $\langle\mathcal O_{i}\rangle
(\g,\q)$ in all the vacua of the theory?

Before we provide a proof, let us illustrate the result in the case of
the pure $\text{SU}(N)$ gauge theory. As explained in the previous
subsection (Ex.\ \ref{algpropEx}), the ideal $\ideal$ is in this case
generated by the polynomial $S^{N}-q$, which yields the algebraic
equation
\be\label{Seqpure} \langle S\rangle^{N} - q = 0\, .\ee
This equation has $N$ solutions associated with the $N$ vacua of the
theory,
\be\label{Ssolpure} \langle k|S|k\rangle = q^{1/N}e^{2i\pi k/N}\, ,\ee
and this yields indeed the full solution of the model. Let us
emphasize that this result strongly depends on the precise definition
of $\ring$ and in particular of the ring $\pring$. For example, if
instead of $\pring =\mathbb C[q]$ we had used $\mathbb C[q^{2}]$, then
the only relation that could be considered would be $S^{2N}=q^{2}$,
and this has unphysical solutions.

The fundamental ingredient in proving that the algebraic equations
with coefficients in $\pring$ \eqref{eqgenerators} give enough
information to determine the expectation values is of course the
existence of the polynomial equations described in Section
\ref{poleqSec}. Theorem \ref{thmPolynomial} implies that for any
chiral operator $\mathcal O$, there exists a degree $v$ polynomial
$P_{\mathcal O}\in\pring[X]$ such that
\be\label{PO5} P_{\mathcal O}(\mathcal O) = 0\ee
is an operator relation of the form \eqref{oprelpol1}. These
polynomials (or more precisely the polynomials $P_{\mathcal
O}\circ\rho_{\mathcal O}$ obtained after expressing $\mathcal O$ in
terms of the generators $\mathcal O_{1},\ldots,\mathcal O_{n}$ as in
\eqref{COpgen}) are thus automatically in the ideal $\ideal$, i.e.\
are linear combinations with coefficients in
$\pring[X_{1},\ldots,X_{n}]$ of the polynomials $R_{i}$ appearing in
\eqref{eqgenerators}. In particular, from the equations
\eqref{eqgenerators} one can derive the condition
\be\label{POvev4} P_{\mathcal O}(\langle\mathcal O\rangle) = 0\, .\ee
Thus all we need to do is to prove the Prop.\ \ref{solPolProp} of
Section \ref{powerSec}.

\begin{rem} Assume that one has a set of polynomials
$P_{a}\in\pring[X_{1},\ldots,X_{n}]$ such that the equations $P_{a}=0$
determine completely all the chiral operators expectation values. The
equations $P_{a}=0$ thus encode all the physical information about the
theory. Let $I=(P_{a})$ be the ideal generated by the polynomials
$P_{a}$. Then, in general, $I$ is not equal to $\ideal$ and the
quotient ring $\pring[X_{1},\ldots,X_{n}]/I$ is not equal to the
chiral ring. All that can be said is that $I\subset\ideal$, however
$I$ does not need to be a radical ideal. Physically speaking, this
means that $I$ and $\pring[X_{1},\ldots,X_{n}]/I$ contain in general
unphysical information depending on arbitrary choices. On the other
hand, one always has $r(I)=\ideal$ as a consequence of Hilbert's
nullstellansatz, and thus the chiral ring is obtained from
$\pring[X_{1},\ldots,X_{n}]/I$ by setting to zero all the nilpotent
elements.

For example, in the case of the pure $\text{SU}(N)$ gauge theory, one
could replace the algebraic equation \eqref{Seqpure} by
$(S^{N}-q)^{2}=0$. Clearly, the ideal $((S^{N}-q)^{2})$ is strictly
included in $\ideal = (S^{N}-q)$, and the associated ring $\mathbb
C[q,S]/((S^{N}-q)^{2})$ has a nilpotent element.

\end{rem}
\subsubsection{The power of the polynomial equations, again}
\label{powerbisSec}

Strictly speaking, to prove Prop.\ \ref{BasicProp} we actually don't
need the full power of Prop.\ \ref{solPolProp}, but only the fact that
the full set of polynomial equations (which is infinite) determines
unambiguously all the expectation values. So let us start by analysing
this weaker statement.

The idea is to consider the operator $\mathcal O_{z_{1},\ldots,z_{n}}$
defined by
\be\label{Ozdef}\mathcal O_{z_{1},\ldots,z_{n}} =
\sum_{\alpha=1}^{n}z_{\alpha}\mathcal O_{\alpha}\, ,\ee
where the $\mathcal O_{1},\ldots,\mathcal O_{n}$ form a set of
generators of $\ring$. The $z_{\alpha}$ in \eqref{Ozdef} are arbitrary
complex numbers. From the polynomial $P_{\mathcal
O_{z_{1},\ldots,z_{n}}}\in\pring[X]$, we can derive the expectation
values $\langle i|O_{z_{1},\ldots,z_{n}}|i\rangle$ in all the vacua
$|i\rangle$. The important point is that, by continuity in the
$z_{\alpha}$, there is no ambiguity in labeling the vacua for
different values of the $z_{\alpha}$. One can then deduce the
expectation values of all the generators from
\be\label{genvevder} \langle i|\mathcal O_{\alpha}|i\rangle =
\frac{\partial \langle i|O_{z_{1},\ldots,z_{n}}|i\rangle}{\partial
z_{\alpha}}\,\cdotp\ee
Since the most general chiral operator is of the form \eqref{COpgen},
its expectation value is straightforwardly obtained from
\eqref{genvevder} as well.

To complete the proof of Prop.\ \ref{solPolProp}, we need to show that
actually only a finite number of polynomial equations is needed (in
the above argument, we used an infinite set of such equations, labeled
by the variables $z_{\alpha}$). This follows immediately from the fact
that polynomial rings are noetherian, which can be summarized in the
following lemma.
\begin{lem}\label{lemNoether} Let $\mathscr P=(P_{a})_{a\in\mathscr
A}$ be an arbitrary family of polynomials in
$\pfield[X_{1},\ldots,X_{n}]$. Then there always exists a finite
number of polynomials $P_{i}$, $1\leq i\leq p$, $P_{i}\in\mathscr P$,
such that any $P\in\mathscr P$ can be written as $P=\sum_{i=1}^{p}
a_{i} P_{i}$ for some $a_{i}\in\pfield[X_{1},\ldots,X_{n}]$.
\end{lem}
\noindent We refer the reader to standard textbooks \cite{books} for a
proof. In our case, the family of polynomials that we consider is
formed by all the polynomials of the form $P_{\mathcal
O}\circ\rho_{\mathcal O}$, for all the chiral operators $\mathcal O$, 
with $\rho_{\mathcal O}$ defined by \eqref{COpgen}. 

\subsection{The chiral ring and operator mixing}
\label{opmixSec}

In this subsection, we are going to illustrate, using very simple
examples, the fact that all the physics of the theory in encoded in
the chiral ring $\ring$ and that any additional piece of information
must be unphysical (i.e.\ corresponds to arbitrary choices). All we
say is very elementary, yet it clarifies many confusions and correct
some errors that are commonly found in the literature. As we shall
see, an important source of confusion comes from the possibility to
define in different ways some composite operators. This ambiguity is a
non-perturbative version of the ambiguity associated to a choice of
scheme in ordinary perturbative quantum field theory. It is directly
related to the freedom one has in performing field redefinitions.
Field redefinitions do not change the physics nor the chiral ring, but
they can drastically change the way the solution of the model is
presented.

\begin{exmp}\label{pureN1} Let us start by considering once more the
case of the pure gauge theory, but this time with gauge group $\uN$
instead of $\text{SU}(N)$. This yields the following
puzzle.\footnote{I would like to thank Mina Aganagic for bringing this
puzzle to my attention.} The solution of the model is still given by
\eqref{Ssolpure}, which is often summarized by saying that the
effective quantum glueball superpotential is given by the
Veneziano-Yankielowicz formula,
\be\label{WVY} W(S) = -S\ln\frac{S^{N}}{e^{N}q}\,\cdotp\ee
It is indeed straightforward to check that the equations $W'(S)=0$
yield the solutions \eqref{Ssolpure}. Let us now consider the case
$N=1$. On the one hand, since the gauge theory is in this case a free
$\u$ theory, we do not expect any non-trivial quantum correction.
However, we still have a non-trivial glueball superpotential
\eqref{WVY} and a non-trivial gluino condensate
\be\label{SU1}\langle S\rangle = q\, .\ee
How is this possible?

One interpretation, advocated in \cite{AV}, is that to any classical
super Yang-Mills theory is associated an infinite number of physically
inequivalent quantum theories with the same classical limit. The $\u$
theory with the condensate \eqref{SU1} would then correspond to a
non-standard way to quantize the abelian gauge theory (or to a
non-standard UV completion in the language of \cite{AV}), which would
yield a non-trivial quantum abelian gauge theory.

We do not subscribe to this interpretation. Actually, we shall make
clear that there is always a unique quantum supersymmetric gauge
theory associated with a given classical supersymmetric gauge theory
and that the ambiguities described in \cite{AV} correspond to field or
parameter redefinitions.

To understand how this works for our simple $\u$ example, let us
compute the chiral ring. At the perturbative level, the $\u$ theory
has no non-trivial chiral operator except of course the identity
$\mathbb I$ and the perturbative chiral ring is given by
\be\label{RpertN1}\ring_{\text{pert}} = \mathbb C[q]\, .\ee
On the other hand, the quantum chiral ring associated with \eqref{SU1}
is given by (this is simply \eqref{Rpure} for $N=1$)
\be\label{RpureN1} \ring = \mathbb C[q,S]/(S-q)\, .\ee
The rings $\ring_{\text{pert}}$ and $\ring$ are clearly isomorphic,
\be\label{RRpertequal}\ring=\ring_{\text{pert}}\, .\ee
From the discussion in previous Sections, we know that this implies
that the $\u$ theory does not have any non-trivial quantum
corrections. In particular, the result \eqref{SU1} and the glueball
superpotential \eqref{WVY} for $N=1$ are completely unphysical.

These statements might still appear surprising, so let us spell their
meaning in a very concrete way. When one makes the claim that the $\u$
theory has a non-trivial condensate \eqref{SU1}, one actually has
forgotten to analyse precisely the definition of the operator $S$ in
the quantum theory. As we shall explain in details below, the operator
$S$ (as many other commonly used operators in supersymmetric gauge
theories) is ambiguous in the non-perturbative $\u$ quantum theory.
This ambiguity is very similar to the ambiguity (scheme-dependence)
one encounters in defining composite operators in ordinary
perturbative quantum field theory. In the $\u$ theory, the operator
$S$ can mix with the operator $q\mathbb I$ (that we note simply by
$q$) because their $\u_{\text R}$ charges \eqref{asign1} turn out to
be the same (equal to three) when $N=1$. Eq.\ \eqref{SU1} simply means
that we have chosen a scheme in which in the quantum theory the
operator $S$ is \emph{defined} to be $q\mathbb I$. The condensate
\eqref{SU1} is thus completely fake, it comes from a mixing with the
identity operator!

We hope that the above example, though essentially trivial, already
shows the interest in working with the chiral ring. The main lesson is
that the commonly used tools, like the effective superpotentials, can
contain a lot of redundant and completely unphysical information that
obscure the physics, which is unlike the chiral ring $\ring$. The $\u$
theory is of course extreme; in this case the superpotential
\eqref{WVY} is totally arbitrary and entirely without physical
content.
\end{exmp}
\begin{exmp}\label{Adjarbit} Let us now look at the $\uN$ gauge theory
with one adjoint $\phi$ already discussed in Section
\ref{numbervacSec}, Ex.\ \ref{Ex3}. We can decompose the chiral ring
of this model according to the grading associated to the $\u_{\text
R}'$ symmetry \eqref{asign2},
\be\label{gradedRX} \ring = \bigoplus_{n\in\mathbb N}\ring_{n}\, .\ee
Let us discuss the subring $\ring_{0}$ \cite{CRcons}. It is generated
by the operators
\be\label{ukdef} u_{k} = \Tr\phi^{k}\, .\ee
Because $\phi$ is a $N\times N$ matrix, the $u_{k}$ are not all
independent. There exists polynomial constraints of the form
\be\label{R0C1} u_{N+p} = Q_{p}(u_{1},\ldots,u_{N})\, ,\quad p\geq 1\ee
that show that only $u_{1},\ldots,u_{N}$ are independent.

In the literature, it is often claimed that the relations
\eqref{R0C1}, which are trivial classical identities, ``are corrected
by instantons.'' The quantum relations would then take a corrected
form,
\be\label{R0C2} u_{N+p} = \tilde Q_{p}(u_{1},\ldots,u_{N};q)\, ,\quad
p\geq 1\, ,\ee
where now the polynomials $\tilde Q_{p}$ depend non-trivially on $q$
and coincide with the $Q_{p}$ when $q=0$.

The question we would like to answer is: are the ``quantum
corrections'' that appear in \eqref{R0C2} genuine, unambiguous
physical quantum corrections? From our previous discussions, it should
be clear that the answer is \emph{no}. The chiral ring $\ring_{0}$
\emph{does not depend on the form of the relations \eqref{R0C2}}. In
all cases, $\ring_{0}$ is isomorphic to a simple polynomial ring
\be\label{R0form}\ring_{0} = \mathbb C[q,X_{1},\ldots,X_{N}]\, ,\ee
where the $X_{i}$ are as usual algebraically independent variables
(identified here with the $u_{i}$). The relations \eqref{R0C2} are
mere \emph{definitions} of what we mean by $u_{k}$ for $k>N$ in the
quantum theory. These definitions can of course be totally arbitrary.
They are only restricted by the $\u$ symmetries of the theory (in the
case at hand, the $\u_{\text R}$ symmetry \eqref{asign1} implies that
$\tilde Q_{p}=Q_{p}$ for $p<2N$). Clearly, and contrary to the standard
claims, the relations \eqref{R0C2}, being arbitrary, cannot be
computed in any well-defined sense in the quantum gauge theory.

Since this is at the origin of considerable confusion, let us give
more concrete details. Imagine that you want to compute the
expectation value $\langle u_{k}\rangle$, or any chiral correlator
containing the operator $u_{k}$, in the quantum gauge theory, using a
microscopic first principle approach as in \cite{mic1,mic2,mic3}. A
crucial part of the calculation involves integrating over the moduli
space of instantons. The instanton moduli space has singularities
corresponding to instantons with vanishing size. When $k<2N$, these
singularities are integrable, i.e.\ the integral over the moduli space
with the insertion of the operator $u_{k}$ is well-defined. However,
when $k\geq 2N$, the singularities are no longer integrable. Typically
one finds a result of the form $\infty\times 0$, the $\infty$ coming
from the integration over the instanton size and the $0$ coming from a
Grassmann integral. This phenomenon is described in details in a
special case for example in Section VII.2 of \cite{instrev}.

From our previous discussion, it should not be surprising that the
correlators involving $u_{k}$ for $k\geq 2N$ are ill-defined. The
ambiguity we find is simply the ambiguity associated with a choice of
the polynomials $\tilde Q_{p}$ in \eqref{R0C2}. In instanton calculus,
one usually proceeds by regularizing the instanton moduli space. There
is an infinite number of possible inequivalent regularizations. Once
regularized, the moduli space integrals are all well-defined and we
find a definite answer for the correlators. To each regularization is
associated a particular definition of the operators $u_{k}$ for $k>N$,
i.e.\ a particular choice for the polynomials $\tilde
Q_{p}$.\footnote{So there is an injective map between the space of
polynomials $\tilde Q_{p}$ and the space of regularizations of the
instanton moduli space. We do not know if this map is an isomorphism.}

In essence, the above phenomenon is the same as the one encountered in
perturbation theory when one defines composite operators. The
definition depends on the scheme. In our case, we are dealing with
chiral operators which are unambiguous at the perturbative level, but
a regularization is needed at the non-perturbative level.

Of course, the physics of the gauge theory is independent of the
particular regularization of the instanton moduli space that one uses.
This translates in the fact that the ring \eqref{R0form} is
independent of the precise form of the polynomials $\tilde Q_{p}$.

Usually, one uses the non-commutative deformation to regularize the
instanton moduli space. In this case the generating function
\be\label{defF} F(z) = z^{N}\exp\Bigl(-\sum_{k\geq
1}\frac{u_{k}}{kz^{k}}\Bigr)\ee
satisfies the constraint
\be\label{Fcons} F(z) + \frac{q}{F(z)} = P(z)\, ,\ee
for a certain degree $N$ monic polynomial $P(z)$. The condition
\eqref{Fcons} is \emph{equivalent} to a particular choice for the
relations \eqref{R0C2} and the polynomials $\tilde Q_{q}$ can be
computed recursively by expanding the left-hand side of \eqref{Fcons}
at large $z$ and using the fact that the terms with negative powers of
$z$ must vanish. Equation \eqref{Fcons} can be easily solved and
yields
\begin{align} \label{NCf1} F(z) &=\frac{1}{2}\Bigl( P(z) +
\sqrt{P(z)^{2} - 4 q}\Bigr)\, ,\\ \label{NCf2}
R(z) &= \frac{F'(z)}{F(z)} = \sum_{k\geq 0}\frac{u_{k}}{z^{k+1}} =
\frac{P'(z)}{\sqrt{P(z)^{2} - 4 q}}\,\cdotp\end{align}
These formulas for the generating functions are of course
well-known. They imply that $R(z)$ and $F(z)$ are well-defined
meromorphic functions on the Seiberg-Witten curve
\be\label{SWcurve} y^{2} = P(z)^{2} - 4 q\, .\ee
What is usually not appreciated is that this result is a consequence
of an arbitrary choice for the relations \eqref{R0C2} and does not
contain any non-trivial physical information. Other choices for the
relations are possible. For example, it is perfectly sensible to
make the choice $\tilde Q_{p}=Q_{p}$, in which case one finds that
$F(z)$ is simply a polynomial and $R(z)$ a rational function with
simple poles,
\begin{align}\label{RFcl} F(z) &= P(z)=\prod_{i=1}^{N}(z-z_{i})\, \\
R(z) & =\frac{P'(z)}{P(z)} =
\sum_{i=1}^{N}\frac{1}{z-z_{i}}\,\cdotp\end{align}
The interest in making the choices that lead to \eqref{NCf1} and
\eqref{NCf2} is that the solution of the model can then be presented
in an elegant way. This will be made clear in Section \ref{CoulSec}.
\end{exmp}
\begin{exmp}
As a last simple example of the use of the chiral ring, let us analyse
in more details the ``ambiguities'' pointed out in \cite{AV}. The
puzzle can be presented in the following way. In supersymmetric gauge
theories, there exists operators that vanish at the perturbative level
but do not at the quantum level. For example, in the pure
$\text{SU}(N)$ gauge theory, $S^{k}=0$ in perturbation theory as soon
as $k\geq N$ whereas $S^{k}\not = 0$ in the full quantum theory. Let
$\mathcal O$ be such an operator. Imagine that we add $\mathcal O$ to
the tree-level superpotential \eqref{treeW},
\be\label{wtreeQ}\wt\longrightarrow\tilde W_{\text{tree}}=\wt +
g\mathcal O\, .\ee
Clearly, at the classical level, the theories described by $\wt$ and
$\tilde W_{\text{tree}}$ are the same. However, they look different at
the quantum level. It might seem that we have an amgiguity in
quantizing the classical theory and that new types of theories,
corresponding to different ``UV completions''  in the language of
\cite{AV}, can be defined.

How do we solve the puzzle using the notion of the chiral ring? Any
operator $\mathcal O$ can be written in the form \eqref{COpgen}. The
fact that the operator vanishes simply means that the polynomial
$\rho_{\mathcal O}$ is proportional to the instanton factors
$q_{\alpha}$. Thus adding the term $g\mathcal O$ to $\wt$ is
\emph{equivalent} to adding a term $g\rho_{\mathcal O}(\mathcal
O_{1},\ldots,O_{n})$, which simply amounts to a $\q$-dependent
redefinition of some of the couplings $g_{k}$ appearing in the
standard classical tree level superpotential \eqref{treeW}. So the
theory with $\tilde W_{\text{tree}}$ is not a new theory. It is simply
a standard theory written in terms of an unusual parametrization, for
which the tree-level couplings depend artificially on the instanton
factors.

For example, in the pure $\text{SU}(N)$ gauge theory, the most general
classical tree-level superpotential that can be considered is
\be\label{Wtpureex} \wt = \sum_{k=1}^{N-1}g_{k}S^{k}\, .\ee
Taking into account \eqref{Rpure}, we see that adding a term of the
form $g S^{rN + s}$ with $0\leq s<N$ in the quantum theory is simply
equivalent to redefining $g_{k}\rightarrow g_{k}+q^{r}g\delta_{ks}$ in
\eqref{Wtpureex}.
\end{exmp}
\section{The chiral ring and phases}
\label{CRPSec}

We now have all the necessary tools to study the phases of the super
Yang-Mills theories. An interesting feature that was pointed out in
\cite{csw1} is that in a given phase, there are new relations between
chiral operators that come on top of the operator relations that we
have discussed in Section \ref{CRsSec}. The authors of \cite{csw1}
proposed that these phase-dependent relations may be used to
distinguish the phases. One of our goal in the following is to make
this idea precise. We shall see that indeed, individual phases are
\emph{characterized} by a set of phase-dependent relations. Quite
remarkably, there are priviliged operators in each phase, that we call
primitive operators, such that the full set of relations in a phase
can be reduced to a \emph{single} polynomial equation satisfied by any
of the primitive operator.

We start in \ref{anaphaSec} by giving a physically-motivated
definition of what is meant by ``being in the same phase.'' We then
proceed in \ref{IrrSec} and \ref{PrimarySec} to study the mathematical
consequences, making a direct link between the decomposition of the
polynomial equations of Th.\ \ref{thmPolynomial} into irreducible
components and the existence of distinct phases. Eventually, we are
led to a very simple description of the individual phases in terms of
primitive operators which is explained in \ref{PrimitiveSec}. All
these results have a very natural geometric interpretation discussed
in \ref{geomSec}.

For the study of the phases it is simpler mathematically to use the
chiral ring $\hring$ defined in \eqref{CRform2} instead of $\ring$ and
thus we shall do so in the following unless explicitly stated
otherwise.

\subsection{Phases and analytic continuations}
\label{anaphaSec}

Phases are characterized by the fact that they cannot change under a
smooth deformation. In other words, if we start with some given
parameters $\g$ and $\q$ and in a given vacuum $|i\rangle$, then by
smoothly varying the parameters we must remain in the same phase. By
allowing the most general analytic continuations, we can then explore
the full phase diagram of the theory. As explained in Section
\ref{monodSec}, an analytic continuation can induce a permutation of
the vacua. 
\begin{defn}\label{DefMonod} The \emph{monodromy group} of the theory 
is the group generated by the permutations of the vacua obtained by
performing analytic continuations along arbitrary closed loop in the
theory space parametrized by $(\g,\q)$.
\end{defn}
\begin{defn}\label{DefPhases}
A \emph{phase} of a supersymmetric gauge theory is defined to be an
orbit of the monodromy group acting on the set of vacua.
\end{defn}
\noindent We thus have the following decomposition,
\be\label{orbitDec} \bigl\{ |i\rangle\, ,\ 1\leq i\leq v\bigr\} =
\bigcup_{p=1}^{\Phi} |p)\, , \ee
where we have denoted by $|p)$ the orbits.

The Def.\ \ref{DefPhases} is \emph{analytic} in nature. As already
emphasized in Section \ref{IntroSec}, using a direct analytic approach
to compute the phase structure of the theory is in general extremely
difficult because the analytic structure of the expectation values
$\langle\mathcal O\rangle(\g,\q)$, for which explicit formulas are
usually not known, can be very complicated. Our goal in the following
is to develop an \emph{algebraic} point of view which turns out
to be very powerful.

\subsection{Irreducible polynomials and phases}
\label{IrrSec}
\subsubsection{The fundamental example}
\label{IrrSecf}

Let us assume for the moment that the chiral ring is generated by a
single operator $\mathcal O$. This might seem to be a gross
oversimplification, but it will become clear in \ref{PrimitiveSec}
that this is not so and that most of the relevant features can be
described by making this assumption. The chiral ring is thus of the
form
\be\label{CRexone} \hring = \pfield[\mathcal O] = \pfield [X]/\ideal\,
.\ee
The ideal $\ideal$ is always generated by a single polynomial in this
case (one says that the ring $\pfield[X]$ is principal) which is
obviously the degree $v$ polynomial $P_{\mathcal O}$ of Th.\
\ref{thmPolynomial},
\be\label{Iprinc} \ideal = (P_{\mathcal O})\, .\ee
The expectation values $\langle i|\mathcal O|i\rangle$ in the $v$
vacua of the theory correspond to the $v$ roots of the equation
\be\label{POeqsol} P_{\mathcal O}(z) = \sum_{k=0}^{v}a_{k}(\g,\q)
z^{v-k} = 0\, .\ee

According to Def. \ref{DefPhases}, the phase structure of the theory
can be computed by finding how the roots of the polynomial
\eqref{POeqsol} are permuted when the parameters $\g$ and $\q$ are
varied arbitrarily. However, instead of focusing on these
\emph{analytic} properties, it turns out to be much more fruitful to
study the \emph{arithmetic} properties of the polynomial $P_{\mathcal
O}$. 

Let us start with a basic definition. A polynomial $P\in\pfield [X]$
is said to be \emph{irreducible} if it cannot be written as the
product of two other non-trivial polynomials in $\pfield [X]$. In
other words, if $P = R S$ with $R,S\in\pfield[X]$ then either $R$ or
$S$ must be in $\pfield$. Let us note that the property of
irreducibility strongly depends on the base field $\pfield$.

Any polynomial in $\pfield [X]$ has a prime decomposition. In
particular, the polynomial $P_{\mathcal O}$ can be decomposed in a
unique way (up to trivial multiplications by non-zero elements of
$\pfield$) as the product of relatively prime irreducible polynomials
$P_{p}\in\pfield[X]$ of degree $v_{p}\geq 1$,
\be\label{Podec} P_{\mathcal O} = \prod_{p=1}^{\Phi}P_{p}^{n_{p}}\, .\ee
The integers $n_{p}$ must be equal to one, since otherwise
$\prod_{p=1}^{\Phi}P_{p}$ would be a nilpotent element of $\hring$, in
contradiction with the discussion of Section \ref{AlgpropSec}. This
decomposition in irreducible parts is of fundamental interest to us
because of the following Theorem.
\begin{thm}\label{IrrThm} Each phase of the theory is associated with
an irreducible factor $P_{p}$ in the prime decomposition of the
polynomial $P_{\mathcal O}$ over the field $\pfield$. In particular,
in \eqref{orbitDec} the phase $|p)$ contains the vacua associated with
the roots of the polynomial $P_{p}$.
\end{thm}
\noindent This result shows that one can use algebraic techniques to
study the phases of the gauge theories. This is extremely useful
because in many cases it is much easier to prove that a polynomial is
irreducible, or to find the decomposition into irreducible factors,
than to study the analytic properties of the roots.

Let us prove Th.\ \ref{IrrThm}. To solve the equation \eqref{POeqsol},
we can solve the $\Phi$ algebraic equations
\be\label{Ppeqsol} P_{p}(z) = \sum_{k=0}^{v_{p}}a_{p,k}(\g,\q)
z^{v_{p}-k} = 0\ee
independently. Let us decompose the set of vacua as
\be\label{vacdecppp} \bigl\{|i\rangle\, ,\ 1\leq i\leq v\bigr\} =
\bigcup_{p=1}^{\Phi} \bigl\{|p,i\rangle\, ,\ 1\leq i
\leq v_{p}\bigr\}\, , \ee
in such a way that the expectations values $\langle p,i|\mathcal
O|p,i\rangle=\mathcal O_{p,i}(\g,\q)$ be the roots of $P_{p}$. Let us
now perform an analytic continuation of $\mathcal O_{p,i}(\g,\q)$
along an arbitrary closed loop in the $(\g,\q)$-space. \emph{Because
the coefficients $a_{p,k}$ in \eqref{Ppeqsol} are in $\pfield$,} they
are single-valued functions of the parameters. Thus after the analytic
continuation the polynomial $P_{p}$ remains the same. This implies
that the analytic continuation of $\mathcal O_{p,i}$ must still be a
root of $P_{p}$: the monodromy group acts by permuting the roots of
the individual irreducible factors $P_{p}$, but \emph{cannot mix the
roots of different factors.} In other words, vacua $|p,i\rangle$ and
$|p',i'\rangle$ for $p\not= p'$ must be in different phases.

Conversely let us show that all the vacua $|p,i\rangle$, for $1\leq
i\leq v_{p}$, are in the same phase. If this were not the case, then
the monodromy group would have distinct orbits when acting on the
roots of $P_{p}$. To each orbit, one can associate a polynomial whose
roots correspond to the vacua in the orbit. Using an argument along
the lines of Section \ref{PSec}, one can show that these polynomials
are in $\pfield[X]$. They would thus provide a non-trivial
decomposition of $P_{p}$ over $\pfield$, which is impossible.

All the explicit examples we shall be dealing with in the present
paper correspond to $\pfield = \mathbb C(\g,\q)$. One useful
elementary tool to study irreducibility properties of polynomial over
this field is to use the following lemma.
\begin{lem}\label{testLem} Let $\pring = \mathbb C[\g,\q]$ and
$\pfield = \mathbb C(\g,\q)$. Then if $P\in\pring[X]$, $\deg P\geq 1$,
is irreducible over $\pring$ it is also irreducible over $\pfield$.
\end{lem}
The proof can be found in standard textbooks. The result is used as
follows. Imagine that you want to prove the irreducibility of
$P\in\pfield [X]$ over $\pfield$. We can always factorize
$P(X)=a\tilde P(X)$ with $a\in\pfield$ and $\tilde P\in\pring[X]$,
where the coefficients of $\tilde P$ are relatively prime in $\pring$.
Clearly the irreducibility of $P$ over $\pfield$ is equivalent to the
irreducibility of $\tilde P$ over $\pfield$. We thus have to study the
possible factorizations $\tilde P = QR$ over $\pfield[X]$. The lemma
shows that the coefficients of $Q$ and $R$ can be restricted to be in
$\pring$ instead of $\pfield$ without loss of generality.

\begin{exmp} Let us study the phase structure of the pure
$\text{SU}(N)$ gauge theory, for which the chiral ring is generated by
a single field $S$. From \eqref{PS} we know that the relevant
polynomial is $P_{S}(X) = X^{N}-q$. We have to study the possible
factorizations $P_{S} = QR$ in $\mathbb C(q)[X]$. From the Lem.\
\ref{testLem}, we can assume that $Q$ and $R$ are in $\mathbb
C[q][X]$. Since $P_{S}$, viewed as a polynomial in $q$, is of degree
one, either $R$ or $Q$, say $R$, does not depend on $q$. By setting
$q=0$ in the factorization condition, one sees that $R$ is necessarily
proportional to $X^{r}$ for some $r\geq 0$ and that actually $r=0$
since zero is not a root of $P_{S}$. Thus $R\in\mathbb C$ which proves
that $P_{S}$ is irreducible. Thus the pure gauge theory has only one
phase.

Of course the result in this case follows trivially from the analytic
method, because the equation $P_{S}=0$ can be solved explicitly, see
\eqref{Ssolpure}. All the vacua $|k\rangle$ can be smoothly connected
to each other by analytic continuation: $|k\rangle\rightarrow
|k+s\rangle$ by encircling $s$ times the origin in the $q$-plane,
$q\rightarrow e^{2i\pi s}q$. The algebraic approach is useful when
explicit formulas for the roots do not exist (or are too complicated),
see Sections \ref{HiggsConfSec} and \ref{CoulSec}.
\end{exmp}
\subsubsection{Operator relations in a phase}

The description of the phases given by Th.\ \ref{IrrThm} in terms of
the decomposition $P_{\mathcal O}=\prod_{p}P_{p}$ has an interesting
consequence. The expectation values of $\mathcal O$ satisfy
\be\label{POOb} P_{\mathcal O}(\langle\mathcal O\rangle) = 0\ee
in all the vacua of the theory and we thus have an operator relation
$P_{\mathcal O}(\mathcal O)=0$ in the sense of Def.\ \ref{oprelDef} in
Section \ref{CRdefSec}. On the other hand, in the phase $|p)$, the
expectation values $(p|\mathcal O|p)$ (by which we mean the
expectation values in any of the vacua belonging to the phase $|p)$)
satisfy the \emph{stronger} constraint
\be\label{Oprelphasep} P_p\bigl((p|\mathcal O|p)\bigr) =
0\, .\ee
This naturally leads to the following definitions.
\begin{defn}\label{orelphaseDef} Let $\mathcal O^{(1)},\ldots,\mathcal
O^{(p)}$ be chiral operators. An \emph{operator relation in a phase
$|\varphi)$} is a polynomial equation of the form
\be\label{oprelpolpahse} P(\mathcal O^{(1)},\ldots,\mathcal O^{(p)}) =
0\, ,\quad P\in\pring[X_{1},\ldots,X_{p}]\, ,\ee
such that $P(\langle i|\mathcal O^{(1)}|i\rangle,\ldots,\langle
i|\mathcal O^{(p)}|i\rangle)$ identically vanishes in all the vacua
$|i\rangle$ belonging to the phase $|\varphi)$.
\end{defn}
\begin{defn}\label{CRphaseDef}
Let $\ideal_{|\varphi)}$ be the ideal generated by all the operator
relations in the phase $|\varphi)$. Clearly,
$\ideal\subset\ideal_{|\varphi)}$ and thus $\ideal_{|\varphi)}$ can be
seen as an ideal of the chiral ring $\ring$ defined in \eqref{CRform}
(or of $\hring$ defined in \eqref{CRform2}). The quantum chiral rings
in the phase $|\varphi)$ are then defined by
\be\label{CRphasedef} \ring_{|\varphi)}=\ring/\ideal_{|\varphi)}\,
,\quad \hring_{|\varphi)}=\hring/\ideal_{|\varphi)}\, .\ee
\end{defn}
\noindent The rings $\ring_{|\varphi)}$ and $\hring_{|\varphi)}$ have
remarkable properties that are discussed in the following Sections.

\subsubsection{On the use of irreducible polynomials}

Consider now a general supersymmetric gauge theory. Imagine that we
want to demonstrate that two vacua $|i\rangle$ and $|j\rangle$ belong
to the same phase. For example, in theories with fundamentals, we
would like to show that the ``confining'' and the ``Higgs'' vacua are
in the same phase. The discussion in the previous subsections suggests
the following strategy: find an operator $\mathcal O$ such that the
vacua $|i\rangle$ and $|j\rangle$ are associated with two roots of the
\emph{same} irreducible factor in the decomposition of $P_{\mathcal
O}$. This approach turns out to be a very efficient way to make the
proof.

Let us be more precise in the case of the theory \eqref{wtex4}. As we
have already explained, this model is the natural arena to study the
possible transitions from the Higgs to the confining regime. The claim
is that all the vacua of rank one of the model should be in the same
phase, irrespective of the pattern of gauge symmetry breaking. This is
a direct consequence of the following result.

\begin{lem}\label{CFThm} When $\Nf<N$, the polynomial $P_{S}$ for the
glueball superfield $S$ in the model \eqref{wtex4} is irreducible over
$\mathbb C[\mu,m_{1},\ldots,m_{\Nf},q]$. When $\Nf\geq N$,
$P_{S}(X)=X^{\binom{\Nf}{N}}\tilde P_{S}(X)$, where $\tilde P_{S}$ is
irreducible over $\mathbb C[\mu,m_{1},\ldots,m_{\Nf},q]$. The factor
$X^{\binom{\Nf}{N}}$ corresponds to a purely classical part associated
with the vacua of rank zero and the other factor $\tilde P_{S}$ to the
Higgs/confining vacua of rank one.
\end{lem}
A reader that would be interested specifically in the problem of the
equivalence between the Higgs and confining regimes may now jump to
Section \ref{HiggsConfSec} where an explicit construction of the
polynomial $P_{S}$ and the proof of Lem.\ \ref{CFThm} can be found.

\subsection{The prime decomposition}
\label{PrimarySec}

In Section \ref{IrrSecf} we assumed that the chiral ring were
generated by a single operator $\mathcal O$. The phase structure of
the model is then given by the decomposition of the polynomial
$P_{\mathcal O}$ into irreducible factors.

How can we generalize this result to the generic case with a finite
number of generators $\mathcal O_{1},\ldots,\mathcal O_{n}$?

\subsubsection{Phases and operator relations}

First, we have the analogue of Th.\ \ref{thmPolynomial} for a given
phase.

\begin{prop}\label{Proppoly} Let $|\varphi)$ be a phase that contains
$v_{\varphi}$ vacua. Then any chiral operator $\mathcal O$
satisfies a degree $v_{\varphi}$ operator relation in the phase
$|\varphi)$ of the form $P_{\mathcal O}^{|\varphi)}(\mathcal O) = 0$,
$P_{\mathcal O}^{|\varphi)}\in\pring[X]$.
\end{prop}
The proof is strictly similar to the proof of Th.\ \ref{thmPolynomial}
and we let the details to the reader. From Prop.\ \ref{Proppoly} one
can directly derive the analogue of Prop.\ \ref{BasicProp} in Section
\ref{CRsSec}.
\begin{prop}\label{BPProp} The full solution of the theory in the
chiral sector in the phase $|\varphi)$ is coded in the chiral ring
$\ring_{|\varphi)}$ (or $\hring_{|\varphi)}$) in the phase
$|\varphi)$, i.e.\ we can compute the expectation values
$\langle\mathcal O\rangle$ in any vacuum belonging to the phase
$|\varphi)$ and for any chiral operator $\mathcal O$ from the
knowledge of the ring $\ring_{|\varphi)}$ (or $\hring_{|\varphi)}$).
\end{prop}
This result makes very precise the idea proposed in \cite{csw1}. If
$\ideal_{|\varphi)} =\ideal$ then clearly the theory has only one
phase. However, in general one has a strict inclusion
$\ideal\subsetneq \ideal_{|\varphi)}$ and there are new operator
relations valid only in the phase $|\varphi)$. Moreover these new
relations completely determine the expectation values in the phase.

One may ask if the chiral ring $\hring_{|\varphi)}$ (or equivalently
the operator relations in the phase $|\varphi)$) could be considered
to be like an ``order parameter'' characterizing the phase in some
fundamental way. The answer to this question is no. This is best
illustrated on an example, so let us consider the $\uN$ theory with
one adjoint $\phi$ and tree-level superpotential \eqref{wtex3}. If $p$
is the degree of $W'$, as in \eqref{Wder}, then the theory has
$v_{\varphi}=pN$ vacua of rank one (see \eqref{vrquan} and
\eqref{vrhform}) corresponding to an unbroken gauge group. It is not
difficult to show that all these vacua are in the same phase
$|\varphi)$ (see Section \ref{CoulSec}). This phase does not depend on
the value of $p$: increasing $p$ amounts to turning on some couplings
and the new vacua that then appear can be smoothly connected to the
old vacua. Physically, this phase simply corresponds to the standard
confining phase of the pure super Yang-Mills theory. On the other
hand, the structure of $\hring_{|\varphi)}$ does depend on $p$. This
can be seen, for example, from the fact that the dimension of
$\hring_{|\varphi)}$ viewed as a $\pfield$ vector space is equal to
the number $v_{\varphi}$ of vacua in the phase $|\varphi)$ (this is a
general result that will be derived in Section \ref{PrimitiveSec}) and
that this number depends on $p$.

The lesson is that it is not trivial to obtain new kinds of order
parameters that can help in distinguishing the phases at a fundamental
level. In particular, the chiral ring itself is not a good candidate,
because physically equivalent phases can have distinct chiral rings.
Nevertheless, our formalism can be used to shed an interesting new
light on this question, using Galois theory. This is explained in
details in a separate publication \cite{Galois}.

\subsubsection{The chiral field}

The chiral ring in a phase has a crucial property that plays in
particular a prominent r\^ole in \cite{Galois}.

\begin{prop}\label{CRFieldThm} Let $|\varphi)$ be a phase. The ideal
$\ideal_{|\varphi)}\subset\ring$ is prime. Equivalently, the ring
$\ring_{|\varphi)}$ is an integral domain.
\end{prop}
When $\ideal_{|\varphi)}$ is generated by a single polynomial $P$ (as
in the case studied in Section \ref{IrrSecf}), the condition that
$\ideal_{|\varphi)}$ is prime is equivalent to the condition that the
polynomial $P$ is irreducible. In general, it means that if
$RS\in\ideal_{|\varphi)}$, then either $R\in\ideal_{|\varphi)}$ or
$S\in\ideal_{|\varphi)}$. Clearly this is equivalent to the fact that
$A_{|\varphi)}$, which is isomorphic to
$\pring[X_{1},\ldots,X_{n}]/\ideal_{|\varphi)}$, is an integral
domain: in $\ring_{|\varphi)}$, $\mathcal A\mathcal B = 0$ implies
that either $\mathcal A=0$ or $\mathcal B=0$.

It is not difficult to understand why $A_{|\varphi)}$ must be an
integral domain. Pick two operators
$\mathcal A$ and $\mathcal B$ such that $\mathcal A\mathcal B=0$.
This is equivalent to the fact that the expectation value of $\mathcal
A\mathcal B$ in any vacuum belonging to the phase $|\varphi)$
vanishes,
\be\label{proprime}(\varphi|\mathcal A\mathcal B|\varphi) =
0=(\varphi|\mathcal A|\varphi)(\varphi|\mathcal B|\varphi)\, .\ee
If $\mathcal A$ and $\mathcal B$ are both zero in $\ring_{|\varphi)}$
then there is nothing to prove. Let us thus assume that $\mathcal
A\not = 0$ and let us prove that this implies that $\mathcal B=0$. The
condition $\mathcal A\not = 0$ in $\ring_{|\varphi)}$ means that there
exists at least one vacuum $|i\rangle$ in the phase $|\varphi)$ such
that $\langle i|\mathcal A|i\rangle\not = 0$. Equation
\eqref{proprime} then automatically implies that $\langle i|\mathcal
B|i\rangle = 0$. Let now $|j\rangle$ be an arbitrary vacuum in
$|\varphi)$. Because $|i\rangle$ and $|j\rangle$ are in the same
phase, the expectation value $\langle j|\mathcal B|j\rangle$ can be
obtained by analytic continuation from $\langle i|\mathcal B|i\rangle
= 0$ and is thus automatically zero. The conclusion is that the
expectation value of $\mathcal B$ vanishes in all the vacua of the
phase $|\varphi)$, i.e.\ that $\mathcal B=0$ in $\ring_{|\varphi)}$.

It is important to realize that this property is very special to the
chiral rings \emph{in a given phase} and that it is not shared by the
chiral ring $\ring$ (or $\hring$) in general. For example, in the case
studied in \ref{IrrSecf}, $P_{\mathcal O} = \prod_{p}P_{p} = 0$ in
$\ring$, but the individual irreducible factors $P_{p}$ are all
non-zero in $\ring$ if there is more than one phase. It is actually
not difficult to show that in general $\ring$ is an integral domain if
and only if the theory is realized in a single phase.

An even stronger property is true for the ring $\hring_{|\varphi)}$.
\begin{thm}\label{CRFieldThm2} Let $|\varphi)$ be a phase. The ideal
$\ideal_{|\varphi)}\subset\hring$ is maximal. Equivalently, the ring
$\hring_{|\varphi)}$ is a field, which is the field of fractions of
$\ring_{|\varphi)}$.
\end{thm}
Being a field is a very remarkable property for an algebra of
operator. It means that every non-zero operator has an inverse. Very
concretely, if $\mathcal O=\rho_{\mathcal O}(\mathcal
O_{1},\ldots,\mathcal O_{n})$, $\rho_{\mathcal
O}\in\pfield[X_{1},\ldots,X_{n}]$, is an arbitrary non-zero operator,
then it is always possible to find another non-zero operator $\mathcal
O'=\rho_{\mathcal O'}(\mathcal O_{1},\ldots,\mathcal O_{n})$,
$\rho_{\mathcal O'}\in\pfield[X_{1},\ldots,X_{n}]$, such that
$\mathcal O\mathcal O' = \rho_{\mathcal O}(\mathcal
O_{1},\ldots,\mathcal O_{n})\rho_{\mathcal O'}(\mathcal
O_{1},\ldots,\mathcal O_{n}) = 1$ in $\hring_{|\varphi)}$. In other
words, thank's to the additional operator relations that are satisfied
in a given phase, an arbitrary rational function in the generators
$\mathcal O_{1},\ldots,\mathcal O_{n}$ can always be shown to be equal
to a particular polynomial.
\begin{exmp}\label{FieldpureEx} Before we proceed to the proof, let us
illustrate this result for the pure $\text{SU}(N)$ gauge theory. This
theory is realized in a single phase and thus Th.\ \ref{CRFieldThm2}
implies that the chiral ring
\be\label{CRsimplefield} \hring = \mathbb C(q)[S]/(S^{N}-q)\ee
itself should be a field. Indeed, using the operator relation
$S^{N}=q$, it is clear that the inverse of the glueball operator is
simply given by $S^{-1} = S^{N-1}/q$. The inverse of an arbitrary
operator of the form $\rho(S)$ for $\rho\in\mathbb C(q)[X]$ can also
be constructed straightforwardly using the euclidean division
algorithm.
\end{exmp}

The simplest proof of Th.\ \ref{CRFieldThm2} relies on the fact that
$\hring_{|\varphi)}$ is a finite dimensional $\pfield$ vector space.
Indeed $\hring$ itself is finite dimensional, as explained in Section
\ref{AlgpropSec}. Assume then that $\mathcal O\in\hring_{|\varphi)}$
is non zero and consider the $\pfield$-linear map $\mathcal
O'\mapsto\mathcal O\mathcal O'$. This map is injective because
$\hringp$ is an integral domain (using exactly the same argument that
shows that $\ringp$ is an integral domain). Being a linear map of a
finite dimensional vector space, it must also be surjective and thus
in particular its image contains the identity. This implies that
$\mathcal O$ has an inverse as was to be shown.

The ring $\ringp$, being an integral domain, has a field of fractions
$\text{Frac}(\ringp)$ which is the smallest field containing $\ringp$
and which is built by considering fractions of the elements of
$\ringp$. Clearly $\hringp\subset\text{Frac}(\ringp)$ and since
$\hringp$ is a field the inclusion must be an equality.

We shall have more to say about the chiral field $\hringp$ in Section 
\ref{PrimitiveSec}.

\subsubsection{The prime decomposition}

We have seen that phases are characterized by prime ideals
$\ideal_{|\varphi)}\supset\ideal$ describing the operator relations in
the given phase. When there is only one generator $\ideal =
(P_{\mathcal O})$ as in \eqref{Iprinc}, these prime ideals are
generated by the irreducible factors of $P_{\mathcal O}$. In the
general case, the decomposition of a given polynomial into irreducible
factors is replaced by the decomposition of a given radical ideal into
prime ideals.
\begin{thm}\label{primedecTh} Let $\ideal$ be the ideal of operator
relations. Then one can write in a unique way
\be\label{decideal}\ideal=\bigcap_{p=1}^{\Phi}\ideal_{|p)}\ee
where the $\ideal_{|p)}$ are prime ideals such that
$\ideal_{|p)}\not\subset\ideal_{|p')}$ if $|p)\not =|p')$. The
decomposition \eqref{decideal} corresponds to the decomposition
\eqref{orbitDec} into phases, the ideal $\ideal_{|p)}$ being generated
by the operator relations in the phase $|p)$.
\end{thm}
This theorem provides a completely general algebraic method to obtain
the phase structure of a given model. Of course, computing the prime
decomposition of a radical ideal $\ideal$ is non-trivial. It is one of
the basic problem of computational commutative algebra. A very useful
fact is that sophisticated algorithms that perform this decomposition
have been implemented on computer algebra systems like
\textsc{Singular} \cite{singular} that are heavily used in Section
\ref{CoulSec} to study the phases of the model \eqref{wtex3}.

The Th.\ \ref{primedecTh} is standard and a proof can be found in the
textbooks \cite{books}. However, since this is a fundamental result
for us, and also because the textbooks usually deal with the most
general case of the primary decomposition of an arbitrary ideal
instead of the simpler prime decomposition of a radical ideal that we
need, let us briefly sketch the argument. If $\ideal$ is prime then
the theory has only one phase and there is nothing to do. If $\ideal$
is not prime, then we can find an operator relation of the form
$ab\in\ideal$ but with $a\not\in\ideal$ for example. It is then
natural to impose new operator relations corresponding to $a=0$ or
$b=0$, which are associated with the radical ideals
$\ideal_{1}=r(\ideal + (a))$ and $\ideal_{2} = r(\ideal + (b))$. Using
the fact that $\ideal$ is radical, it is not difficult to check that
$\ideal = \ideal_{1}\cap\ideal_{2}$. If $\ideal_{1}$ and $\ideal_{2}$
are prime, then we have finished. Otherwise, we can repeat the above
argument and further decompose the ideals $\ideal_{1}$ and/or
$\ideal_{2}$. Eventually, this process must terminate because $\ring$
is noetherian and we find the decomposition \eqref{decideal}. If we
had two decompositions for $\ideal$ based on prime ideals
$(\ideal_{|p)})$ and $(\mathscr J_{|q)})$, then it is easy to see that
$\cap_{p}\ideal_{|p)}=\cap_{q}\mathscr J_{|q)}$ implies that, for any
$p$, $\ideal_{|p)} = \cap_{q}(\ideal_{|p)}+\mathscr J_{|q)})$. Because
$\ideal_{|p)}$ is prime this implies that there exists $q$ for which
$\ideal_{|p)} = \ideal_{|p)} + \mathscr J_{|q)}$, i.e.\ $\mathscr
J_{|q)}\subset\ideal_{|p)}$. Similarly $\ideal_{|p')}\subset\mathscr
J_{|q)}$ for some $p'$. The requirement that
$\ideal_{|p)}\not\subset\ideal_{|p')}$ if $|p)\not =|p')$ then shows
that $p=p'$ and $\ideal_{|p)}=\mathscr J_{|q)}$, proving the
uniqueness of the decomposition.

\subsection{Primitive operators}
\label{PrimitiveSec}

We are now going to complete our toolkit with a remarkable result that
drastically simplifies the description of individual phases.

\subsubsection{The structure of the chiral field in a phase}
\begin{thm}\label{PrimOpThm} Let $|\varphi)$ be a phase that contains
$v_{\varphi}$ vacua. The associated chiral ring $\hringp$ is generated
by a single operator $\mathcal O_{\varphi}$, called a primitive
operator for the phase $|\varphi)$. This operator satisfies an
operator relation in the phase $|\varphi)$ of the form $P_{\mathcal
O_{\varphi}}^{|\varphi)}(\mathcal O_{\varphi}) = 0$, where
$P_{\mathcal O_{\varphi}}^{|\varphi)}\in\pfield [X]$ is irreducible
and of degree $v_{\varphi}$. In particular,
\be\label{fieldpform} \hringp = \pfield[X]/(P_{\mathcal
O_{\varphi}}^{|\varphi)})\ee
and $\dim_{\pfield}\hringp = v_{\varphi}$.
\end{thm}
This theorem shows that the physics of a given phase is always
entirely coded in the expectation value of a $\emph single$ chiral
operator $\mathcal O_{\varphi}$. All we need to know is the
irreducible polynomial equation satisfied by this expectation value.
All the other expectation values in the phase are simple polynomials
in $\langle\mathcal O_{\varphi}\rangle$ with coefficients in
$\pfield$.

\begin{exmp}\label{HCPrimEx} Let us consider a gauge theory that is
realized in a single phase. Then Th.\ \ref{PrimOpThm} implies that the
chiral ring of such a theory is generated by a single operator, as in
the case of the pure gauge theory. This is clearly an extremely
powerful result. For example, from Lem.\ \ref{CFThm} we can deduce that
the glueball operator $S$ is a primitive operator for the model
\eqref{wtex4} when $\Nf<N$. In particular, this implies that all the
operators of the form $\Tr\phi^{k}$, $\Tr
W^{\alpha}W_{\alpha}\phi^{k}$ and $\tilde Q^{f}\phi^{k}Q_{f'}$ for any
$k$, are actually simple polynomials in $S$! We shall see this
explicitly in Section \ref{HiggsConfSec}.
\end{exmp}

The Th.\ \ref{PrimOpThm} is a direct consequence of the Primitive
Element Theorem whose proof (which requires some technology that we
have not introduced) can be found in standard textbooks \cite{books}.
The somewhat simplified version that we need is as follows.
\begin{lem}\label{PELem} Let $\pfield$ be a field of characteristic
zero. Let $\mathsf K\supset\pfield$ be a finitely generated and
algebraic field extension (an algebraic extension is such that any
element of $\mathsf K$ satisfies an algebraic equation with
coefficients over $\pfield$). Then there always exists an element
$\alpha\in\mathsf K$ such that $\mathsf K=\pfield(\alpha)$ is the
field generated by $\alpha$ over $\pfield$.
\end{lem}
In our case $\pfield$ is the field of parameters which is always of
characteristic zero since it contains $\mathbb C$ as a subfield. The
extension field we consider is $\hringp\supset\pfield$. It is finitely
generated since the chiral ring $\hring$ itself is and it is algebraic
by Prop.\ \ref{Proppoly}.

\subsubsection{A simple test for a primitive operator}

There are in general many primitive operators in a given phase. More
precisely, we have the following Proposition.
\begin{prop}\label{PrimOpexis} Let $\mathcal O$ be a chiral operator
such that $P_{\mathcal O}^{|\varphi)}(\mathcal O)=0$ in the phase
$|\varphi)$ (see Prop.\ \ref{Proppoly}). Then $\mathcal O$ is a
primitive operator in the phase $|\varphi)$ if and only if
$P_{\mathcal O}^{|\varphi)}$ is irreducible.
\end{prop}
This is an easy consequence of Th.\ \ref{PrimOpThm}. Indeed, if we
denote by $\pfield(\mathcal O)$ the subfield of $\hringp$ generated by
$\mathcal O$ over $\pfield$, then $\dim_{\pfield}\pfield(\mathcal O) =
\deg P_{\mathcal O}^{|\varphi)}$ because $P_{\mathcal O}^{|\varphi)}$
is irreducible. This shows that $\dim_{\pfield}\pfield(\mathcal O) =
v_{\varphi}=\dim_{\pfield}\hringp$ and thus that $\pfield(\mathcal O)
= \hringp$.

Physically, the primitive operators are the operators that
``distinguish'' all the vacua of the phase: by analytic continuation
their expectation value can have $v_{\varphi}$ distinct semi-classical
expansions. 
\begin{prop}\label{PrimTest} Let $\mathcal O$ be a chiral operator and
$|\varphi)$ be a phase containing $v_{\varphi}$ vacua. Assume that for
some given values of the parameters, the $v_{\varphi}$ expectations
values $\langle i|\mathcal O|i\rangle$ for $|i\rangle\in|\varphi)$ are
distinct complex numbers. Then $\mathcal O$ is a primitive operator in
the phase $|\varphi)$.
\end{prop}
This result provides a simple numerical test to show that an operator
is primitive in a given phase, because the expectation values $\langle
i|\mathcal O|i\rangle$ for some given parameters $\g$ and $\q$ can be
found by solving numerically the system of algebraic equations that
corresponds to the prime ideal $\mathscr I_{|\varphi)}$ defining the
phase $|\varphi)$.

\subsubsection{The quantum effective superpotential}

A very natural way to construct a primitive operator is as follows.
The quantum effective superpotential $\weff^{|i\rangle}(\g,\q)$ (also
often denoted as $\wl$ in the literature) is defined by performing the
path integral in a given vacuum $|i\rangle$ and extracting the
$F$-terms from the resulting effective action for the background
chiral superfields $\g$ and $\q$. The fundamental property of $\weff$
is that its derivatives with respect to the couplings yield the
associated expectation values. For example, with a tree-level
superpotential \eqref{treeW},
\be\label{wleq}\frac{\partial\weff^{|i\rangle}}{\partial g_{k}} =
\langle i|\mathcal O_{k}|i\rangle\ee
and we also have
\be\label{wleq2}\frac{\partial\weff^{|i\rangle}}{\partial \ln
q_{\alpha}} = \langle i| S_{\alpha}|i\rangle\ee
where $S_{\alpha}$ is the glueball operator in the simple factor
$\mathfrak g_{\alpha}$ of the gauge group (see Section \ref{GeneSec}).
If couplings to all the generators of the chiral ring are introduced,
as we assume in this subsection, then clearly the full solution of the
theory is encoded in the analytic function $\weff(\g,\q)$.

A very nice property of the analytic function $\weff(\g,\q)$ is that
it is always given by the expectation value of a certain chiral
operator,
\be\label{WeffOprel} \weff(\g,\q) = \langle\mathcal W\rangle\, .\ee
This is a consequence of the Ward identity
\be\label{uRwl} 3\weff =
\sum_{k}[g_{k}]g_{k}\frac{\partial\weff}{\partial g_{k}} +
\sum_{\alpha}[q_{\alpha}]q_{\alpha}\frac{\partial\weff}{\partial
q_{\alpha}}\,\cvp \ee
which follows from the $\u_{\text R}$ symmetry \eqref{asign1}. In
\eqref{uRwl}, we have denoted by $[g_{k}]$ and $[q_{\alpha}]$ the
$\u_{\text R}$ charges of the various couplings. Using \eqref{wleq}
and \eqref{wleq2}, we then obtain \eqref{WeffOprel} with
\be\label{WOpform} \mathcal W =\frac{1}{3}\Bigl(
\sum_{k}[g_{k}]g_{k}\mathcal O_{k} +\sum_{\alpha}[q_{\alpha}]
S_{\alpha}\Bigr)\in\hring\, .\ee
As for any other chiral operator, $\mathcal W$ satisfies a degree $v$
operator relation $P_{\mathcal W}(\mathcal W) = 0$. The phase
structure of the theory can then always be obtained from the
factorization of $P_{\mathcal W}$ into irreducible factors over
$\pfield$. Moreover, in each phase, $\mathcal W$ is a primitive
operator. In particular, in a given phase, any chiral operator
expectation value is always given in terms of the effective
superpotential by a simple (phase-dependent) \emph{polynomial}
expression,
\be\label{PWOPrel} (\varphi|\mathcal O|\varphi) =
T^{|\varphi)}_{\mathcal O}(\weff^{|\varphi)})\, ,\quad
T^{|\varphi)}_{\mathcal O}\in\pfield[X]\, . \ee
Note that an expression of the form \eqref{PWOPrel} would be valid for
\emph{any} primitive operator in each individual phases. In this
sense, the notion of a primitive operator is an algebraic
generalization of the notion of the quantum effective superpotential.

To finish this subsection, let us mention that the Lemma \ref{PELem}
can be refined \cite{books}. If a set of generators for the field
$\mathsf K$ over $\pfield$ is known, then it can be shown that the
primitive element can always be chosen to be a linear combination with
coefficients in $\pfield$ of these generators. Eq.\ \eqref{WOpform}
shows that $\mathcal W$ is precisely of this form.

\subsection{The geometric picture}
\label{geomSec}

Up to now, we have emphasized the algebraic point of view, because
this is how the calculations are done in practice. However, there is a
standard and elegant geometric interpretation of the results. For
simplicity, let us consider the case where $\pring=\mathbb C[\g,\q]$.
The operator relations that generate the ideal $\ideal$ can be
interpreted as the defining equations for an affine algebraic variety
$\mathscr M$. If $v$ is as usual the number of vacua of the theory,
this ``quantum space of parameters'' is a $v$-fold cover of the
$(\g,\q)$-plane. The chiral ring $\ring$ defined in \eqref{CRform} is
simply the ring of regular functions on $\mathscr M$, often called the
coordinate ring of the variety $\mathscr M$ in the literature. The
decomposition of the set of vacua into phases \eqref{orbitDec}, or
equivalently the prime decomposition of the ideal $\ideal$
\eqref{decideal}, corresponds to the decomposition of the variety
$\mathscr M$ into irreducible components,
\be\label{Mdec} \mathscr M = \bigcup_{p=1}^{\Phi}\mathscr M_{|p)}\,
.\ee
The ring $\ring_{|p)}$ defined in \eqref{CRphasedef} is the ring of
regular functions on the irreducible variety $\mathscr M_{|p)}$. On
the other hand, $\hring_{|p)}$ corresponds to the field of rational
functions on $\mathscr M_{|p)}$. 

The existence of primitive operators (Section \ref{PrimitiveSec}) has
also a nice geometrical interpretation. The fact that $\hring_{|p)} =
\pfield [X]/(P^{|p)})$ for a certain polynomial $P^{|p)}$ (see
\eqref{fieldpform}) shows that the variety $\mathscr M_{|p)}$ can be
described by the single equation $P^{|p)}=0$. This result corresponds
to a standard theorem in algebraic geometry: any irreducible affine
variety is birationally equivalent to a hypersurface.

\subsection{Phase transitions}
\label{phasetransSec}

The various irreducible components of $\mathscr M$ may intersect
non-trivially. The variety $\mathscr M_{|p)}\cap\mathscr M_{|p')}$,
with associated ideal of operator relation $r(\mathscr
I_{|p)}+\mathscr I_{|p')})$, describes the phase transition between
$|p)$ and $|p')$. Physically, these phase transitions are associated
with the appearance of new massless degrees of freedom that often
correspond to non-trivial IR fixed points of the gauge theory. It is
actually natural to consider that the intersections between distinct
phases correspond to new phases of the gauge theory. The variety
$\mathscr M_{|p)}\cap\mathscr M_{|p')}$ itself can have a non-trivial
decomposition in terms of irreducible components, corresponding to the
prime decomposition of $r(\mathscr I_{|p)}+\mathscr I_{|p')})$. These
irreducible components can themselves intersect, etc\ldots One can
also consider the intersections between more than two phases. In
general, a very complex nested structure of phases and phase
transitions can emerge in this way, associated with families of
non-trivial superconformal fixed points. Even though this is beyond
the goals of the present work, it is clear that our approach and the
tools we are using are perfectly appropriate for a systematic study of
this structure.

\subsection{Chiral duality}
\label{CSDSec}

At the classical level, a gauge theory is characterized by its gauge
group, its matter content and its tree-level superpotential. At the
quantum level, things are much more interesting. On the one hand, only
gauge invariant operators make sense and thus the gauge group is no
longer directly visible (the gauge group is not a physical symmetry
but a redundancy in the description of the physics). On the other
hand, the equations of motion derived from the tree-level
superpotential are quantum corrected. The result is that two
completely different looking classical theories may correspond to
physically equivalent quantum theory. One then says that the theories
are ``dual'' to each other. A weaker but very useful statement is that
two theories can be physically equivalent below a certain energy
scale, i.e.\ two distinct theories in the UV may flow to the same
theory in the IR. This kind of equivalence is usually called a
``Seiberg duality.''

In the context of the present paper, it is very natural to study
dualities between theories that have physically equivalent chiral
sectors. We call a duality of this type a \emph{chiral duality}. In
general, a chiral duality is not the same as a Seiberg duality, since
it also applies to cases where all the fields are massive. However,
when the chiral ring is generated by massless moduli, then clearly
Seiberg duality implies the equivalence of the chiral sectors of the
theories. The case of massive theories can then be obtained by
deformation. In practice, this can yield powerful tests of Seiberg
dualities.

Each individual phase of a given theory can be considered to be a
consistent quantum theory of its own and it is natural to study
dualities between phases rather than between full gauge theories. We
are thus led to the following definition.
\begin{defn}\label{StrongSDDef} A strong chiral duality between two
phases $|p)$ and $|q)$ of two possibly distinct gauge theories is an
isomorphism between the rings $\ring_{|p)}$ and $\ring_{|q)}$.
\end{defn}
\noindent In the geometric language of Section \ref{geomSec}, the
strong chiral Seiberg duality is thus an isomorphism between the
affine algebraic variety $\mathscr M_{|p)}$ and $\mathscr M_{|q)}$.

As we have emphasized many times, it is very natural to allow rational
combinations of the parameters to enter into the definition of the
most general chiral operators. This leads to a weak form of the chiral
duality.
\begin{defn}\label{WeakSDDef} A weak chiral duality (or simply a
chiral duality for short) between two phases $|p)$ and $|q)$ of two
possibly distinct gauge theories is an isomorphism between the fields
$\hring_{|p)}$ and $\hring_{|q)}$.
\end{defn}
\noindent Geometrically, the weak chiral duality between two phases
$|p)$ and $|q)$ is equivalent to the birational equivalence between
the associated irreducible algebraic varieties $\mathscr M_{|p)}$ and
$\mathscr M_{|q)}$. This is weaker than a strong chiral duality
because the invertible birational mapping $\mathscr
M_{|p)}\rightarrow\mathscr M_{|q)}$ can be singular for certain values
of the parameters (at the poles in the denominators). Nevertheless,
the weak chiral duality ensures that the algebras of operators over
$\pfield$ are the same in the two dual theories and thus they cannot
be physically distinguished. As we illustrate below, the standard
examples of Seiberg duality correspond to the weak form of Def.\
\ref{WeakSDDef}.

\begin{exmp}\label{ToyEx} Let us first use a toy example to illustrate
the above concepts. Let us explain how to construct chiral duals to
the pure gauge theory \eqref{CRsimplefield}. A dual must be in a
single phase as is the original theory and thus it is described by a
single primitive operator $s$. The ring isomorphism implies that $s$
can be written as a polynomial in the glueball $S$ with coefficients
in $\mathbb C(q)$. If $s$ then satisfies an \emph{irreducible}
polynomial equation of degree $N$ over $\mathbb C(q)$, then we know
that the rings of the two theories must coincide (their dimensions
over $\mathbb C(q)$ will be the same). In this case $S$ can also be
expressed as a polynomial in $s$ with coefficients in $\mathbb C(q)$,
yielding the birational mapping. For example, in the case $N=3$,
consider $s=S+S^{2}$. The operator $s$ satisfies the degree three
equation
\be\label{toyexeq} s^{3}-3 q s - q - q^{2}=0\ee
as a consequence of $S^{3}=q$. Using the relation between $s$ and $S$,
it is clear that one can interpolate smoothly between the three roots
of \eqref{toyexeq} and thus this equation is irreducible over $\mathbb
C(q)$. This shows immediately that the theories described by
$\eqref{toyexeq}$ and by $S^{3}=q$ are chiral dual. The polynomial
relation giving $S$ as a function of $s$ can be readily obtained,
\be\label{bira} s = S+S^{2} \Longleftrightarrow S =
\frac{1}{1-q}\bigl( 2 q + s - s^{2}\bigr)\, . \ee
This gives the birational isomorphism between the varieties $S^{3}-q=0$
and $s^{3}-3 q s - q - q^{2}=0$ \eqref{toyexeq}.

\end{exmp}
\begin{exmp}\label{SdualEx} Let us now consider the $\text{SU}(N)$
theory with $\Nf$ flavors and tree-level superpotential \eqref{wtex2}
(see also Ex.\ \ref{QCDEx} in Section \ref{seSec}). We assume that
$\Nf>3N/2$ and we limit our discussion to the sector of zero baryonic
charge for simplicity. The chiral ring is then generated by the
mesonic operators $M_{f'}^{\ f}$ and by the glueball $S$. The operator
relations read
\be\label{SDoprel1} m_{f}^{\ f''}M_{f''}^{\ f'} = \Nf S
\delta_{f}^{f'}\, ,\quad S^{N} = q\det m\, .\ee
Let us also consider a different gauge theory, with gauge group
$\text{SU}(\Nf-N)$, $\Nf$ flavors of quarks $q_{f}$ and $\tilde
q^{f}$, one singlet $N_{f}^{\ f'}$ and tree-level superpotential
\be\label{treeSD} \wt = \bigl(q_{f}\tilde q^{f'} + m_{f}^{\
f'}\bigr)N_{f'}^{\ f}\, .\ee
The chiral ring in the zero baryonic charge sector is generated by the
mesons $\hat M_{f}^{\ f'} = \tilde q^{f'} q_{f}$, the singlet
$N_{f'}^{\ f}$ and the glueball $s$. It can be argued (see for
examples \cite{ISlectures}) that the operator relations in the quantum
theory read
\be\label{SDoprel2} \hat M_{f}^{\ f'} = -m_{f}^{\ f'}\, ,\quad \hat
M_{f}^{\ f''}N_{f''}^{\ f'} = \Nf s \delta_{f}^{f'}\, ,\quad s^{N} =
(-1)^{\Nf}\frac{\det m}{\hat q}\, \cvp\ee
where $\hat q$ is the instanton factor. From \eqref{SDoprel1} and
\eqref{SDoprel2} it is clear that the two fields $\mathbb C(q,m_{f}^{\
f'})[M_{f}^{\ f'},S]$ and $\mathbb C(\hat q,m_{f}^{\ f'})[\hat
M_{f}^{\ f'},N_{f'}^{\ f},s]$ are isomorphic, with the identifications
\be\label{idSD} M_{f}^{\ f'} = N_{f}^{\ f'}\, ,\quad S = -s\, ,\quad q
= \frac{(-1)^{\Nf-N}}{\hat q}\,\cdotp\ee
The relations \eqref{idSD} give the birational isomorphism between the
varieties defined by \eqref{SDoprel1} and \eqref{SDoprel2}. The
singularity of the mapping at $\hat q=0$ corresponds to the well-known
fact that the model \eqref{treeSD} breaks supersymmetry at tree-level,
see for example \cite{ISS} for an extensive discussion.
\end{exmp}
\subsection{On the semi-classical phase diagram}
\label{semiclSec}

It is often useful to start the analysis of the phase diagram of a
given gauge theory by using the weak coupling approximation. One then
obtains a decomposition of the set of vacua of the theory of the form
\be\label{orbitsc} \bigl\{ |i\rangle\, ,\ 1\leq i\leq v\bigr\} =
\bigcup_{p=1}^{\tilde\Phi} |p)_{\text{s.c.}}\, , \ee
where the ``semi-classical'' phases $|p)_{\text{s.c.}}$ contain vacua
that can be connected to each other in the weak coupling region. In
general, the phases of the full quantum theory appearing in the
decomposition \eqref{orbitDec} can contain several of the
semi-classical phases appearing in \eqref{orbitsc}, since vacua that
cannot be smoothly related at weak coupling may be related by an
analytic continuation that probe the strong coupling regime of the
theory.

Let us note that explicit formulas for the chiral operator expectation
values can be easily obtained at weak coupling and thus in practice
the decomposition \eqref{orbitsc} can be most easily computed using
the standard ``analytic'' approach. Nevertheless, it is interesting to
explain how the semi-classical approximation can be interpreted in the
algebraic language that we have developed so far.

It turns out that the semi-classical decomposition \eqref{orbitsc}
corresponds to a factorization of the polynomial equations of the form
\be\label{factsc} P_{\mathcal O} = \prod_{p=1}^{\tilde\Phi}\tilde
P_{p}\, ,\ee
where now the factors $\tilde P_{p}$ are irreducible polynomials with 
coefficients in
\be\label{asc}\pring_{\text{s.c.}} = \mathbb C[\g]\{\q\}\, ,\ee
which is the ring of arbitrary convergent power series in $\q$ and
polynomials in $\g$. Note the difference with the decomposition
\eqref{Podec} in the full quantum theory, which was over the
polynomial ring $\pring=\mathbb C[\g,\q]$ and not the power series
ring $\mathbb C[\g]\{\q\}$. The polynomials $\tilde P_{p}$ are called
the Weierstrass polynomials in the mathematical literature. Their roots
are given by Puiseux expansions (power series expansions involving in
general fractional powers of the instanton factors) that correspond to
the small $\q$ expansions of the chiral operators expectation values.
It is clear that if we perform analytic continuations along closed
loops in parameter space that remain in the small $\q$ region (staying
within the radius of convergence of the series defining the
coefficients of the polynomials appearing in \eqref{factsc}), the
polynomials $\tilde P_{p}$ remain invariant and thus the roots of two
different factors in \eqref{factsc} cannot be smoothly connected. This
explains the correspondence between \eqref{factsc} and
\eqref{orbitsc}. We also have a nice illustration of the importance of
the base ring: going from the semi-classical approximation to the full
quantum theory amounts to studying factorization properties over a
polynomial ring instead of a power series ring. We shall present an
explicit example in Section \ref{HiggsConfSec}.

\subsection{Summary}

Let us briefly recapitulate what we have done in the previous
Sections.

--- The chiral sector of any supersymmetric gauge theory is described
by a set of polynomial equations with coefficients in a ring of
parameters $\pring$ which in most cases is a simple polynomial ring,
$\pring=\mathbb C[\g,\q]$. In particular, if $v$ is the number of
vacua of the theory, any chiral operator satisfies a degree $v$
algebraic equation with coefficients in $\pring$. The full set of
operator constraints is always generated by a finite subset of
equations.

--- The phases of the gauge theory can be studied by computing the
decomposition of these polynomials in irreducible factors or more
generally the prime decomposition of the ideal of operator relations.

--- A given phase can always be described by a single ``primitive''
operator (which is not unique) that satisfies an irreducible
polynomial equation. All the other operators are given by a polynomial
expression in terms of the primitive operator.

In the next two Sections we are going to apply these ideas to study
two interesting models in details.

\section{Application: Higgs and confinement}
\label{HiggsConfSec}
\subsection{The model and the general theorem}
\label{41Sec}

We now focus on the $\uN$ model with $\Nf$ flavors \eqref{wtex4} or
more generally on
\be\label{gentree} \wt = \Tr W(\phi) + \tilde Q^{f}m_{f}^{\ f'}(\phi) 
Q_{f'}\, ,\ee
with 
\be\label{wgen} W'(z) = \sum_{k=0}^{d} g_{k}z^{k} =
g_{d}\prod_{k=1}^{d}(z-w_{i})\, ,\quad \det m_{f}^{\ f'}(z) 
= \prod_{k=1}^{\Nf}(z-m_{k})\, .\ee
The most general classical vacuum $|N_{i};\nu_{j}\rangle_{\text{cl}}$
is labeled by the numbers of eigenvalues of the matrix $\phi$,
$N_{i}\geq 0$ and $\nu_{j}=0$ or $1$, that are equal to $w_{i}$ and
$m_{j}$ respectively \cite{csw2}. The constraint
\be\label{trivialconst}\sum_{i=1}^{d}N_{i} + \sum_{j=1}^{\Nf}\nu_{j} =
N\ee
must be satisfied. The gauge group $\uN$ is broken down to
$\text{U}(N_{1})\times\cdots\times\text{U}(N_{d})$ in a vacuum
$|N_{i};\nu_{j}\rangle_{\text{cl}}$. As explained in Section
\ref{numbervacSec}, chiral symmetry breaking implies that the quantum
vacua can be labeled as $|N_{i},k_{i};\nu_{j}\rangle$ with
$k_{i}\in\mathbb Z_{N_{i}}$.

\begin{defn} The \emph{rank} $r$ of a vacuum
$|N_{i},k_{i};\nu_{j}\rangle$ is defined to be the number of non-zero
integers $N_{i}$.
\end{defn}
\noindent Taking into account the mass gap in the non-abelian unbroken
factors of the gauge group, the low energy gauge group is $\u^{r}$ and
thus $r$ counts the number of massless photons. This number cannot
change when the parameters are smoothly varied and thus $r$ is a phase
invariant (this can also be trivially checked on the solution of the
model). Let us note that for the model \eqref{wgen}, $r\leq\min(N,d)$.
The fundamental result conjectured in \cite{csw2} that we want to
prove can be summarized as follows.
\begin{thm} \label{GenFund} 
The model \eqref{gentree} has, for a given value of the rank $r$, a
unique phase $|r)$ containing all the vacua of rank $r$.
\end{thm}
This result is equivalent to the fact that one can always interpolate
smoothly between two vacua $|N_{i},k_{i};\nu_{j}\rangle$ and
$|N_{i}',k_{i}';\nu_{j}'\rangle$ that have the same value of $r$. It
encompasses in particular all the possible interpolations between
various ``confining'' and ``Higgs'' vacua.

\subsection{Using the weak coupling approximation}
\label{UsewcSec}
\subsubsection{Semi-classical phases}

The proof of Th.\ \ref{GenFund} can be simplified if one realizes that
many analytic continuations between vacua are trivial, in the sense
that they can be described in the semi-classical regime by computing
explicitly the expectation values in a semi-classical expansion. The
associated irreducible polynomials can of course be written down
straightforwardly, but this is cumbersome and useless in these cases.
The algebraic method will be better used later to deal with the
genuinely quantum interpolations, that cannot be understood
semi-classically.

So let us compute the leading terms in a semi-classical expansion
around an arbitrary vacuum $|N_{i},k_{i};\nu_{j}\rangle$. This
expansion is governed by the gluino condensation in each unbroken
$\text{U}(N_{i})$ factors of the gauge group. For example, the
quantum effective superpotential is given by
\be\label{Weffsemi} \weff^{|N_{i},k_{i};\nu_{j}\rangle} =
\sum_{i=1}^{d}N_{i}W(w_{i}) + \sum_{j=1}^{\Nf}\nu_{j}W(m_{j}) +
\sum_{i=1}^{d}N_{i}\La_{i}^{3} e^{2i\pi k_{i}/N_{i}} + \cdots \ee
where we have neglected subleading terms when $q\rightarrow 0$. The
$\La_{i}$ are the dynamically generated scales for the unbroken gauge 
groups. In terms of the scale $\La$ of the $\uN$ gauge theory, which
is itself related to the instanton factor by the relation \eqref{defq}
\be\label{qLafla} q = \La^{2N-\Nf}\, ,\ee
one has
\be\label{LaiLarel} \La_{i}^{3N_{i}} = q \frac{W''(w_{i})^{N_{i}}
\prod_{j=1}^{\Nf}(w_{i}-m_{j})}{\prod_{j\not =
i}(w_{i}-w_{j})^{2N_{j}}
\prod_{j=1}^{\Nf}(w_{i}-m_{j})^{2\nu_{j}}}\,\cdotp\ee
This formula is obtained by integrating out the various massive
degrees of freedom: the denominator is produced by the $W$ bosons
charged under $\text{U}(N_{i})$ and the numerator comes from the
massive matter fields, adjoint multiplet (term
$W''(w_{i})^{N_{i}}$) and fundamental flavors (term
$\prod_{j=1}^{\Nf}(w_{i}-m_{j})$).

The formulas \eqref{Weffsemi} and \eqref{LaiLarel} immediately show
that:\\
$\bullet$ the vacua that have the same set of integers $\{N_{i}\}$ and
$\{\nu_{j}\}$ can all be smoothly connected to each other. Indeed,
arbitrary permutations of the $N_{i}$ on the one hand and of the
$\nu_{j}$ on the other hand can be obtained by performing an analytic
continuation that induces the same permutations on the parameters
$w_{i}$ and $m_{j}$ respectively. Note that under such an analytic
continuation, the integers $k_{i}$ do not change and remain associated
with the same integers $N_{i}$.\\
$\bullet$ vacua corresponding to fixed values of the $N_{i}$ and
$\nu_{j}$ but arbitrary values of the $k_{i}$ are all smoothly
connected to each other by performing analytic continuations of the
form $w_{i}-m_{j}\mapsto e^{2i\pi}(w_{i}-m_{j})$.\\

This is all we can do at the semi-classical level. The semi-classical
phase diagram \eqref{orbitsc} is thus made up of phases labeled by the
set of integers $\{N_{i}\}$ and $\{\nu_{j}\}$ but it is impossible to
interpolate between vacua that have different values of the $N_{i}$
and the $\nu_{j}$ by staying at weak coupling.

\begin{exmp}\label{scHCEx} To understand clearly what we have done,
let us consider for example the case of the $N=2$, $\Nf=3$ theory,
with $d=2$ in \eqref{wgen}. This theory has fourteen vacua that can be
labeled as $|N_{1},k_{1};N_{2},k_{2};\nu_{1},\nu_{2},\nu_{3}\rangle$.
Three vacua have rank $r=0$ ($|0,0;0,0;1,1,0\rangle$,
$|0,0;0,0;1,0,1\rangle$, $|0,0;0,0;0,1,1\rangle$), ten vacua have rank
$r=1$ ($|2,0;0,0;0,0,0\rangle$, $|2,1;0,0;0,0,0\rangle$,
$|0,0;2,0;0,0,0\rangle$, $|0,0;2,1;0,0,0\rangle$,
$|1,0;0,0;1,0,0\rangle$, $|1,0;0,0;0,1,0\rangle$,
$|1,0;0,0;0,0,1\rangle$, $|0,0;1,0;1,0,0\rangle$,
$|0,0;1,0;0,1,0\rangle$, $|0,0;1,0;0,0,1\rangle$) and one vacuum has
rank $r=2$ ($|1,0;1,0;0,0,0\rangle$). From the semi-classical analysis
only, we know that all the vacua of rank $r=0$ are in the same phase.
The vacuum at $r=2$ yields another phase on its own. At rank $r=1$, we
have two distinct semiclassical phases, corresponding to either a
$\text{U}(2)$ unbroken gauge group (four ``confining'' vacua) or to a
trivial $\u$ unbroken gauge group (six ``Higgs'' vacua). Theorem
\ref{GenFund} implies that, taking into account the strong coupling
quantum effects, these ten vacua are actually in the same phase.
\end{exmp}
\subsubsection{The strongly quantum problem}

The semi-classical analysis of the previous subsection shows that the
non-trivial interpolations correspond to changing the values of the
non-zero integers $N_{i}$ (and thus also of some of the $\nu_{j}$
according to \eqref{trivialconst}). This of course can be done step by
step, and thus it is enough to show that any of the $N_{i}$ can be
changed by one unit as long as it remains non-zero. Since the scales
\eqref{LaiLarel} of the various $\text{U}(N_{i})$ factors can be
separated at will, one can try to study this phenomenon in a limit
where the theory reduces to a $\text{U}(N_{i})$ model of the form
\eqref{wtex4} with one flavor of quark (one flavor is enough to study
changes of the number of colors by one unit). Precisely, if we choose
for example $i=1$, then we can consider the region of parameters where
the $W$ bosons and all the quarks except one are extremely massive,
$w_{j}\rightarrow\infty$ and $m_{j}\rightarrow\infty$ for $j\geq
2$, while $W''(w_{1})=\mu$ and the effective instanton factor
\be\label{scaling}
\frac{\prod_{j=2}^{\Nf}(w_{1}-m_{j})^{1-2\nu_{j}}}
{\prod_{j=2}^{N}(w_{1}-w_{j})^{2N_{j}}}\,q = q_{\text{eff}}\ee
remains constant. Clearly, if the interpolation is possible in this
limit, then it will be possible in the more general cases. Thus we see
that the general Th.\ \ref{GenFund} can be derived from the following
simplified lemma.
\begin{lem}\label{FundLem} The model \eqref{wtex4} with $\Nf=1$ is
realized in only one phase, i.e.\ the $N$ ``confining'' vacua
$|N,k;0\rangle$ for $0\leq k\leq N-1$ and the $N-1$ ``Higgs''
vacua $|N-1,k;1\rangle$ for $0\leq k\leq N-2$ can be smoothly
connected to each other.
\end{lem}
\noindent This statement contains all the relevant strongly quantum
information about the interpolation between Higgs and confining
phases. It will be derived in \ref{firrpolSec} by proving that the
glueball operator satisfies a degree $2N-1$ irreducible polynomial
equation over $\mathbb C[\mu,m,q]$, where $m$ is the mass of the
flavor.

\subsection{The operator relations}
\label{OpeRel1Sec}

The chiral ring of the model \eqref{gentree} is generated by the
operators
\be\label{opedef} u_{k} = \Tr \phi^{k}\,
,\quad v_{k} = -\frac{1}{16\pi^{2}}\Tr W^{\alpha}W_{\alpha}\phi^{k}\,
,\quad t_{f',\, k}^{\ f} = \tilde Q^{f}\phi^{k}Q_{f'}\, .\ee
As usual, it is useful to introduce the generating functions
\be\label{genfgen} R(z) = \sum_{k\geq
0}\frac{u_{k}}{z^{k+1}}\,\cvp\quad S(z) = \sum_{k\geq
0}\frac{v_{k}}{z^{k+1}}\,\cvp\quad G_{f'}^{\ f}(z) = \sum_{k\geq
0}\frac{t_{f',\, k}^{\ f}}{z^{k+1}}\,\cvp\ee
and also the function $F(z)$ defined by \eqref{defF} that satisfies by
construction
\be\label{FRrel} R(z)=\frac{F'(z)}{F(z)}\,\cdotp\ee
When $\Nf\geq N$ there are also baryonic operators, but they will play
no r\^ole in our analysis. Indeed, it is enough to consider the
operators \eqref{opedef} to prove that all the vacua at a given rank
can be smoothly connected. From Section \ref{PrimitiveSec} we then
know that at a given rank the baryonic operators are simple
polynomials in the generators \eqref{opedef}.

When $\Nf<2N$, the ring of parameters of the model is
\be\label{ringfla}\pring = \mathbb
C[g_{0},\ldots,g_{d},m_{1},\ldots,m_{\Nf},q]\, .\ee
When $\Nf=2N$ we must allow arbitrary series in $q$. 

\subsubsection{Kinematical and dynamical relations}
\label{knrelSec}

We now need to write down a full set of operator relations. It is
natural to distinguish ``kinematical'' and ``dynamical'' relations.

The kinematical relations come from the fact that the number of colors
$N$ in the theory is finite. Thus, amongst the generators
\eqref{opedef}, only the $u_{k}$ for $1\leq k\leq N$, the $v_{k}$ and
$t_{f',\, k}^{\ f}$ for $0\leq k\leq N-1$ can be independent. As
explained in \ref{opmixSec}, there is some freedom in defining the
other operators. We choose to define the $u_{k}$ for $k>N$ by imposing
the constraint
\be\label{Fflaeq} F(z) + q U(z)/F(z) = P(z)\, ,\ee
where 
\be\label{Udef} U(z)=\prod_{f=1}^{\Nf}(z-m_{f})\ee
and $P(z)$ is a degree $N$ polynomial. The condition \eqref{Fflaeq}
generalizes the choice \eqref{Fcons} made in the case $\Nf=0$. It is
\emph{equivalent} to relations of the form \eqref{R0C2}, where now the
polynomials $\tilde Q_{p}$ also depend on the completely symmetric
polynomials
\be\label{sigmadef}
\sigma_{i}=\sum_{f_{1}<\cdots<f_{i}}m_{f_{1}}\cdots m_{f_{i}}\ee
in the quark masses,
\be\label{urelfla} u_{N+p}=\tilde
Q_{p}(u_{1},\ldots,u_{N};\sigma_{1},\ldots,\sigma_{\Nf};q)\, .\ee
Similar kinematical constraints for the operators $v_{k}$ and
$t_{f',\, k}^{\ f}$ at $k\geq N$ also exist, but they don't need to be
discussed independently. Indeed, it turns out that they follow from
\eqref{Fflaeq} and from the dynamical relations we now discuss.

The dynamical relations are the famous generalized Konishi anomaly
equations. For our model, we have four infinite families of
equations, labeled by an integer $n\geq -1$,
\begin{gather}\label{a1b}N\sum_{k\geq 0} g_{k}u_{n+k+1} +\sum_{f}
t_{f,\, n+1}^{\ f} -
2\sum_{k_{1}+k_{2}=n}u_{k_{1}}v_{k_{2}} = 0\\ 
\label{a2b} N\sum_{k\geq 0}
g_{k}v_{n+k+1} - \sum_{k_{1}+k_{2}=n} v_{k_{1}}v_{k_{2}} = 0\\
\label{a3b} N\bigl( t_{f'\, n+2}^{\ f} -m_{f}t_{f'\, n+1}^{\
f}\bigr) - v_{n+1}\delta_{f'}^{f} = 0\\ \label{a4b} N\bigl( t_{f'\,
n+2}^{\ f} -m_{f'}t_{f'\, n+1}^{\ f}\bigr) -
v_{n+1}\delta_{f'}^{f} = 0\, .
\end{gather}
In terms of the generating functions \eqref{genfgen}, these equations 
read
\begin{gather}\label{a1} NW'(z)R(z) + N\sum_{f}G_{f}^{\ f}(z) -
2 S(z)R(z) =
N^{2}\Delta_{R}(z) \\ \label{a2}
NW'(z) S(z) - S(z)^{2} = N^{2}\Delta_{S} (z) \\\label{a3}
N(z-m_{f})G_{f'}^{\ f}(z)  - S(z)\delta_{f'}^{f} =
N\Delta_{f'}^{f}(z) \\\label{a4}
N(z-m_{f'})G_{f'}^{\ f}(z)  - S(z)\delta_{f'}^{f} =
N\tilde \Delta_{f'}^{f}(z)\, .
\end{gather}
where the right hand side of the above equations are polynomials.

At the \emph{perturbative} level, the equations
\eqref{a1b}--\eqref{a4b} have been derived in \cite{seifla}. In the
perturbative approach, the kinematical relations are not given by
\eqref{urelfla}, but by their classical counterpart obtained by
setting $q=0$. At the non-perturbative level, the anomaly equations
get non-trivial quantum corrections. However, it turns out that these
corrections can be made implicit for a privileged definition of the
variables, which is precisely the one given by \eqref{Fflaeq}. A proof
of this result in the case of the $\Nf=0$ theory was given in
\cite{mic3} and the case of arbitrary $\Nf$ will appear in
\cite{micV}.

\subsubsection{The ideal of operator relations}

One approach to solve the model, used for example in \cite{csw2}, is
to solve the anomaly equations, then to impose some ad hoc constraints
on the generating functions, and finally to fix the remaining
ambiguity by extremizing a postulated glueball superpotential. This
approach is not appropriate in our framework, since we want to obtain
a completely algebraic description of the solution.

We are going to show that both the ad hoc constraints imposed in
\cite{csw2} and the constraints coming from the glueball
superpotential are automatically implemented when the relations
\eqref{urelfla} are taken into account in addition to the anomaly
equations. Equivalently, the radical of the ideal generated by the
relations \eqref{urelfla}--\eqref{a4b} is the ideal $\ideal$ of
operator relations defined in Section \ref{FoundSec}.\footnote{In all
cases that we have checked explicitly using \textsc{Singular}, the
ideal generated by \eqref{urelfla}--\eqref{a4b} is actually radical
and thus coincides with $\ideal$. We believe that this is true in
general but we have not tried to find a proof, since this result is
not useful for our purposes.} Physically speaking, this means that the
constraint \eqref{Fflaeq} completely fixes the polynomials in the
right hand side of \eqref{a1}--\eqref{a4}, up to a discrete ambiguity
corresponding to a choice of vacuum.

Let us focus on the model \eqref{wtex4} with $\smash{W'(z)=\mu z}$
since we know from the discussion in Section \ref{UsewcSec} that the
study of this case is sufficient for our purposes.\footnote{The
general case can be treated along the same lines, see also
\cite{CRcons}.} There is no difficulty in finding the general solution
to \eqref{a1}--\eqref{a4} taking into account the asymptotic behaviour
of the generating functions. First, by combining \eqref{a3} and
\eqref{a4} and using the large $z$ limit, we find that $t_{f',\, 0}^{\
f}$ must be diagonal,
\be\label{tffd} \langle\tilde Q^{f}Q_{f'}\rangle=t_{f',\, 0}^{\ f} =
t_{f}\delta_{f'}^{f}\, .\ee
The generating functions are then expressed in terms of $v_{0}=S$ and 
the $t_{f}$,
\begin{align}\label{Sform} S(z) & = \frac{N\mu}{2}\Bigl( z -\sqrt{z^{2} - 
4S/\mu}\Bigr)\, \\ \label{Gform} G_{f'}^{\ f}(z) & =
\delta_{f'}^{f}\frac{\frac{1}{N} S(z) + t_{f}}{z-m_{f}}\,\cvp\\
\label{Rform}
R(z) & = \frac{1}{2}\sum_{f}\frac{1}{z-m_{f}} +
\frac{1}{\sqrt{z^{2}-4S/\mu}}\Bigl(N-\frac{1}{2}\sum_{f}\frac{z+
2t_{f}/\mu}{z-m_{f}}\Bigr)\, .
\end{align}
By expanding at large $z$, we see that the formulas
\eqref{Sform}--\eqref{Rform} are equivalent to identities giving the
infinite number of operators in \eqref{opedef} in terms of polynomials
in $S$ and the $t_{f}$ with coefficients in $\pfield =\mathbb
C(q,\mu,m_{1},\ldots,m_{\Nf})$ (the instanton factor $q$ actually does
not enter into these relations). We can thus write the chiral ring as 
the quotient ring
\be\label{Crfla} \hring = \pfield[t_{1},\ldots,t_{\Nf},S]/\ideal\, ,
\ee
where $\ideal$ is now the ideal generated by the set of operator
relations between the generators $S$ and $t_{f}$. This ideal contains 
all the non-trivial quantum information.

The ideal $\ideal$ can be computed in principle as follows. From
\eqref{Rform}, we find polynomial relations of the form
\be\label{ukpolrel} u_{k} = \rho_{u_{k}}(t_{1},\ldots,t_{f},S)\ee
with $\rho_{u_{k}}\in\pfield[X_{1},\ldots,X_{\Nf+1}]$. By plugging
\eqref{ukpolrel} into \eqref{urelfla}, we find in principle an
infinite set of constraints on the generators $S$ and $t_{f}$. By the
noetherian property, we know that only a finite number of these
constraints are independent. It is not difficult to use this method to
study simple cases (in practice it turns out that the first $\Nf+1$
non-trivial equations generate $\ideal$), but it becomes quite
cumbersome for large values of $N$ and $\Nf$, in particular because
the polynomials in \eqref{ukpolrel} and \eqref{urelfla} are quite
complicated. Fortunately, it is possible to find a much simpler set of
generators for the ideal $\ideal$.

\subsubsection{Simplifying the relations}

The generating function $R(z)$ given in \eqref{Rform} is a two-sheeted
analytic function which has generically $2\Nf$ poles located at
$z=m_{f}$ on both sheets. On the other hand, \eqref{Fflaeq} can be
solved explicitly and from \eqref{FRrel} we obtain an alternative
formula for $R(z)$,
\be\label{Rrelform} R(z) = \frac{1}{2}\sum_{f}\frac{1}{z-m_{f}} +
\frac{1}{\sqrt{P(z)^{2} - 4 q U(z)}}\Bigl(P'(z) -
\frac{1}{2}P(z)\sum_{f}\frac{1}{z-m_{f}}\Bigr)\, .\ee
From this formula, we see that $R(z)$ has poles only at $z=m_{f}$,
consistently with \eqref{Rform}, but we find an additional condition:
because $U(m_{f})=0$, the pole is either on the first sheet or on the
second sheet (depending on the sign of the square root) but not on
both. Let us note that this condition has been imposed in an ad hoc
way in the literature \cite{csw2}. In our framework, it is essential
to understand that it follows from the algebraic relations
\eqref{urelfla}, and that no additional ad hoc conditions need to be
imposed.

The total number of poles of $R(z)$ is thus $\Nf$ and not $2\Nf$. This
yields $\Nf$ constraints on \eqref{Rform} and thus on the $t_{f}$ and
$S$. A pole at $z=m_{f}$ on the first sheet (the first sheet is
defined by the condition $R(z)\sim N/z$ at infinity) corresponds to a
vacuum with $\nu_{f}=1$, while a pole at $z=m_{f}$ on the second sheet
corresponds to a vacuum with $\nu_{f}=0$. The residues of the poles at
$z=m_{f}$ can be computed from \eqref{Rform} and are given by
\be\label{residues} \frac{1}{2}\left(1 \mp
\frac{m_{f}+2t_{f}/\mu}{\sqrt{m_{f}^{2}-4S/\mu}}\right)\, ,\ee
with the minus or plus sign corresponding to the poles on the first
and second sheets respectively. The fact that one of these residues
must vanish is thus equivalent to $(m_{f}+2t_{f}/\mu)^{2} =
m_{f}^{2}-4S/\mu$ or
\be\label{tf1} t_{f}^{2}+\mu m_{f} t_{f} + \mu S = 0\, .\ee
This yields $\Nf$ algebraic equations that automatically belong to the
ideal $\ideal$ in \eqref{Crfla}. As we have explained, these equations
are consequences of \eqref{ukpolrel}, but are much simpler and easier
to use.

We need one additional equation (at least) to find a full set of
generators of $\ideal$. This last equation determines the glueball
$S$. In the Dijkgraaf-Vafa matrix model approach, it is found by
extremizing the glueball superpotential. In our approach, we simply
need to use one non-trivial (i.e.\ $q$-dependent) relation of the form
\eqref{urelfla}. If we expand $F(z)$ defined in \eqref{defF} as
\be\label{Fexp} F(z) = z^{N} - \sum_{k\geq 1}F_{k}z^{N-k}\, ,\ee
the simplest relation that follows from \eqref{Fflaeq} is simply
\be\label{lastIeq} F_{2N-\Nf}=q\, .\ee

Equations \eqref{tf1} and \eqref{lastIeq} are in principle all we
need. The claim is that they generate the ideal $\ideal$ and that this
ideal is prime for $\Nf<N$ (meaning that there is only one phase is
this case) or has two components in the prime decomposition
\eqref{decideal} when $\Nf\geq N$ (because in this case we have a
phase with no quantum correction corresponding to a completely broken
gauge group). If we eliminate the variables $t_{f}$ from \eqref{tf1}
and \eqref{lastIeq}, we should find a polynomial equation for $S$
whose degree is equal to the number $v$ of quantum vacua computed in
Section \ref{numbervacSec}. If $\Nf<N$, this polynomial should be
irreducible and if $\Nf\geq N$ it should have two irreducible
components. We shall prove all these properties in full generality in
\ref{firrpolSec}, by simplifying further the set of generators of the
ideal $\ideal$. In particular, we shall be able to find an explicit
formula for the polynomial equation satisfied by $S$. However, before
we tackle the general case, let us first study a simple illustrative
example.

\subsection{A simple case in details}
\label{ExHCSec}

Let us look at the theory with $N=2$ and $\Nf=1$. It is the simplest
non-trivial example, yet it displays all the important qualitative
features that are also found in the most general situation. The model
has three quantum vacua, two ``confining'' $|2,0;0\rangle = |\text
C,1\rangle$ and $|2,1;0\rangle=|\text C,2\rangle$ with unbroken gauge
group and chiral symmetry breaking and one ``Higgs''
$|1,0;1\rangle=|\text H\rangle$. Our main goal is to show that these
three vacua are in the same phase.

We have to implement Eq.\ \eqref{lastIeq} which here reads $F_{3}=q$. 
Expanding \eqref{defF}, it is straightforward to find
\be\label{F3exp} F_{3} = \frac{1}{3} u_{3} - \frac{1}{2} u_{1}u_{2} +
\frac{1}{6} u_{1}^{3} = q\, .\ee
Expanding \eqref{Rform}, we also find
\be\label{u1u2u3} u_{1} = -t/\mu\, ,\ u_{2}=(3S - m t)/\mu\, ,\ u_{3} =
-(2 S t + \mu m S + \mu m^{2} t)/\mu^{2}\ee
where we have noted $m_{1}=m$ and $t_{1}=t=\tilde Q Q$ is the meson
operator. Plugging \eqref{u1u2u3} into \eqref{F3exp} and also taking
into account \eqref{tf1}, we find the two relations that generate the
ideal $\ideal$,
\begin{align}\label{Ieq1} t^{2} + \mu m t + \mu S & = 0\\\label{Ieq2}
t^{3} + \mu(3 m t - 5 S) t + 2\mu^{2} m( m t + S) + 6 \mu^{3} q & =
0\, .\end{align}

We can now illustrate explicitly many properties discussed in Sections
\ref{FoundSec} and \ref{CRPSec}. We are going to check successively
that:\\
(i) $S$ and $t$ both satisfy irreducible degree three polynomial
equations $P_{S}=0$ and $P_{t}=0$ over $\mathbb C[\mu,m,q]$. This will
imply immediately that the Higgs and the two confining vacua belong to
the same phase.\\
(ii) $S$ and $t$ are primitive operators and thus all the operators in
the theory can be written as polynomials in $S$ or in $t$ with
coefficients in $\mathbb C(\mu,m,q)$.\\
(iii) At weak coupling, the Higgs and confining vacua are not
connected. This means that $P_{S}$ and $P_{t}$ actually factorize over
$\mathbb C[\mu,m]\{q\}$.

Point (i) can be checked by eliminating $S$ or $t$ from the two
equations \eqref{Ieq1} and \eqref{Ieq2}. It is trivial to eliminate
$S$ using \eqref{Ieq1} and plugging the result into \eqref{Ieq2} we
find the polynomial equation for $t$,
\be\label{Ptex} P_{t}(t) = t^{3} + \mu m t^{2} + \mu^{3} q  = 0\, .\ee
To find the equation for $S$, we first eliminate $t^{3}$ from
\eqref{Ieq2} by multiplying \eqref{Ieq1} by $t$ and subtracting, and
then we eliminate $t^{2}$ from the resulting equations by using the
same procedure. This yields
\be\label{Strel} S t = \mu^{2} q\ee
and
\be\label{PSex} P_{S}(S) = S^{3} + \mu^{2}mqS + \mu^{3}q^{2} = 0\,
.\ee
Let us now show that $P_{t}$ is irreducible. We write 
\be\label{irrpro1} P_{t}(t,\mu,m,q) = A(t,\mu,m,q)B(t,\mu,m,q)\, .\ee
Since the degree in $q$ of $P_{t}$ is one, either $A$ or $B$ (let us
say $A$) must be independent of $q$. By setting $q=0$ in \eqref{irrpro1}
we thus find
\be\label{irrpr2} t^{2}(t+\mu m) = A(t,\mu,m)B(t,\mu,m,q=0)\, .\ee
But $A$ cannot be a multiple of $t$ or of $t+\mu m$: it would
contradict \eqref{irrpro1} since $P_{t}(t=0,\mu,m,q)\not = 0$ and
$P_{t}(t=-\mu m,\mu,m,q)\not = 0$. Thus \eqref{irrpr2} implies that
$A$ doesn't depend on $t$, proving that $P_{t}$ is irreducible. The
birational equivalence \eqref{Strel} between the two equations
\eqref{Ptex} and \eqref{PSex} also immediately implies that $P_{S}$ is
irreducible as well. \emph{This proves that the confining and Higgs
vacua are in the same phase}.

Since the polynomial equations satisfied by $S$ and $t$ are
irreducible, they both must be primitive operators. From the
discussion in Section \ref{PrimitiveSec}, we know that all the
operators of the theory can then be expressed as polynomials in either
$t$ or $S$. We can now see this explicitly. From
\eqref{Sform}--\eqref{Rform}, it is manifest that all the operators
\eqref{opedef} are polynomials in $t$ and $S$. These immediately yield
polynomials in $t$, since Eq.\ \eqref{Ieq1} shows that $S$ itself is a
polynomial in $t$. They also yield polynomials in $S$, since we can
also express $t$ as a polynomial in $S$ by using \eqref{Strel} and
\eqref{PSex},
\be\label{Spoltrel} t=-\mu m - \frac{S^{2}}{\mu q}\,\cdotp\ee

Let us finally illustrate the relation between the weak coupling
expansion and the full quantum theory, using for example the glueball
superfield $S$. It is not difficult to solve \eqref{PSex} at small
$q$. The three roots, corresponding to the expectation values in the
three vacua, have series expansion of the form
\begin{align}\label{SqWeir1} \langle\text H|S|\text H\rangle &=\mu
m^{2}
\sum_{k\geq 1}h_{k} (q/m^{3})^{k}\\\label{SqWeir2}
\langle\text C,1|S|\text C,1\rangle &= \mu m^{2}\sum_{k\geq
1}c_{k}(q/m^{3})^{k/2}\\\label{SqWeir3}
\langle\text C,2|S|\text C,2\rangle & = \mu m^{2}\sum_{k\geq
1}(-1)^{k}c_{k}(q/m^{3})^{k/2}\, .
\end{align}
The numerical coefficients $h_{k}$, $c_{k}$ can be easily computed,
for example
\be\label{coefeccalc} h_{1} = -1\, ,\ h_{2} = 1\, ,\ h_{3}=-3\, ,\
c_{1}= i\, ,\ c_{2} = 1/2\, ,\ c_{3} = 3i/8\, ,\ldots\ee
The series expansions \eqref{SqWeir1}--\eqref{SqWeir3} clearly show
that the vacua $|\text C,1\rangle$ and $|\text C,2\rangle$ can be
analytically continued into each other at small $q$, but that they are
disconnected from the Higgs vacuum in this approximation.
Algebraically, the polynomial $P_{S}$ factorizes,
\be\label{Weirfactor}P_{S} = \tilde P_{S}^{|\text C\rangle} \tilde
P_{S}^{|\text H\rangle}\, ,\ee
where
\be\label{defPHC} \tilde P_{S}^{|\text C\rangle}(S) = \bigl(S-
\langle\text C,1|S|\text C,1\rangle\bigr) \bigl(S- \langle\text
C,2|S|\text C,2\rangle\bigr)\, ,\ \tilde P^{|\text H\rangle}(S) =
S-\langle\text H|S|\text H\rangle \ee
are the Weierstrass polynomials discussed in \ref{semiclSec} whose
coefficients are arbitrary series in $q$, i.e.\ elements of $\mathbb
C[\mu,m]\{q\}$. Going from the weak coupling approximation to the full
quantum theory is mathematically equivalent to allowing only
polynomials in $q$, and not arbitrary series, for the coefficients of
the polynomial. As we have already shown, a non-trivial decomposition
of the form \eqref{Weirfactor} is then no longer possible: $P_{S}$ is
irreducible over $\mathbb C[\mu,m,q]$, showing that strong coupling
effects make the Higgs and confining phases indistinguishable.

\subsection{The general case}
\label{firrpolSec}

As explained at the end of Section \ref{numbervacSec}, the model
\eqref{wtex4} that we are studying has vacua of rank one and also
vacua of rank zero when $\Nf\geq N$. These vacua of rank zero are
trivial in the sense that they have no quantum correction. They
correspond to a trivial solution of \eqref{Fflaeq} and
\eqref{a1}--\eqref{a4} for which $S(z)=0$ and
$F(z)=\prod_{i=1}^{N}(z-m_{f_{i}})$ is a polynomial dividing $U(z)$.
The $v_{0}=\binom{\Nf}{N}$ rank zero vacua can trivially be connected
to each other by permuting the masses $m_{f}$. The ideal of operator
relations thus decomposes as
\be\label{Idec01}\ideal = \ideal_{|0)}\cap\ideal_{|1)}\, , \ee
where $\ideal_{|0)}$ is the prime ideal of classical relations at rank
zero. All the non-trivial quantum information is included in the
operator relations in the vacua of rank one $\ideal_{|1)}$. Moreover,
note that when $\Nf<N$ there is no vacuum of rank zero and
$\ideal=\ideal_{|1)}$. Thus in all cases, the Th.\ \ref{GenFund} that
we want to prove is equivalent to the fact that $\ideal_{|1)}$ is
prime.

\subsubsection{Simple generators for $\ideal_{|1)}$}

Using \eqref{lastIeq} for general $N$ and $\Nf$ is not very convenient.
To find the general form of the algebraic equation we need, the best
approach is to solve directly the constraint \eqref{Fflaeq}. Moreover,
as explained above, we can focus on the ideal $\ideal_{|1)}$.

First, it will be useful, in an intermediate stage, to solve
explicitly \eqref{tf1} as
\be\label{solvetf} t_{f} = -\frac{\mu}{2}\Bigl( m_{f} +
(2\nu_{f}-1)\sqrt{m_{f}^{2}-4S/\mu}\Bigr)\, .\ee
The integers $\nu_{f}=0$ or 1 correspond to the labels introduced in
\ref{41Sec} to distinguish the various vacua. From \eqref{FRrel} and
\eqref{Rform} it is then straightforward to obtain, by direct
integration, an explicit expression for $F(z)$. Using
\begin{align}\label{Int1} &\int_{\infty}^{z}\d
x\Bigl(\frac{1}{\sqrt{x^{2}-a^{2}}} - \frac{1}{x}\Bigr) =
\ln\frac{z+\sqrt{z^{2}-a^{2}}}{2 z}\\ \label{Int2} &
\int_{\infty}^{z}\frac{\d x}{(x-m)\sqrt{x^{2}-a^{2}}} =
\frac{1}{\sqrt{m^{2}-a^{2}}}\ln\frac{(z-m)(m+\sqrt{m^{2}-a^{2}})}{mz -
a^{2} + \sqrt{(m^{2}-a^{2})(z^{2}-a^{2})}}\,\cvp
\end{align}
we get
\begin{multline}\label{Fintform} F(z) =
\left(\frac{z+\sqrt{z^{2}-4S/\mu}}{2}\right)^{N-\Nf/2}
\prod_{f}(z-m_{f})^{\nu_{f}}\\
\prod_{f}\left(\frac{m_{f}+\sqrt{m_{f}^{2}-4S/\mu}}{m_{f} z - 4s/\mu +
\sqrt{(m_{f}^{2}-4S/\mu)(z^{2}-4S/\mu)}}\right)^{\nu_{f}-1/2}\, .
\end{multline}
Let us now perform an analytic continuation, starting from the sheet
where $F(z)\sim z^{N}$ at infinity and going through the cut of the
square root $\sqrt{z^{2}-4S/\mu}$. Here we assume that $S\not=0$,
i.e.\ that the cut is non-trivial. This means that we exclude the
trivial classical solutions $S=0$ or in other words that we are
looking for operator relations in $\ideal_{|1)}$. The analytic
continuation produces the changes
\be\label{acont1}\sqrt{z+\sqrt{z^{2}-4S/\mu}}\ \longrightarrow\
\sqrt{z-\sqrt{z^{2}-4S/\mu}}\ee
\begin{multline} \label{acont2}
\sqrt{m_{f} z - 4s/\mu +
\sqrt{(m_{f}^{2}-4S/\mu)(z^{2}-4S/\mu)}}\ \longrightarrow\\ -
\sqrt{m_{f} z - 4s/\mu - \sqrt{(m_{f}^{2}-4S/\mu)(z^{2}-4S/\mu)}}
\end{multline}
The global minus sign in \eqref{acont2} comes from crossing part of
the double cut that originates from the double zero of $m_{f} z - 4s/\mu
- \sqrt{(m_{f}^{2}-4S/\mu)(z^{2}-4S/\mu)}$ at $z=m_{f}$. The function 
$F$ thus becomes
\begin{multline}\label{Fcontform} F(z)\ \longrightarrow\ \hat F(z) =
\left(\frac{z-\sqrt{z^{2}-4S/\mu}}{2}\right)^{N-\Nf/2}
\prod_{f}(z-m_{f})^{\nu_{f}}\\ (-1)^{\Nf}
\prod_{f}\left(\frac{m_{f}+\sqrt{m_{f}^{2}-4S/\mu}}{m_{f} z - 4s/\mu -
\sqrt{(m_{f}^{2}-4S/\mu)(z^{2}-4S/\mu)}}\right)^{\nu_{f}-1/2}\, .
\end{multline}
On the other hand, \eqref{Fflaeq} implies that
\be\label{Fcontform2} \hat F(z) = qU(z)/F(z)\, .\ee
Comparing \eqref{Fcontform} and \eqref{Fcontform2}, using a few simple
algebraic manipulations including the identity
\begin{align}\label{algiden}\left( m_{f}+\sqrt{m_{f}^{2} -
4S/\mu}\right)^{1-2\nu_{f}} &=
\left(\frac{\mu}{4S}\right)^{\nu_{f}}\Bigl( m_{f} +
(1-2\nu_{f})\sqrt{m_{f}^{2}-4S/\mu}\Bigr)\\ &
= -\left(\frac{\mu}{4S}\right)^{\nu_{f}}\frac{2S}{t_{f}}\, \cvp
\end{align}
we obtain a necessary and sufficient condition for \eqref{Rform} and
\eqref{Fflaeq} to be simultaneously satisfied,
\be\label{Strelgen} S^{N-\Nf} \prod_{f=1}^{\Nf}t_{f} = \mu^{N}q\, .\ee
This equation generalizes \eqref{Strel} to arbitrary $N$ and $\Nf$.
Together with \eqref{tf1}, we have obtained a simple set of generators
for the ideal $\ideal_{|1)}$ of operator relations,
\be\label{idealgen}\boxed{\ideal_{|1)} = \Bigl(t_{1}^{2}+\mu m_{1}t_{1} + \mu
S,\ldots, t_{\Nf}^{2}+\mu m_{\Nf}t_{\Nf} + \mu S,S^{N-\Nf}
\prod_{f=1}^{\Nf}t_{f} - \mu^{N}q\Bigr)}\,.\ee
\subsubsection{The polynomial equations for $S$}

From \eqref{Idec01}, we know that the polynomial $P_{S}$ for the
glueball $S$ is of the form
\be\label{PS11} P_{S}(S) = S^{\binom{\Nf}{N}}P_{S}^{|1)}(S)\, ,\ee
where conventionally we set $\binom{\Nf}{N}=0$ if $\Nf<N$. The
polynomial $P_{S}^{|1)}$ must be of degree $v_{1}$ given by
\eqref{vex4}. It can be constructed in principle by eliminating the
variables $t_{f}$ from the relations defining $\ideal_{|1)}$.

This is extremely elementary when $\Nf=1$. In this case, noting
$m_{1}=m$ and $t_{1}=t$, the relations are simply
\begin{align}\label{Nf11} t^{2}+\mu m t + \mu S & = 0\\
\label{Nf12} S^{N-1}t - \mu^{N} q & = 0\, .\end{align}
Solving \eqref{Nf12} for $t$ and plugging the result in \eqref{Nf11}
we find
\be\label{PSNf1} \boxed{P_{S}(S) = S^{2N-1} + \mu^{N} m q S^{N-1} +
\mu^{2N-1}q^{2}}\quad \text{for}\ \Nf =1\, .\ee
This equation generalizes \eqref{PSex} to arbitrary $N$. The case
$\Nf=2$ is a little bit more tedious but the calculation is still
tractable and yields
\begin{multline}\label{PSNf2} P_{S}^{|1)}(S) = S^{4N-4} - \mu^{N}
m_{1}m_{2}q S^{3N-4} - 2 \mu^{2N-2} q^{2} S^{2N-2} + \mu^{10N-1}
(m_{1}^{2}+m_{2}^{2})q^{2}S^{2N-3}\\ - \mu^{3N-2} m_{1}m_{2} q^{3}
S^{N-2} + \mu^{4N-4} q^{4} \quad\text{for}\ \Nf=2\, .\end{multline}
For $\Nf\geq 3$ the calculations become daunting. In particular, the
degree of $P_{S}^{|1)}$ grows exponentially. As a last example, we
indicate the solution for $N=2$ and $\Nf=3$,
\begin{multline}\label{N2Nf3ex} P_{S}^{|1)} = S^{5} + 4 \mu q^{2} S^{4}
+ \mu^{2}\bigl[ 6 q^{3} + m_{1}m_{2}m_{3} -
2(m_{1}^{2}+m_{2}^{2}+m_{3}^{2})q\bigr]q S^{3}\\ + \mu^{3}\bigl[
4q^{4} - 5 m_{1}m_{2}m_{3} q + m_{1}^{2}m_{2}^{2} + m_{1}^{2}m_{3}^{2}
+ m_{2}^{2}m_{3}^{2} -
4(m_{1}^{2}+m_{2}^{2}+m_{3}^{2})q^{2}\bigr]q^{2}S^{2}\\ +\mu^{4}\bigl[
q^{5} - 5 m_{1}m_{2}m_{3} q^{2} -
2(m_{1}^{2}+m_{2}^{2}+m_{3}^{2})q^{3} +
m_{1}m_{2}m_{3}(m_{1}^{2}+m_{2}^{2}+m_{3}^{2}) \\+ (m_{1}^{4} +
m_{2}^{4} + m_{3}^{4}) q \bigr] q^{3}S +\mu^{5}\bigl[
m_{1}m_{2}m_{3} q^{3} + m_{1}^{2}m_{2}^{2}m_{3}^{2} +
m_{1}m_{2}m_{3}(m_{1}^{2}+m_{2}^{2}+m_{3}^{2})q \\+ (m_{1}^{2}m_{2}^{2}
+ m_{1}^{2}m_{3}^{2} + m_{2}^{2}m_{3}^{2})q^{2}\bigr]
q^{4}\quad\text{for}\ N=2\ \text{and}\ \Nf=3\, .
\end{multline}

Interestingly, it is actually possible to give a general formula for
$P_{S}^{|1)}$. We claim that
\be\label{PSG1} P_{S}^{|1)}(S) =
\prod_{f=1}^{\Nf}\prod_{\nu_{f}=0}^{1}\left[ S^{N-\Nf/2} -
\frac{\mu^{N}q}{(4 S)^{\Nf/2}}\prod_{f'=1}^{\Nf}\left( - m_{f'} +
(2\nu_{f'}-1)\sqrt{m_{f'}^{2}-4S/\mu}\right)\right]\ee
for $\Nf\leq N$. From \eqref{solvetf} and \eqref{Strelgen} it is clear
that $P_{S}^{|1)}(S)=0$. The formula is single-valued by construction
and thus, by an argument already used many times, we know that the
right hand side of \eqref{PSG1} must be a rational function. This
means that when we expand \eqref{PSG1}, all the square roots
automatically cancel. Actually, we have chosen the powers of $S$ in
\eqref{PSG1} such that, for $\Nf\leq N$, only positive powers of $S$
enter in $P_{S}^{|1)}$, with
\be\label{PSexp} P_{S}^{|1)}(S) = S^{(2N-\Nf)2^{\Nf-1}} +
\cdots +\mu^{(2N-\Nf)2^{\Nf-1}} q^{2^{\Nf}}\, .\ee
When $\Nf>N$, the small $S$ behaviour is no longer necessarily
dominated by the second terms in the bracket in \eqref{PSG1} and there
are thus negative powers of $S$ in \eqref{PSG1}. It is not difficult
to see that by multiplying by a suitable power of $S$ we obtain a
polynomial with the correct degree \eqref{vex4}. For example, one can
derive Eq.\ \eqref{N2Nf3ex} most efficiently using this method.

\subsubsection{The irreducibility of $P_{S}^{|1)}$}

Let us finally prove that the ideal \eqref{idealgen} is prime. From
the analysis in Section \ref{UsewcSec}, we know that if the ideal is
prime in the case $\Nf=1$, it will automatically be prime for all
values of $\Nf$.

We thus consider the degree $2N-1$ polynomial \eqref{PSNf1}. To prove
the irreducibility, we can proceed for example as in \ref{ExHCSec}
below Eq.\ \ref{irrpro1}. Let us assume that
\be\label{Nf1sp} P_{S}(S,\mu,m,q) = A(S,\mu,m,q)B(S,\mu,m,q)\ee
where $A$ and $B$ are polynomials in $S$ with coefficients in $\mathbb
C[\mu,m,q]$. Assume that $A$ and $B$ both depend on $q$. Then their
degree in $q$ must be one. This is possible if and only if the roots
of $P_{S}$, viewed as a degree two polynomial in $q$, are rational
functions of $S$, $\mu$ and $m$. But this is not so, because the
discriminant
\be\label{Delta} \Delta = \mu^{2N-1}S^{2N-2}(m^{2}\mu - 4 S)\ee
is not a perfect square. We can thus assume that $A$, for example, is
independent of $q$. Eq.\ \eqref{Nf1sp} for $q=0$ then implies that
$S^{2N-1}=A(S,\mu,m)B(S,\mu,m,q=0)$ and thus $A(S,\mu,m) = S^{p}\tilde
A(\mu,m)$ for some $p\geq 0$. But $P_{S}(S=0)\not = 0$ and thus
necessarily $p=0$ and $A$ does not depend on $S$. This completes the
proof: \emph{there is no distinction between Higgs and confining vacua
in our theory.}

The above reasoning also shows that $S$ is a primitive operator in the
case $\Nf=1$. Actually, from the small $q$ expansion and using Prop.\
\ref{PrimTest}, it is very simple to show that $S$ is a primitive
operator for all $\Nf$. In particular, Prop.\ \ref{PrimOpexis} then
implies that $P_{S}^{|1)}$ given by \eqref{PSG1} is irreducible for
all $\Nf$, a rather non-trivial algebraic result.

\section{On the phases of the theory with one adjoint}
\label{CoulSec}

We now focus on the theory with only one adjoint chiral superfield
\eqref{wtex3}. When only adjoint fields are present, the screening
mechanism, which is responsible for the equivalence between Higgs and
confinement in theories with fundamentals, cannot occur. As a result,
the phase structure of the model is much more intricate
\cite{fer1,fer2,csw1}.

We are going to use the algebraic techniques introduced in the
previous Sections coupled with the computer algebra systems
\textsc{Singular} and PHC \cite{singular,PHC} to compute the full
phase diagram for all gauge groups $\uN$ with $2\leq N\leq 7$ (the
cases $2\leq N\leq 4$ were already worked out in \cite{fer1,fer2} and
some phases at $N=5$ and $N=6$ were also discussed in
\cite{fer2,csw1}). One of our goal is to present several non-trivial
examples of irreducible polynomial equations satisfied by primitive
operators.

\subsection{The operator relations}

The chiral ring is generated by the operators
\be\label{opedef2} u_{k} = \Tr \phi^{k}\, ,\quad v_{k} =
-\frac{1}{16\pi^{2}}\Tr W^{\alpha}W_{\alpha}\phi^{k}\, .\ee
As in \ref{OpeRel1Sec}, we introduce the generating functions
\be\label{genfgen2} R(z) = \sum_{k\geq
0}\frac{u_{k}}{z^{k+1}}=\frac{F'(z)}{F(z)}\,\cvp\quad S(z) = \sum_{k\geq
0}\frac{v_{k}}{z^{k+1}}\,\cdotp\ee
The field of parameters of the model is given by
\be\label{ringadj}\pfield = \mathbb C(g_{0},\ldots,g_{d},q)\, ,\ee
where $d$ is the degree of the derivative $W'(\phi)$ of the tree-level
superpotential. We shall always assume that $d\leq N$, since higher
values of the degree do not yield new phases.

As in \ref{OpeRel1Sec}, we have kinematical and dynamical operator
relations. We have already studied the kinematical relations in
Section \ref{opmixSec}, Ex.\ \ref{Adjarbit}. They are of the form
\eqref{R0C2} and are equivalent to the constraint \eqref{Fcons}. The
dynamical relations, on the other hand, are special cases of
\eqref{a1b} and \eqref{a2b} in which the fundamentals are integrated
out. The full set of relations thus read
\begin{gather} \label{rel1}  u_{N+p} = \tilde
Q_{p}(u_{1},\ldots,u_{N};q) \\ \label{rel2} 
N\sum_{k=0}^{d}g_{k}u_{n+k+1} - 2 \sum_{k_{1}+k_{2}=n}u_{k_{1}}v_{k_{2}}
= 0\\ \label{rel3}  N\sum_{k=0}^{d}g_{k}v_{n+k+1} - 
\sum_{k_{1}+k_{2}=n}v_{k_{1}}v_{k_{2}} = 0
\end{gather}
for any $p\geq 1$ and $n\geq -1$, or equivalently in terms of the
generating functions
\begin{gather}\label{Rel1}  F(z) + q/F(z) = P(z)\\ \label{Rel2} 
NW'(z) R(z) - 2R(z) S(z) = N^{2}\Delta_{R}(z)\\ \label{Rel3} 
NW'(z) S(z) - S(z)^{2} = N^{2}\Delta_{S}(z)\,\end{gather}
where $P$, $\Delta_{R}$ and $\Delta_{S}$ are polynomials. Eq.\
\ref{rel3} can be solved to express all the $v_{k}$ for $k\geq d$ in
terms of $v_{0},\ldots,v_{d-1}$. Eq.\ \ref{rel2} can then be used to
express all the $u_{k}$ for $k\geq d$ in terms of
$u_{1},\ldots,u_{d-1}$ and $v_{0},\ldots,v_{d-1}$. This can be made
explicit by solving \eqref{Rel2} and \eqref{Rel3},
\begin{align}\label{Sadj1} S(z) & = \frac{N}{2}\Bigl( W'(z) -
\sqrt{W'(z)^{2} - 4\Delta_{S}(z)}\Bigr)\\ \label{Sadj2}
R(z) & = \frac{N\Delta_{R}(z)}{\sqrt{W'(z)^{2} -
4\Delta_{S}(z)}}\,\cdotp\end{align}
The above formulas give all the operators $u_{k}$ and $v_{k}$ in terms
of the coefficients of the polynomials
\begin{align}\label{DR}\Delta_{R}(z) &=
g_{d}z^{d-1}+\sum_{k=0}^{d-2}a_{k}z^{k}\\ \label{DS}\Delta_{S}(z) &=
\sum_{k=0}^{d-1}b_{k}z^{k}\, .\end{align}
There is a simple linear mapping betweem the coefficients
$a_{0},\ldots,a_{d-2},b_{0},\ldots,b_{d-1}$ and the operators
$u_{1},\ldots,u_{d-1},v_{0},\ldots,v_{d-1}$ given by
\be\label{Deltaform} \Delta_{R}(z) = \frac{1}{N}\Tr\frac{W'(z) -
W'(\phi)}{z-\phi}\,\cvp\quad \Delta_{S}(z) = -\frac{1}{16\pi^{2}N}\Tr
W^{\alpha}W_{\alpha}\frac{W'(z)-W'(\phi)}{z-\phi}\,\cdotp\ee
The chiral ring can thus be expressed as
\be\label{adjCR}\hring =
\pfield[a_{0},\ldots,a_{d-2},b_{0},\ldots,b_{d-1}]/\ideal\,
,\ee
where the ideal $\ideal$ is generated by the relations obtained by
using \eqref{rel1}. From the noetherian property, we know that only a 
finite number of relations is required. Indeed, we have the following 
simple lemma.
\begin{lem}\label{AdjLem} The ideal $\ideal$ in \eqref{adjCR} is
generated by the relations \eqref{rel1} for $1\leq p\leq N+2d -2$.
\end{lem}
\noindent Indeed, the hypothesis of the lemma is equivalent to the
condition
\be\label{Fascond} F(z) + q/F(z) = P(z) + \mathcal O(z^{-N-2d +1})\,
.\ee
Using $R=F'/F$ and \eqref{Sadj2}, this yields
\be\label{Rascond} R(z) = \frac{P'(z)}{\sqrt{P(z)^{2} - 4 q}} +
\mathcal O(z^{-2N-2d})= \frac{N\Delta_{R}(z)}{\sqrt{W'(z)^{2} -
4\Delta_{S}(z)}}\, .\ee
Squaring this equality and multiplying by the denominators we find
\be\label{polkascond} P'(z)^{2}\bigl(W'(z)^{2}-4\Delta_{S}(z)\bigr) -
N^{2}\Delta_{R}^{2}(z)\bigl(P(z)^{2}-4q\bigr)=\mathcal O(z^{-1})\,
.\ee
Since the left hand side of this equality is a polynomial, it must
identically vanish. Working backward and using the asymptotics at
infinity $R(z)\sim N/z$, we deduce that \eqref{NCf2} and thus by
integration \eqref{NCf1} are valid. Equivalently, the full set of
equations \eqref{rel1} follows.

\subsection{Methodology}
\label{methSec}
\subsubsection{SINGULAR and PHC}

\textsc{Singular} \cite{singular} is a symbolic computer software for
commutative algebra and algebraic geometry. It implements rigorous and
powerful algorithms that can compute, amongst many other things, the
primary decomposition \eqref{decideal}. In principle, we can put the
explicit formulas for the generators of the ideal $\ideal$ given by
the Lem.\ \ref{AdjLem} in \textsc{Singular} and obtain as the output
the full phase diagram with explicit formulas for the generators of
the operator relations in each phase. Using the same algorithms,
\textsc{Singular} can also factorize complicated polynomials and we
have used it heavily below to prove the irreducibility of our
polynomial equations.

PHC is a numerical software for algebraic geometry that can also
compute (with a certain degree of certainty) the decomposition of an
affine variety into irreducible components. The algorithms in PHC
(which means Polynomial Homotopy Continuation) are very much in line
with the analytic approach to compute the phase diagram, Section
\ref{anaphaSec}. The software computes the intersection points (called
``witness points'') of the variety under study with generic
hyperplanes and study the permutations that these points undergo when
the hyperplanes are moved randomly. The orbits of the permutation
group acting on the witness points yield the irreducible components of
the variety. One loophole is that one can never be sure to obtain all
the possible permutations between the witness points, since the number
of random loops in hyperplane space that the computer can sample is
always finite. Nevertheless, the program can be used with confidence
to prove the irreducibility of a given component, by finding enough
permutations to ensure that the action of the permutation group is
transitive.

The simultaneous use of both PHC and \textsc{Singular} can be quite
effective. In particular, it occurs frequently that one programme is
much more efficient in terms of CPU time than the other, depending on
the details of the particular case under study. However, because only
\textsc{Singular} provides fully rigorous results, we have actually
double-checked all our calculations in the present paper using both
softwares.

\subsubsection{Some phase invariants}
\label{PinvSec}

There exists a few simple quantities that must be phase invariants
\cite{csw1}. These invariants are very useful and simplify the
computation of the phase diagram.

\paragraph{The rank}

The formulas \eqref{NCf2} and \eqref{Sadj2} are compatible only if the
following standard factorization equations are satisfied,
\begin{align}\label{FactCa} &P(z)^{2} - 4 q = M_{N-r}(z)^{2}C_{2r}(z)\\
\label{FactCb}&  W'(z)^{2} - 4\Delta_{S}(z) =
g_{d}^{2}N_{d-r}(z)^{2}C_{2r}(z)\, ,\end{align}
where $r$ is an integer satisfying $1\leq r\leq\min(d,N)=d$ and
$M_{N-r}$, $N_{d-r}$ and $C_{2r}$ are monic polynomials of degree
$N-r$, $d-r$ and $2r$ respectively. These conditions show that the
generating functions $R$ and $F$ defined in \eqref{genfgen2} and
\eqref{defF} are both single valued on the genus $r-1$ hyperelliptic
curve
\be\label{defcurve}\mathcal C_{r}:\ y^{2} = C_{2r}(z)\, .\ee
Clearly, the integer $r$ cannot change by analytic continuation and
thus it is a phase invariant. By looking at the classical limit, it is
straightforward to check that $r$ corresponds to the rank of the
vacua, defined in Section \ref{numbervacSec}, Ex.\ \ref{Ex3}.

\paragraph{A refinement of the rank}

Let us note that the polynomials
\be\label{Ppmdef} P_{\pm}(z) = P(z)\mp 2q^{1/2}\ee
cannot have common roots. Since $P^{2}-4q=P_{+}P_{-}$, \eqref{FactCa}
implies that
\be\label{spsmfact} P_{\pm}(z) = M_{\pm}(z)^{2}C_{\pm}(z)\ee
where $M_{\pm}$ and $C_{\pm}$ are polynomials of degrees $s_{\pm}$
and $N-2s_{\pm}$ respectively, with
\be\label{spsmr} s_{+}+s_{-} = N-r\, ,\quad s_{\pm}\leq N/2\, ,\ee
and
\be M_{N-r}=M_{+}M_{-}\, ,\quad C_{2r}= C_{+}C_{-}\, .\ee
When $q\mapsto e^{2i\pi}q$, the integers $s_{+}$ and $s_{-}$ are
permuted, but clearly the unordered set of integers
$\{s_{+},s_{-}\}=\{s_{-},s_{+}\}$ cannot change by analytic
continuation and is thus a phase invariant. Note that unlike the rank,
there is no clear physical interpretation of the integers $s_{+}$ and
$s_{-}$. We shall call the set $\{s_{+},s_{-}\}$ the \emph{refined
rank}.

It is actually easy to write down explicitly operator relations valid
at a given rank or for given $\{s_{+},s_{-}\}$ using the notion of
subdiscriminants, see Appendix A.

\paragraph{The confinement index}

The fact that both $R$ and $F$, $F'/F=R$, are single valued on the
same curve \eqref{defcurve} implies that the period integrals of the
one-form $R\d z$ must be integers. As is well-known, these integers
are identified with the integers $N_{i}$ and $k_{i}$ that label the
vacua $|N_{1},k_{1};\ldots;N_{d},k_{d}\rangle$ of the theory (these
vacua were dicussed in Section \ref{numbervacSec}, Ex.\ \ref{Ex3}).

Let us now consider the greatest common divisor of the compact periods
of $R\d z$ in a given vacuum of rank $r$ for which the integers
$N_{i_{1}},\ldots,N_{i_{r}}$ are non-zero,
\be\label{tdef} t = N_{i_{1}}\wedge\cdots\wedge N_{i_{r}}\wedge
(k_{i_{1}}-k_{i_{2}})\wedge\cdots\wedge (k_{i_{1}}-k_{i_{r}})\, .\ee
The periods of $\frac{1}{t}R\d z$ are thus also integers and this
implies that not only $F$ but also $F^{1/t}$ will be single-valued on 
the curve \eqref{defcurve}. Thus there exists an analytic function
$\varphi$ defined on the curve \eqref{defcurve} such that
\be\label{Ftcond} F(z) = \varphi(z)^{t}\, .\ee
Clearly, $t$ cannot change by analytic continuation and is thus a new
phase invariant. The integer $t$ can be given a nice physical
interpretation \cite{csw1}: it is the smallest positive integer such
that the $t^{\text{th}}$ tensor product of the fundamental
representation does not confine. For this reason, $t$ is usually
called the confinement index. Note that $1\leq t\leq N$ and that $t$
always divides $N$.

\subsubsection{Semi-classical interpolations}
\label{sc2Sec}

One can, as in \ref{UsewcSec}, easily find the possible semiclassical
interpolations between the vacua of our model. The quantum effective
superpotential is a special case of \eqref{Weffsemi}
\be\label{Weffsemi2} \weff^{|N_{i},k_{i}\rangle} =
\sum_{i=1}^{d}N_{i}W(w_{i}) +
\sum_{i=1}^{d}N_{i}\La_{i}^{3} e^{2i\pi k_{i}/N_{i}} + \cdots \ee
with
\be\label{LaiLarel2} \La_{i}^{3N_{i}} = q \frac{W''(w_{i})^{N_{i}}}
{\prod_{j\not = i}(w_{i}-w_{j})^{2N_{j}}}\,\cdotp\ee
These formulas show that:\\
$\bullet$ the vacua $|N_{i},k_{i}\rangle$ and $|N_{i},k_{i}+1\rangle$
are smoothly connected at weak coupling by performing the analytic
continuation $q\mapsto e^{2i\pi}q$.\\
$\bullet$ the vacua
$|\ldots;N_{i},k_{i};\ldots;N_{j},k_{j};\ldots\rangle$ and
$|\ldots;N_{j},k_{j};\ldots;N_{i},k_{i};\ldots\rangle$ are permuted
when $w_{i}$ and $w_{j}$ are permuted.\\
These are the only possible smooth interpolations between vacua at
weak coupling.

\subsection{The phase diagram}
\label{diagSec}

From the above discussion, we can deduce that the ideal of operator
relations can be decomposed as
\be\label{idealrank}\ideal = \bigcap_{\substack{1\leq s_{+},s_{-}\leq
N/2\\0\leq s_{+}+s_{-}\leq N-1}} \bigcap_{\substack{1\leq t\leq N\\
t|N}}\ideal_{\{s_{+},s_{-}\},\,t}\, ,\ee
where $\ideal_{\{s_{+},s_{-}\},\,t}$ is the ideal of operator relations 
satisfied in the phases having a given $\{s_{+},s_{-}\}$ and $t$. It
is natural to make the following conjecture.

\medskip

\noindent\textbf{Conjecture.} There is a unique phase for given
refined rank $\{s_{+},s_{-}\}$ and confinement index $t$. In other
words, the ideals $\ideal_{\{s_{+},s_{-}\},\,t}$ are prime and
\eqref{idealrank} gives the full phase structure of the model.

\medskip

It is plausible that a general mathematical proof of this conjecture
could be given. Our goal, which is to illustrate in some cases the
concepts developed in Sections \ref{FoundSec} and \ref{CRPSec}, is
more modest and we shall give a proof only when $2\leq N\leq 7$.

To study the phases at rank $r$, we always consider a tree level
superpotential of degree $d+1=r+1$. This is the minimal degree that
allows the realization of these phases. The phases then also contain
the minimal number of vacua \eqref{vrhform}. Considering $d>r$ does
not yield any new non-trivial structure; there are more vacua
\eqref{vrquan} but not more phases. The new permutations between vacua
that one needs to consider are generated by trivial classical
permutations of the roots $w_{i}$ in \eqref{Wder}. We shall also
always set $g_{d}=g_{r}=1$ for simplicity (this can be achieved by a
simple rescaling of the fields).

\subsubsection{Some simple cases in general}
\label{simcaSec}

A few phases can be easily studied for any $N$.

\paragraph{Phase of rank one} This case can be studied by considering
a quadratic tree-level superpotential $W(\phi)=\frac{1}{2}m\phi^{2}$.
There are $\hat v_{1}(N) = N$ vacua $|N,k\rangle$ with unbroken gauge
group $\uN$ that all have $t=N$ and $\{s_{+},s_{-}\}=\{N/2,N/2-1\}$ is
$N$ is even or $\{s_{+},s_{-}\}=\{(N-1)/2,(N-1)/2\}$ is $N$ is odd. It
is straightforward to find the explicit solution and to show that
$\langle N,k|S|N,k\rangle = m q^{1/N}e^{2i\pi k/N}$. All the vacua are
thus trivially related to each other by analytic continuation and thus
there is a unique phase at this rank (this also follows from the
analysis at weak coupling in \ref{sc2Sec}). This phase is of course
the same as the confining phase of the pure gauge theory
\eqref{CRsimplefield}, which can be obtained my sending $m$ to
infinity.
\paragraph{Phase of rank $N$} This is the Coulomb phase with $\hat
v_{N}(N)=1$ vacuum $|1,0;\ldots;1,0\rangle$, $t=1$ and
$\{s_{+},s_{-}\}=\{0,0\}$. The unbroken gauge group is $\u^{N}$. Note
that with one vacuum there can be only one phase. The solution to the
constraints \eqref{FactCa} and \eqref{FactCb} is simply (since
$d=r=N$)
\be\label{solNCo} W'=g_{N}P\, ,\quad \Delta_{S}=g_{N}^{2}q\, .\ee
\paragraph{Phase of rank $N-1$} There are $\hat v_{N-1}(N) = 2N-2$
vacua in this case, labeled as $|1,0;\ldots;
1,0;2,k;1,0;\ldots;1,0\rangle$, $0\leq k\leq 1$, with unbroken gauge
group $\u^{N-2}\times\text{U}(2)$. All these vacua have $t=1$ and
$\{s_{+},s_{-}\}=\{1,0\}$. There is only one phase because all the
vacua can be smoothly related at weak coupling as explained in
\ref{sc2Sec}.
\paragraph{Phases with $\{s_{+},s_{-}\}=\{0,N-r\}$} These phases can
exist at ranks $r\geq N/2$. They generalize the phases of rank $N$ and
$N-1$ discussed previously. As noticed in \cite{csw1}, the solution to
the constraints \eqref{FactCa}, \eqref{FactCb} and \eqref{spsmfact}
has a simple form. One immediately gets $M_{-}=M_{N-r}$ and
\begin{align}\nonumber C_{2r}& =P_{+}C_{-} = (P_{-}-4q^{1/2})C_{-}=
(M_{N-r}^{2}C_{-}-4 q^{1/2})C_{-}\\\label{C2rcalc}
&=(M_{N-r}C_{-})^{2} - 4 q^{1/2} C_{-}\, .\end{align}
Since $d=r$, one also has 
\be\label{C2rother} g_{r}^{2}C_{2r}=W'^{2}-4\Delta_{S}\, .\ee
Comparing \eqref{C2rcalc} and \eqref{C2rother}, we get
\be\label{C2C2}(W'-g_{r}M_{N-r}C_{-})(W'+g_{r}M_{N-r}C_{-}) =
4(\Delta_{S}-g_{r}^{2}q^{1/2}C_{-})\, .\ee
Let us assume now that $r\leq N-1$ (the solution for $r=N$ is given by
\eqref{solNCo}). This condition ensures that $\deg C_{-}=2r-N\leq
r-1$ and thus the degree of the right hand side of \eqref{C2C2} is
less than or equal to $r-1$. Since $\deg(W'+g_{r}M_{N-r}C_{-})=r$,
\eqref{C2C2} implies that
\be\label{spsolu} W'=g_{r}M_{N-r}C_{-}\, ,\quad \Delta_{S}=g_{r}^{2}q^{1/2}C_{-}\,
,\quad P=M_{N-r}^{2}C_{-}- 2 q^{1/2}\, .\ee
The first equation in \eqref{spsolu} fixes the polynomials $M_{N-r}$
and $C_{-}$. There is a $\binom{r}{N-r}$-fold degeneracy corresponding
to the choice of the $N-r$ roots of $M_{N-r}$ amongst the $r$ roots of
$W'$. The second equation fixes the glueball operators and adds a
twofold degeneracy corresponding to the choice of sign for the square
root of $q$. Overall, the solution thus describes $2\binom{r}{N-r}$
vacua. The third equation fixes the scalar operators and is also very
convenient to study the classical limit. The unbroken gauge group is
clearly $\u^{2r-N}\times\text{U}(2)^{N-r}$ and, by computing the first
semi-classical corrections, it is straightforward to check that the
$2\binom{r}{N-r}$ vacua are of the form
$|2,k;\ldots;2,k;1,0;\ldots;1,0\rangle$ for $0\leq k\leq 1$, with
$N-r$ slots $2,k$ and $2r-N$ slots $1,0$ that can be permuted in all
possible ways. All these vacua can be smoothly connected at weak
coupling and thus there is only one phase of this type for any given
$r\geq N/2$. Note finally that the confinement index is always $t=1$,
except in the case $N$ even and $r=N/2$ for which $t=2$.

The full classification of the phases for the gauge groups
$\text{U}(2)$ and $\text{U}(3)$ immediately follows from the above
discussion. \\
$\bullet$ The $\text{U}(2)$ theory can have the Coulomb phase of rank
two and confinement index one corresponding to the vacuum
$|1,0;1,0\rangle$ and the confining phase of rank one and confinement
index two corresponding to the vacua $|2,0\rangle$ and
$|2,1\rangle$.\\
$\bullet$ The $\text{U}(3)$ theory has the Coulomb phase of rank three
and confinement index one (vacuum $|1,0;1,0;1,0\rangle$), the
confining phase of rank one and confinement index three (vacua
$|3,0\rangle$, $|3,1\rangle$ and $|3,2\rangle$) and the phase of rank
two with $\{s_{+},s_{-}\}=\{0,1\}$, $t=1$ and vacua $|2,0;1,0\rangle$,
$|2,1;1,0\rangle$, $|1,0;2,0\rangle$ and $|1,0;2,1\rangle$.\\
$\bullet$ In the case of $\text{U}(4)$, we get immediately the phases
at rank four (the Coulomb phase with vacuum
$|1,0;1,0;1,0;1,0\rangle$), three (one phase containing the six vacua
$|1,0;1,0;2,0\rangle$, $|1,0;1,0;2,1\rangle$ and permutations of the
slots) and one (the confining phase with vacua $|4,k\rangle$ for
$0\leq k\leq 3$). At rank two, we have the phase
$\{s_{+},s_{-}\}=\{0,2\}$ with $t=2$ containing the two vacua
$|2,0;2,0\rangle$ and $|2,1;2,1\rangle$. There remains eight vacua at
rank two, $|2,0;2,1\rangle$, $|2,1;2,0\rangle$, $|3,k;1,0\rangle$ and
$|1,0;3,k\rangle$ for $0\leq k\leq 2$, all having $t=1$ and
$\{s_{+},s_{-}\}=\{1,1\}$. All these vacua were shown to be in the
same phase in \cite{fer2}. This gives the simplest example of a smooth
interpolation between different gauge groups, here
$\text{U}(2)\times\text{U}(2)$ and $\u\times\text{U}(3)$ \cite{csw1}.

\subsubsection{The case of $\text{U}(5)$}

The phases of ranks one, four and five have already been studied in
\ref{simcaSec}. At rank two, there are 20 vacua all having
$\{s_{+},s_{-}\}=\{2,1\}$, $t=1$ and unbroken gauge groups
$\u\times\text{U}(4)$ (eight vacua) or $\text{U}(2)\times\text{U}(3)$
(twelve vacua). These 20 vacua belong to the same phase as shown in
\cite{csw1,fer2}.

At rank three, we have six $\u\times\text{U}(2)^{2}$ vacua in the
phase $\{s_{+},s_{-}\}=\{0,2\}$. The remaining rank three vacua
correspond to the other six $\u\times\text{U}(2)^{2}$ vacua, given by
$|2,1;2,0;1,0\rangle$, $|2,0;2,1;1,0\rangle$, $|2,0;1,0;2,1\rangle$,
$|2,1;1,0;2,0\rangle$, $|1,0;2,0;2,1\rangle$, $|1,0;2,1;2,1\rangle$,
and the nine $\u^{2}\times\text{U}(3)$ vacua. These fifteen vacua all
have $t=1$ and $\{s_{+},s_{-}\}=\{1,1\}$. We have worked out the
degree fifteen polynomial equation satisfied by the operator
$x=v_{0}/5$ in these vacua,
\begin{multline}\label{PN5311} P(x)=x^{15}+\left(3 q g_1-q
g_2^2\right) x^{12}+15 q^2 x^{10}+\left(12 q^3 g_1-4 q^3 g_2^2\right)
x^7\\+\left(-4 g_1^3 q^3-4 g_0 g_2^3 q^3-27 g_0^2 q^3+g_1^2 g_2^2 q^3+18
g_0 g_1 g_2 q^3\right) x^6+48 q^4 x^5+\\\left(-4 g_2^4 q^4-36 g_1^2
q^4+24 g_1 g_2^2 q^4\right) x^4+\left(32 q^5 g_2^2-96 q^5 g_1\right)
x^2-64 q^6=0\, .\end{multline}
Note that the coefficients of the polynomial are in $\mathbb
C[g_{0},g_{1},g_{2},q]$ as they should. We have shown using PHC and
\textsc{Singular} that $P$ is irreducible over $\mathbb
C[g_{0},g_{1},g_{2},q]$. This implies that the fifteen vacua under
consideration are in the same phase.

\subsubsection{The case of $\text{U}(6)$}

Again, the phases of rank one, six and five are already known.
\paragraph{Rank two} At rank two, there are 35 vacua that can have
either $t=1$, $t=2$ or $t=3$. Thus there must be at least three
distinct phases. The three vacua $|3,k;3,k\rangle$, $0\leq k\leq 2$ at
$t=3$ are connected semi-classically, and thus must form a unique
phase. The eight vacua at $t=2$ correspond to
$\{s_{+},s_{-}\}=\{3,1\}$ and can all be obtained by semi-classical
interpolations starting for example from $|2,0;4,0\rangle$. They are
thus also trivially forming a unique phase.

The case of the 24 vacua having $t=1$ is more interesting. They all
have $\{s_{+},s_{-}\}=\{2,2\}$, so we have studied the polynomial
equations satisfied by the chiral operators in this case. In
particular, we have found using \textsc{Singular} that when
$\{s_{+},s_{-}\}=\{2,2\}$ the operator $x=v_{1}/6$ satisfies a degree
27 equation that factorizes into two irreducible pieces of degrees 3
and 24. The degree 3 part is simply $x^{3}-q$ and is associated with
the $t=3$ vacua. The degree 24 part is given by
\begin{multline}\label{P6222}
P(x) = x^{24}-8 q x^{21}+\left(14 q g_1^4-61 q g_0 g_1^2+38 q
g_0^2\right) x^{20}\\+\left(-q g_1^8+11 q g_0 g_1^6-41 q g_0^2 g_1^4+56
q g_0^3 g_1^2-16 q g_0^4\right) x^{19}+16 q^2 x^{18}\\+\left(103 q^2
g_1^4-158 q^2 g_0 g_1^2-80 q^2 g_0^2\right) x^{17}\\+\left(-13 q^2
g_1^8+115 q^2 g_0 g_1^6-307 q^2 g_0^2 g_1^4+153 q^2 g_0^3 g_1^2+367
q^2 g_0^4\right) x^{16}\\+\bigl(q^2 g_1^{12}-15 q^2 g_0 g_1^{10}+87 q^2
g_0^2 g_1^8-237 q^2 g_0^3 g_1^6+262 q^2 g_0^4 g_1^4\\+64 q^2 g_0^5
g_1^2-288 q^2 g_0^6+16 q^3\bigr) x^{15}+\bigl(64 q^2 g_0^8-48 q^2
g_1^2 g_0^7\\+12 q^2 g_1^4 g_0^6-q^2 g_1^6 g_0^5-272 q^3 g_0^2+424 q^3
g_1^2 g_0+145 q^3 g_1^4\bigr) x^{14}\\+\left(-24 q^3 g_1^8+62 q^3 g_0
g_1^6+266 q^3 g_0^2 g_1^4-1384 q^3 g_0^3 g_1^2+1584 q^3 g_0^4\right)
x^{13}\\+\bigl(-2 q^3 g_1^{12}+26 q^3 g_0 g_1^{10}-147 q^3 g_0^2
g_1^8+519 q^3 g_0^3 g_1^6-1390 q^3 g_0^4 g_1^4\\+2518 q^3 g_0^5
g_1^2-1740 q^3 g_0^6-56 q^4\bigr) x^{12}+\bigl(-5 q^3 g_0^3
g_1^{10}+79 q^3 g_0^4 g_1^8-477 q^3 g_0^5 g_1^6+1352 q^3 g_0^6
g_1^4\\+282 q^4 g_1^4-1744 q^3 g_0^7 g_1^2+792 q^4 g_0 g_1^2+768 q^3
g_0^8+240 q^4 g_0^2\bigr) x^{11}+\bigl(-128 q^3 g_0^{10}+352 q^3
g_1^2 g_0^9\\-280 q^3 g_1^4 g_0^8+98 q^3 g_1^6 g_0^7-16 q^3 g_1^8
g_0^6+q^3 g_1^{10} g_0^5+1536 q^4 g_0^4-1412 q^4 g_1^2 g_0^3\\+525 q^4
g_1^4 g_0^2-129 q^4 g_1^6 g_0+47 q^4 g_1^8\bigr) x^{10}+\bigl(q^4
g_1^{12}-16 q^4 g_0 g_1^{10}\\+114 q^4 g_0^2 g_1^8-103 q^4 g_0^3
g_1^6-688 q^4 g_0^4 g_1^4+2140 q^4 g_0^5 g_1^2-2528 q^4 g_0^6-32
q^5\bigr) x^9\\+\bigl(10 q^4 g_0^3 g_1^{10}-113 q^4 g_0^4 g_1^8+328 q^4
g_0^5 g_1^6+128 q^4 g_0^6 g_1^4-49 q^5 g_1^4-1558 q^4 g_0^7 g_1^2\\+1076
q^5 g_0 g_1^2+1583 q^4 g_0^8+728 q^5 g_0^2\bigr) x^8+\bigl(-480 q^4
g_0^{10}+448 q^4 g_1^2 g_0^9-22 q^4 g_1^4 g_0^8-75 q^4 g_1^6 g_0^7\\+23
q^4 g_1^8 g_0^6-2 q^4 g_1^{10} g_0^5+48 q^5 g_0^4-1456 q^5 g_1^2
g_0^3+412 q^5 g_1^4 g_0^2+232 q^5 g_1^6 g_0-10 q^5 g_1^8\bigr)
x^7\\+\bigl(64 q^4 g_0^{12}-48 q^4 g_1^2 g_0^{11}+12 q^4 g_1^4
g_0^{10}-q^4 g_1^6 g_0^9-912 q^5 g_0^6+944 q^5 g_1^2 g_0^5\\-165 q^5
g_1^4 g_0^4-286 q^5 g_1^6 g_0^3-27 q^5 g_1^8 g_0^2+5 q^5 g_1^{10}
g_0+64 q^6\bigr) x^6\\+\bigl(-5 q^5 g_0^3 g_1^{10}+34 q^5 g_0^4
g_1^8+68 q^5 g_0^5 g_1^6+80 q^6 g_1^4-274 q^5 g_0^6 g_1^4\\-136 q^5
g_0^7 g_1^2+584 q^6 g_0 g_1^2+664 q^5 g_0^8+416 q^6 g_0^2\bigr)
x^5\\+\bigl(-186 q^5 g_0^{10}-89 q^5 g_1^2 g_0^9+128 q^5 g_1^4 g_0^8-9
q^5 g_1^6 g_0^7-7 q^5 g_1^8 g_0^6+q^5 g_1^{10} g_0^5\\-296 q^6
g_0^4-1080 q^6 g_1^2 g_0^3-350 q^6 g_1^4 g_0^2-48 q^6 g_1^6 g_0+q^6
g_1^8\bigr) x^4\\+\bigl(16 q^5 g_0^{12}+24 q^5 g_1^2 g_0^{11}-15 q^5
g_1^4 g_0^{10}+2 q^5 g_1^6 g_0^9-48 q^6 g_0^6\\+632 q^6 g_1^2 g_0^5+362
q^6 g_1^4 g_0^4+80 q^6 g_1^6 g_0^3+64 q^7\bigr) x^3\\+\left(80 q^6
g_0^8-132 q^6 g_1^2 g_0^7-57 q^6 g_1^4 g_0^6-53 q^6 g_1^6 g_0^5+64 q^7
g_0^2+16 q^7 g_1^2 g_0-8 q^7 g_1^4\right) x^2\\+\left(-16 q^6
g_0^{10}+18 q^6 g_1^2 g_0^9-15 q^6 g_1^4 g_0^8+13 q^6 g_1^6 g_0^7-64
q^7 g_0^4-88 q^7 g_1^2 g_0^3+8 q^7 g_1^4 g_0^2\right) x\\+q^6
g_0^{12}+16 q^8+8 q^7 g_0^6-q^6 g_0^9 g_1^6+3 q^6 g_0^{10} g_1^4-q^7
g_0^4 g_1^4-3 q^6 g_0^{11} g_1^2+20 q^7 g_0^5=0\, .
\end{multline}
This is a rather non-trivial example of a polynomial equation. Its
irreducibility, proven using PHC and \textsc{Singular}, implies that
the 24 vacua at $t=1$ form a unique phase. In particular, the eight
$t=1$, $\text{U}(2)\times\text{U}(4)$ vacua, the ten
$\text{U}(5)\times\u$ vacua and the six $t=1$, $\text{U}(3)^{2}$ vacua
can all be smoothly analytically continued into each other.
\paragraph{Rank three} There are $\hat v_{3}(6)=56$ vacua at rank
three. The vacua $|2,0;2,0;2,0\rangle$ and $|2,1;2,1;2,1\rangle$ form
the $\{0,3\}$ phase with $t=2$ and unbroken gauge group
$\text{U}(2)^{3}$. There remains 54 vacua that all have $t=1$ and
$\{s_{+},s_{-}\}=\{2,1\}$, with patterns of gauge symmetry breaking
$\text{U}(4)\times\u^{2}$ (12 vacua),
$\text{U}(3)\times\text{U}(2)\times\text{U}(1)$ (36 vacua) and
$\text{U}(2)^{2}$ (6 vacua at $t=1$). We have been able to show with
\textsc{Singular} and PHC that these 54 vacua form a unique phase and
that the glueball operator $x=v_{0}/6$ is primitive. The
polynomial equation satisfied by $x$ is of the form
\be\label{P6321one} P(x) = A_{+}(x)A_{-}(x)=0\, .\ee
The $A_{\pm}$ are polynomials of degree 27 that are irreducible over
$\mathbb C[g_{0},g_{1},g_{2},q^{1/2}]$. The factors $A_{+}$ and
$A_{-}$ are permuted into each other when $q^{1/2}\mapsto -q^{1/2}$,
making the polynomial $P$ irreducible over $\mathbb
C[g_{0},g_{1},g_{2},q]$. Explicitly, one has
\begin{multline}\label{P6321} A_{+}(x) = x^{27}-\sqrt{q} \left(3
g_1-g_2^2\right) x^{25}+18 q x^{24}-2 q^{3/2} \left(3 g_1-g_2^2\right)
x^{22}\\-q^{3/2} \left(-27 g_0^2+2 \left(9 g_1 g_2-2 g_2^3\right)
g_0+g_1^2 \left(g_2^2-4 g_1\right)-36 \sqrt{q}\right) x^{21}-6 q^2
\left(g_2^2-3 g_1\right){}^2 x^{20}\\-86 q^{5/2} \left(g_2^2-3
g_1\right) x^{19}-3 q^{5/2} \left(27 g_0^2+\left(4 g_2^3-18 g_1
g_2\right) g_0+g_1^2 \left(4 g_1-g_2^2\right)+128 \sqrt{q}\right)
x^{18}\\+15 q^3 \left(g_2^2-3 g_1\right){}^2 x^{17}-164 q^{7/2} \left(3
g_1-g_2^2\right) x^{16}\\+q^{7/2} \left(g_2^6-9 g_1 g_2^4+21 g_1^2
g_2^2-3 g_1^3+162 g_0^2-12 g_0 \left(9 g_1 g_2-2 g_2^3\right)+684
\sqrt{q}\right) x^{15}\\-68 q^{9/2} \left(g_2^2-3 g_1\right)
x^{13}\\-q^{9/2} \left(2 g_2^6-18 g_1 g_2^4+47 g_1^2 g_2^2-26 g_1^3+189
g_0^2-14 g_0 \left(9 g_1 g_2-2 g_2^3\right)+576 \sqrt{q}\right)
x^{12}\\-15 q^5 \left(g_2^2-3 g_1\right){}^2 x^{11}-28 q^{11/2}
\left(g_2^2-3 g_1\right) x^{10}\\+q^{11/2} \left(g_2^6-9 g_1 g_2^4+21
g_1^2 g_2^2-3 g_1^3+162 g_0^2-12 g_0 \left(9 g_1 g_2-2
g_2^3\right)+336 \sqrt{q}\right) x^9\\+6 q^6 \left(g_2^2-3
g_1\right){}^2 x^8-28 q^{13/2} \left(3 g_1-g_2^2\right) x^7\\-3 q^{13/2}
\left(27 g_0^2+\left(4 g_2^3-18 g_1 g_2\right) g_0+g_1^2 \left(4
g_1-g_2^2\right)+48 \sqrt{q}\right) x^6\\-16 q^{15/2} \left(g_2^2-3
g_1\right) x^4-q^{15/2} \left(-27 g_0^2+2 \left(9 g_1 g_2-2
g_2^3\right) g_0+g_1^2 \left(g_2^2-4 g_1\right)-36 \sqrt{q}\right)
x^3\\-4 q^{17/2} \left(3 g_1-g_2^2\right) x-8 q^9\, .
\end{multline}
\paragraph{Rank four} Of the 36 vacua at rank four, 12 belong to the
$\{0,2\}$ phase with unbroken gauge group
$\u^{2}\times\text{U}(2)^{2}$. These vacua can all be obtained by
semi-classical interpolations starting for example from
$|1,0;1,0;2,0;2,0\rangle$. The non-trivial case concerns the 24 vacua
having $\{s_{+},s_{-}\}=\{1,1\}$. There are twelve
$\text{U}(3)\times\u^{3}$ vacua and twelve
$\u^{2}\times\text{U}(2)^{2}$ vacua of this sort. We have shown with
PHC and \textsc{Singular} that they all belong to the same phase, the
glueball operator $x=v_{0}/6$ being primitive with degree 24
irreducible equation given by
\begin{multline}\label{P6411} P(x) = x^{24}+\left(q g_3^3-4 q g_2
g_3+8 q g_1\right) x^{21}\\+\left(4 q^2 g_2^3-q^2 g_3^2 g_2^2-16 q^2 g_0
g_2-14 q^2 g_1 g_3 g_2+3 q^2 g_1 g_3^3+18 q^2 g_1^2+6 q^2 g_0
g_3^2\right) x^{18}\\+\left(15 q^3 g_3^4-80 q^3 g_2 g_3^2+56 q^3 g_1
g_3+88 q^3 g_2^2-224 q^3 g_0\right) x^{16}+\left(640 q^4 g_2-240 q^4
g_3^2\right) x^{14}\\+\bigl(12 g_1 g_3^4 q^4-4 g_2^2 g_3^3 q^4-48 g_0
g_3^3 q^4-32 g_1 g_2^2 q^4\\-48 g_1 g_2 g_3^2 q^4-384 g_0 g_1 q^4+16
g_2^3 g_3 q^4+96 g_1^2 g_3 q^4+192 g_0 g_2 g_3 q^4\bigr)
x^{13}\\+\bigl(2176 q^5-27 g_1^4 q^4+16 g_0 g_2^4 q^4-27 g_0^2 g_3^4
q^4+256 g_0^3 q^4-4 g_1^2 g_2^3 q^4-4 g_1^3 g_3^3 q^4\\+18 g_0 g_1 g_2
g_3^3 q^4-128 g_0^2 g_2^2 q^4-4 g_0 g_2^3 g_3^2 q^4-6 g_0 g_1^2 g_3^2
q^4+g_1^2 g_2^2 g_3^2 q^4\\+144 g_0^2 g_2 g_3^2 q^4+144 g_0 g_1^2 g_2
q^4-80 g_0 g_1 g_2^2 g_3 q^4-192 g_0^2 g_1 g_3 q^4+18 g_1^3 g_2 g_3
q^4\bigr) x^{12}\\+\left(48 g_3^5 q^5-320 g_2 g_3^3 q^5+384 g_1 g_3^2
q^5-1024 g_1 g_2 q^5+512 g_2^2 g_3 q^5\right) x^{11}\\+\bigl(16 g_2^5
q^5-36 g_1^2 g_3^4 q^5+128 g_0 g_2^3 q^5+24 g_1 g_2^2 g_3^3 q^5+72 g_0
g_1 g_3^3 q^5\\+864 g_0 g_1^2 q^5+72 g_1^2 g_2^2 q^5-4 g_2^4 g_3^2
q^5+288 g_0^2 g_3^2 q^5-24 g_0 g_2^2 g_3^2 q^5\\+168 g_1^2 g_2 g_3^2
q^5-768 g_0^2 g_2 q^5-216 g_1^3 g_3 q^5-104 g_1 g_2^3 g_3 q^5-480 g_0
g_1 g_2 g_3 q^5\bigr) x^{10}\\+\left(-448 g_3^3 q^6-3584 g_1 q^6+1792
g_2 g_3 q^6\right) x^9\\+\bigl(-96 g_1 g_3^5 q^6+272 g_2^4 q^6+32 g_2^2
g_3^4 q^6+96 g_0 g_3^4 q^6+608 g_1 g_2 g_3^3 q^6-1792 g_0^2 q^6+384
g_0 g_2^2 q^6\\-192 g_2^3 g_3^2 q^6-400 g_1^2 g_3^2 q^6-512 g_0 g_2
g_3^2 q^6+768 g_1^2 g_2 q^6-864 g_1 g_2^2 g_3 q^6+896 g_0 g_1 g_3
q^6\bigr) x^8\\+\bigl(-64 g_3^6 q^7+512 g_2 g_3^4 q^7+1792 g_2^3
q^7+512 g_1 g_3^3 q^7\\+1664 g_1^2 q^7-1536 g_2^2 g_3^2 q^7-384 g_0
g_3^2 q^7+1024 g_0 g_2 q^7-1920 g_1 g_2 g_3 q^7\bigr) x^6\\+\left(768
g_3^4 q^8+5632 g_2^2 q^8-4096 g_2 g_3^2 q^8+2048 g_0 q^8-512 g_1 g_3
q^8\right) x^4\\+\left(8192 q^9 g_2-3072 q^9 g_3^2\right) x^2+4096
q^{10}
=0\, .
\end{multline}
\subsubsection{The case of $\text{U}(7)$}

This is the most complex case that we are going to study. Note that
because $N=7$ is prime, all the phases with $r\geq 2$ have $t=1$.
Again, the phases of rank one, six and seven, as well as some phases
at ranks four and five, have been studied in \ref{simcaSec}.
\paragraph{Rank two} At rank two, we have twelve $\u\times\text{U}(6)$
vacua, twenty $\text{U}(2)\times\text{U}(5)$ vacua and twenty-four
$\text{U}(3)\times\text{U}(4)$ vacua, for a total of 56 vacua. All
these vacua have the same phase invariants: $r=2$, $t=1$,
$\{s_{+},s_{-}\}=\{3,2\}$. One thus could expect to have a unique
phase containing all these vacua. We have found the degree 56
polynomial equation satisfied by the glueball operator $x=v_{0}/7$. It
has the form \eqref{P6321one}, where now the factors $A_{\pm}$ are
polynomials of degree 28 over $\mathbb C[g_{0},g_{1},q^{1/2}]$ that
are permuted when $q^{1/2}\mapsto -q^{1/2}$. Explicitly,
\begin{multline}\label{P7232} A_{+}(x) = x^{28}+2 \sqrt{q}
\left(g_1^2-4 g_0\right) x^{25}+q \left(g_1^2-4 g_0\right){}^2
x^{22}+46 q^{3/2} x^{21}+4 q^2 \left(4 g_0-g_1^2\right) x^{18}\\-2
q^{5/2} \left(g_1^2-4 g_0\right){}^2 x^{15}-21 q^3 x^{14}+q^3 \left(4
g_0-g_1^2\right){}^3 x^{12}+36 q^{7/2} \left(4 g_0-g_1^2\right)
x^{11}\\+3 q^4 \left(g_1^2-4 g_0\right){}^2 x^8+102 q^{9/2} x^7+3 q^5
\left(4 g_0-g_1^2\right) x^4+q^6\, .
\end{multline}
We have shown using PHC and \textsc{Singular} that $A_{+}$ is
irreducible over $\mathbb C[g_{0},g_{1},q^{1/2}]$, which implies
immediately that $A_{+}A_{-}$ is irreducible over $\mathbb
C[g_{0},g_{1},q]$: the 56 vacua are indeed in the same phase.
\paragraph{Rank three} Since there are 126 rank three vacua, all the
chiral operators satisfy a polynomial equation of degree 126 with
coefficients in $\mathbb C[g_{0},g_{1},g_{2},q]$. We have found this
equation for various chiral operators. In particular, we have shown
with \textsc{Singular} that the equation for the glueball operator
$x=v_{0}/7$ factorizes into two irreducible pieces of degree 42 and 84
associated with two phases $|42)$ and $|84)$,
\be\label{P73gen} P(x) = P_{42}(x)P_{84}(x) = 0\, .\ee
Moreover, $P_{42}$ factorizes over $\mathbb
C[g_{0},g_{1},g_{2},q^{1/2}]$ into two degree 21 factors $A_{+}$ and
$A_{-}$ that are exchanged under $q^{1/2}\mapsto - q^{1/2}$. This
shows that $P_{42}$ corresponds to the $\{s_{+},s_{-}\}=\{3,1\}$
vacua, which therefore must all be in $|42)$. The other 84 vacua thus
all have $\{s_{+},s_{-}\}=\{2,2\}$ and must all be in $|84)$.

It is easy to identify the possible unbroken gauge groups in each
phase, for example by looking at the classical limit of the polynomial
equations for the operators $u_{k}$. It is more difficult to compute
the integers $k_{i}$ for each vacua of the form
$|N_{1},k_{1};N_{2},k_{2};N_{3},k_{3}\rangle$ in a given phase. To do so,
we have computed numerically the gluino condensates $s_{i}$ in the
unbroken $\text{U}(N_{i})$ factors of the gauge group by computing the
relevant contour integrals of the generating function $S(z)$ given in
\eqref{Sadj2}. This calculation, that must be repeated in each
individual vacua, can be easily implemented on Mathematica. The
integers $k_{i}$ can then be extracted from the small $q$ behaviour
$s_{i}\simeq\La_{i}^{3}e^{2i\pi k_{i}/N_{i}}$, where $\La_{i}$ is
given by \eqref{LaiLarel2}. One can also extract the $k_{i}$ from some
contour integrals of the generating function $R(z)$ (see
\eqref{Sadj1}), and we have double-checked the results in this way.

It turns out that the phase $|42)$ contains the twenty-four
$\u\times\text{U}(2)\times\text{U}(4)$ vacua that can be obtained from
$|1,0;2,0;4,0\rangle$ by semi-classical interpolations. It also
contains the eighteen $\text{U}(2)^{2}\times\text{U}(3)$ vacua that
can be obtained from $|2,0;2,0;3,0\rangle$ by semi-classical
interpolations. For completeness, we also give the formula for the
degree 21 polynomial $A_{+}$ in this case,
\begin{multline}\label{P7331} A_{+}(x)= x^{21}+2 q \left(3
g_1-g_2^2\right) x^{17}-3 q^2 x^{14}+q^2 \left(g_2^2-3 g_1\right){}^2
x^{13}\\+q^3 \left(g_2^2-3 g_1\right) x^{10}+q^3 \left(27 g_0^2+\left(4
g_2^3-18 g_1 g_2\right) g_0+g_1^2 \left(4 g_1-g_2^2\right)\right)
x^9\\-3 q^{7/2} \left(2 g_2^3-9 g_1 g_2+27 g_0\right) x^8+57 q^4 x^7+q^5
\left(g_2^2-3 g_1\right) x^3-q^6\, .
\end{multline}

The phase $|84)$ contains all the rank three vacua that are not in
$|42)$, which includes twenty-four
$\u\times\text{U}(2)\times\text{U}(4)$ vacua, eighteen
$\text{U}(2)^{2}\times\text{U}(3)$ vacua, fifteen
$\u^{2}\times\text{U}(5)$ vacua and twenty-seven
$\u\times\text{U}(3)^{2}$ vacua. The polynomial $P_{84}$ is extremely
complicated. It turns out that if we set $g_{1}=g_{2}=0$, the
polynomial remains irreducible (this of course implies that the
polynomial is irreducible in the general case). It is thus enough to
present $P_{82}$ in this special case,
\begin{multline}\label{P7322} P(x) = x^{84}-81 q g_0^2 x^{79}+3699 q^2
x^{77}+2187 q^2 g_0^4 x^{74}+254367 q^3 g_0^2 x^{72}+3413310 q^4
x^{70}\\-19683 q^3 g_0^6 x^{69}-667035 q^4 g_0^4 x^{67}-7708608 q^5
g_0^2 x^{65}-13620477 q^6 x^{63}-708588 q^5 g_0^6 x^{62}\\-5226930 q^6
g_0^4 x^{60}+43654221 q^7 g_0^2 x^{58}-1062882 q^6 g_0^8
x^{57}-70179075 q^8 x^{56}\\-35783694 q^7 g_0^6 x^{55}-496822977 q^8
g_0^4 x^{53}-2358810882 q^9 g_0^2 x^{51}-20726199 q^8 g_0^8
x^{50}\\-1698777354 q^{10} x^{49}-193523256 q^9 g_0^6 x^{48}+940766481
q^{10} g_0^4 x^{46}-14348907 q^9 g_0^{10} x^{45}\\-505521243 q^{11}
g_0^2 x^{44}-117979902 q^{10} g_0^8 x^{43}-4394981908 q^{12}
x^{42}-117920853 q^{11} g_0^6 x^{41}\\-1101683754 q^{12} g_0^4
x^{39}+19938963645 q^{13} g_0^2 x^{37}-129671604 q^{12} g_0^8
x^{36}-20347899486 q^{14} x^{35}\\-814226661 q^{13} g_0^6
x^{34}+4435334415 q^{14} g_0^4 x^{32}-8334566253 q^{15} g_0^2
x^{30}+2904592227 q^{16} x^{28}\\+277451568 q^{15} g_0^6
x^{27}-866557197 q^{16} g_0^4 x^{25}+684971721 q^{17} g_0^2
x^{23}+154884143 q^{18} x^{21}\\-20016153 q^{18} g_0^4 x^{18}+62696268
q^{19} g_0^2 x^{16}\\-24397098 q^{20} x^{14}+367389 q^{21} g_0^2
x^9-9885 q^{22} x^7-q^{24}\, .
\end{multline}
\paragraph{Rank four} At rank four, we have a simple $\{3,0\}$ phase
which contains the eight vacua obtained from $|2,0;2,0;2,0;1,0\rangle$
by semi-classical interpolations. The other 112 vacua all have
$\{s_{+},s_{-}\}=\{2,1\}$. The glueball operator $x=v_{0}/7$ satisfies
a degree 112 irreducible equation of the form \eqref{P6321one}, where
now $A_{+}$ is of degree 56. It turns out that $P$ remains irreducible
is we set $g_{1}=g_{2}=g_{3}=0$, so we can restrict ourselves to this
case for which
\begin{multline}\label{P7421} A_{+}(x) =x^{56}+304 q^{3/2} g_0
x^{51}-5532 q^{5/2} x^{49}+256 q^2 g_0^3 x^{48}-4000 q^3 g_0^2
x^{46}+27312 q^4 g_0 x^{44}\\+128262 q^5 x^{42}+768 q^{9/2} g_0^3
x^{41}-64064 q^{11/2} g_0^2 x^{39}-301616 q^{13/2} g_0 x^{37}+8448 q^6
g_0^4 x^{36}\\-674364 q^{15/2} x^{35}+92928 q^7 g_0^3 x^{34}+100704 q^8
g_0^2 x^{32}-667440 q^9 g_0 x^{30}+24576 q^{17/2} g_0^4 x^{29}\\-13439
q^{10} x^{28}-111616 q^{19/2} g_0^3 x^{27}-1355520 q^{21/2} g_0^2
x^{25}+73728 q^{10} g_0^5 x^{24}-582656 q^{23/2} g_0 x^{23}\\+307200
q^{11} g_0^4 x^{22}+247776 q^{25/2} x^{21}-747008 q^{12} g_0^3
x^{20}-1274368 q^{13} g_0^2 x^{18}+196608 q^{25/2} g_0^5 x^{17}\\-531328
q^{14} g_0 x^{16}-65536 q^{27/2} g_0^4 x^{15}-179840 q^{15}
x^{14}-679936 q^{29/2} g_0^3 x^{13}+65536 q^{14} g_0^6 x^{12}\\-679936
q^{31/2} g_0^2 x^{11}+131072 q^{15} g_0^5 x^{10}-411648 q^{33/2} g_0
x^9+98304 q^{16} g_0^4 x^8-149504 q^{35/2} x^7\\+65536 q^{17} g_0^3
x^6+36864 q^{18} g_0^2 x^4+8192 q^{19} g_0 x^2+4096 q^{20}\, .
\end{multline}
Thus we can interpolate smoothly between the twenty-four
$\u\times\text{U}(2)^{3}$ vacua, sixteen $\u^{3}\times\text{U}(4)$
vacua and seventy-two $\u^{2}\times\text{U}(2)\times\text{U}(3)$ vacua
of the phase.
\paragraph{Rank five} The twenty vacua that can be obtained by
semi-classical interpolations from $|2,0;2,0;1,0;1,0;1,0\rangle$ form
the phase $\{s_{+},s_{-}\}=\{0,2\}$. The remaining thirty-five vacua
(twenty $\u^{3}\times\text{U}(2)^{2}$ and fifteen
$\u^{4}\times\text{U}(3)$) all have $\{s_{+},s_{-}\}=\{1,1\}$ and form
a unique phase. Indeed, $x=v_{0}/7$ satisfies a degree 35 polynomial
equation. For $g_{4}=0$ (this can always be achieved by a simple shift
in the tree-level superpotential), this equation reads
\begin{multline}\label{P7511} P(x) = x^{35}-q \bigl(4 g_3^2-15
g_1\bigr) x^{32}-65 q^2 g_3 x^{30}\\-q^2 \bigl(-4 g_3^4+36 g_1 g_3^2-27
g_2^2 g_3-80 g_1^2+50 g_0 g_2\bigr) x^{29}-705 q^3 x^{28}+100 q^3
\bigl(g_3^3-4 g_1 g_3+4 g_2^2\bigr) x^{27}\\-q^3 \bigl(27 g_2^4+4 g_3^3
g_2^2+40 g_0 g_3^2 g_2-160 g_1^3+88 g_1^2 g_3^2-125 g_0^2 g_3+3 g_1
\bigl(-4 g_3^4-39 g_2^2 g_3+100 g_0 g_2\bigr)\bigr) x^{26}\\-4 q^4
\bigl(1140 g_1-529 g_3^2\bigr) x^{25}-q^4 \bigl(-60 g_3^5-353 g_2^2
g_3^2+160 g_1^2 g_3+3350 g_0 g_2 g_3-3750 g_0^2\\-2 g_1 \bigl(705
g_2^2-92 g_3^3\bigr)\bigr) x^{24}+21760 q^5 g_3 x^{23}\\-4 q^5
\bigl(760 g_1^2+1158 g_3^2 g_1+6650 g_0 g_2-47 \bigl(9 g_3^4+28 g_2^2
g_3\bigr)\bigr) x^{22}\\+4 q^5 \bigl(12 g_1 g_3^5-4 g_2^2 g_3^4-128
g_1^2 g_3^3+117 g_1 g_2^2 g_3^2-27 g_2^4 g_3+320 g_1^3 g_3-180 g_1^2
g_2^2\\+30640 q+625 g_0^2 \bigl(4 g_1-g_3^2\bigr)+10 g_0 g_2 \bigl(8
g_3^3-40 g_1 g_3+45 g_2^2\bigr)\bigr) x^{21}\\-q^5 \bigl(256 g_1^5-128
g_3^2 g_1^4+16 \bigl(g_3^4+9 g_2^2 g_3\bigr) g_1^3-\bigl(27 g_2^4+4
g_3^3 g_2^2\bigr) g_1^2+62400 q g_3 g_1+3125 g_0^4\\-3750 g_0^3 g_2
g_3+80 q \bigl(5 g_2^2-306 g_3^3\bigr)+g_0^2 \bigl(108 g_3^5+825
g_2^2 g_3^2+2000 g_1^2 g_3+450 g_1 \bigl(5 g_2^2-2
g_3^3\bigr)\bigr)\\+2 g_0 g_2 \bigl(54 g_2^4+8 g_3^3 g_2^2-800
g_1^3+280 g_1^2 g_3^2-9 g_1 \bigl(4 g_3^4+35 g_2^2
g_3\bigr)\bigr)\bigr) x^{20}\\+16 q^6 \bigl(880 g_1^3-184 g_3^2
g_1^2+\bigl(-73 g_3^4+116 g_2^2 g_3+600 g_0 g_2\bigr) g_1+3 \bigl(4
g_3^6+32 g_2^2 g_3^3+10 g_0 g_2 g_3^2\\-500 g_0^2 g_3+33
g_2^4\bigr)\bigr) x^{19}-4 q^6 \bigl(27 g_2^6+4 g_3^3 g_2^4-198 g_1
g_3 g_2^4-24 g_1 g_3^4 g_2^2-40 g_1^3 g_2^2+434 g_1^2 g_3^2 g_2^2\\-6250
g_0^3 g_2+10 g_0 \bigl(-33 g_1 g_3^3+6 g_2^2 g_3^2+20 g_1^2 g_3-45 g_1
g_2^2\bigr) g_2+36 g_1^2 g_3^5-224 g_1^3 g_3^3\\+320 g_1^4 g_3+32 q
\bigl(2095 g_1-1292 g_3^2\bigr)+25 g_0^2 \bigl(27 g_3^4-150 g_1
g_3^2+120 g_2^2 g_3+200 g_1^2\bigr)\bigr) x^{18}\\-64 q^7 \bigl(-67
g_3^5+558 g_1 g_3^3-436 g_2^2 g_3^2-680 g_1^2 g_3-300 g_0 g_2 g_3+1875
g_0^2-95 g_1 g_2^2\bigr) x^{17}\\+16 q^7 \bigl(-400 g_1^4+360 g_3^2
g_1^3-5 \bigl(-3 g_3^4+192 g_2^2 g_3+200 g_0 g_2\bigr) g_1^2+2
\bigl(-12 g_3^6\\-70 g_2^2 g_3^3+825 g_0 g_2 g_3^2+2500 g_0^2 g_3+135
g_2^4\bigr) g_1-40 g_0 g_2 g_3 \bigl(g_3^3+15 g_2^2\bigr)\\-125 g_0^2
\bigl(14 g_3^3+15 g_2^2\bigr)+g_3 \bigl(45 g_3 g_2^4+8 g_3^4
g_2^2+28960 q\bigr)\bigr) x^{16}\\+256 q^8 \bigl(129 g_3^4-988 g_1
g_3^2+736 g_2^2 g_3+640 g_1^2+350 g_0 g_2\bigr) x^{15}\\+64 q^8
\bigl(-4 g_3^7-106 g_1 g_3^5+7 g_2^2 g_3^4+308 g_1^2 g_3^3-612 g_1
g_2^2 g_3^2+102 g_2^4 g_3+80 g_1^3 g_3-570 g_1^2 g_2^2\\+9360 q+1250
g_0^2 \bigl(3 g_1-2 g_3^2\bigr)-10 g_0 g_2 \bigl(8 g_3^3-215 g_1
g_3+70 g_2^2\bigr)\bigr) x^{14}\\+10240 q^9 \bigl(11 g_3^3-75 g_1
g_3+40 g_2^2\bigr) x^{13}-256 q^9 \bigl(-53 g_2^4+52 g_3^3 g_2^2+240
g_0 g_3^2 g_2\\+40 g_1^3-522 g_1^2 g_3^2+g_1 \bigl(194 g_3^4+708 g_2^2
g_3-950 g_0 g_2\bigr)+25 \bigl(g_3^6+70 g_0^2 g_3\bigr)\bigr)
x^{12}\\-4096 q^{10} \bigl(235 g_1-46 g_3^2\bigr) x^{11}\\-1024 q^{10}
\bigl(66 g_3^5+118 g_2^2 g_3^2-360 g_1^2 g_3+400 g_0 g_2 g_3+625
g_0^2+4 g_1 \bigl(49 g_3^3+65 g_2^2\bigr)\bigr) x^{10}\\+245760 q^{11}
g_3 x^9-4096 q^{11} \bigl(95 g_3^4+124 g_1 g_3^2+92 g_2^2 g_3-95
g_1^2+200 g_0 g_2\bigr) x^8+327680 q^{12} x^7\\-81920 q^{12} \bigl(16
g_3^3+10 g_1 g_3+5 g_2^2\bigr) x^6\\-65536 q^{13} \bigl(39 g_3^2+10
g_1\bigr) x^4-2621440 q^{14} g_3 x^2-1048576 q^{15} = 0\, .
\end{multline}
The above polynomial can be shown to be irreducible over $\mathbb
C[g_{0},g_{1},g_{2},g_{3},q]$ using both PHC and \textsc{Singular}.
Let us spell out, for the last time, the two basic consequences of the
irreducibility. First, the 35 vacua that correspond to the 35 roots of
the polynomial can all be smoothly connected to each other by analytic
continuations in the parameters. Second, the operator $x$, or $v_{0}$,
is a primitive operator. Thus \emph{any} chiral operator in \emph{any}
of the 35 vacua of the phase is given by a simple polynomial in $x$.

\subsubsection{Summary}

In the following table we give, for each value of $N$, the total
number $v$ of vacua and the total number $\varphi$ of distinct phases
in the model $d=N$, which is the simplest model that realizes all the
possible phases.

\be\nonumber
\begin{matrix}
N &\vline & 1 & 2 & 3 & 4 & 5 & 6 & 7 \\\hline
v &\vline & 1 & 5 & 22 & 101 & 476 & 2282 & 11075\\
\varphi &\vline & 1 & 2 & 3 & 5 & 6 & 10 & 10
\end{matrix}
\ee

\section{Conclusion}

In the present paper, we have used the language of algebraic geometry,
at an elementary level, to formulate and analyse the exact solutions
to $\nn=1$ supersymmetric gauge theories. We have demonstrated that
this approach is completely general and has many practical advantages.
It eliminates confusing points appearing in other approaches, allows
for an elegant global description of the quantum phases and can be
efficiently implemented on the computer. It also provides a precise
formulation of Seiberg dualities. We believe that this is the most
appropriate language in which to discuss the quantum supersymmetric
theories.

Of course there are many possible applications of the formalism and
many open problems could be fruitfully studied along the lines of our
work. An outstanding example is the $\nn=1^{*}$ theory, which is a
deformation of $\nn=4$ in which supersymmetry is broken down to
$\nn=1$ by turning on a tree-level superpotential for the three
adjoints $X$, $Y$ and $Z$ of the form $\frac{1}{2}\Tr (m_{Y} Y^{2} +
m_{Z}Z^{2} + V (X))$, where $V$ is an arbitrary polynomial. Almost
nothing is known about the phase structure of this model beyond the
case of the massive phases \cite{DW}, which are the analogues of the
rank one phases studied in \ref{simcaSec}. A particularly interesting
feature of the $\nn=1^{*}$ model is that it inherits the S-duality of
the $\nn=4$ theory and thus the S-duality group has a non-trivial
action on the vacua of the theory.

Another important problem that we have only skimmed over in
\ref{phasetransSec} is the study of the possible phase transitions.
Phase transitions can be associated with non-trivial superconformal
fixed points and an interesting physics. For example, standard cases
involve the condensation of monopoles, and many more exotic phenomena
can be expected. The methods of the present paper are very well suited
to make a systematic study of these transitions, for example in the
models that we have discussed in Sections \ref{HiggsConfSec} and
\ref{CoulSec}.

Another very natural arena to apply our methods is the landscape of
supersymmetric vacua in string or M theory. Can we find in this
context simple models where a full analysis can be performed? What are
the irreducible components of the space of vacua? Can we obtain a full
description of the possible phase transitions? What is the r\^ole
played by gravity in shaping the structure of the phase diagram? What
are the consequences of the existence of distinct phases (as opposed
to distinct vacua) when one tries to use statistical methods to study
the landscape?

An important lesson that we have learned is that the notion of phase
is a much more fundamental concept than the notion of vacuum in a
fully quantum treatment of the supersymmetric theories. The phases are
the basic, irreducible, building blocks of the quantum theory.
This has interesting consequences for the landscape of possible
universes. For example, the existence of a given vacuum implies, by
quantum consistency, the existence of all the other vacua in the same
phase. In our framework, this simply follows from the fact that the
semiclassical expansion of any given root of an irreducible polynomial
characterizes completely the irreducible polynomial and thus all the
other roots.

Another interesting remark is that it is clearly much more convenient
and natural to work with the irreducible polynomials themselves than
with the series expansions. This feature is in tension with the
standard approach to quantum theory based on the quantization of
classical systems and suggests that a better formulation of quantum
theory might exist.

\subsection*{Acknowledgements}

I would like to thank the Isaac Newton Institute for Mathematical
Sciences, and particularly the organizers of the programme ``Strong
Fields, Integrability and Strings,'' Prof.\ Nick Dorey, Prof.\ S.\
Hands and Dr.\ N.\ MacKay, for providing an extremely stimulating
scientific environment in which important parts of the research
presented in this paper were done.

This work is supported in part by the belgian Fonds de la Recherche
Fondamentale Collective (grant 2.4655.07), the belgian Institut
Interuniversitaire des Sciences Nucl\'eaires (grant 4.4505.86), the
Interuniversity Attraction Poles Programme (Belgian Science Policy)
and by the European Commission FP6 programme MRTN-CT-2004-005104 (in
association with V.\ U.\ Brussels). The author is on leave of absence
from Centre National de la Recherche Scientifique, Laboratoire de
Physique Th\'eorique de l'\'Ecole Normale Sup\'erieure, Paris, France.

\renewcommand{\thesection}{\Alph{section}}
\renewcommand{\thesubsection}{\arabic{subsection}}
\renewcommand{\theequation}{A.\arabic{equation}}
\setcounter{section}{0}
\section{Appendix}

It is possible to write down explicit operator relations valid for
given values of the rank $r$ or of the integers $s_{+}$ and $s_{-}$
that correspond to the factorization conditions \eqref{FactCa} or
\eqref{spsmfact}. The general problem is as follows: given a certain
polynomial
\be\label{polexample} H(z) = \sum_{k=0}^{n}c_{k} z^{n-k} =
c_{0}\prod_{i=1}^{n}
(z-h_{i})\,
,\ee
what are the conditions on the coefficients $c_{k}$ for $Q$ to have
$p$ double roots? The answer to this question, in the case $p=1$, is
well-known. One introduces the discriminant of $H$,
\be\label{discdef} \Delta_{H}^{(0)}
=c_{0}^{n-1}\prod_{i<j}(h_{i}-h_{j})^{2}\, .\ee
Clearly, $\Delta_{H}^{(0)}=0$ if and only if $H$ has a double root.
Moreover, $\Delta_{H}^{(0)}$ is a symmetric polynomial in the roots
$h_{i}$ and can thus be written as a polynomial in the coefficients
$c_{k}$. The algebraic equation
\be\label{Disccond} \Delta_{H}^{(0)}(c_{0},\ldots,c_{n}) = 0\ee
gives the necessary and sufficient condition for $H$ to have a double 
root.

For example, the ideal corresponding to the rank $N-1$ vacua is
generated by the polynomial $\Delta^{(0)}_{P^{2}-4q}$. In the notation
of \eqref{idealrank}, this ideal corresponds to
$\ideal_{\{N/2,N/2-1\},1}$ if $N$ is even or to
$\ideal_{\{(N-1)/2,(N-1)/2\},1}$ if $N$ is odd. As explained in
\ref{diagSec}, these ideals are prime, and thus the polynomials
$\Delta^{(0)}_{P^{2}-4q}$ are irreducible.

Assume now that $H$ has one double root. Can we find an additional
condition on the coefficients $c_{k}$ that would ensure that $H$
actually has two double roots (or one triple root)? This condition is 
not difficult to guess. Consider
\be\label{DelH1} \Delta_{H}^{(1)} =
c_{0}^{n-2}\sum_{k=1}^{n}\prod_{\substack{i<j\\ i,j\not =
k}}(h_{i}-h_{j})^{2}\, .\ee
If, for example, $h_{1}=h_{2}$, then $\Delta^{(1)}_{H} =
c_{0}^{n-2}\prod_{2\leq i<j}(h_{i}-h_{j})^{2}$. Imposing
$\Delta^{(1)}_{H}=0$ thus clearly does the job. Note also that
$\Delta^{(1)}$ is completely symmetric in the roots and can be
expressed as a polynomial in the coefficients as required. More
generally, one has the following standard definitions and theorems.
\begin{defn} The $k^{\text{th}}$ \emph{subdiscriminant}, $0\leq k\leq
n-2$, of the polynomial $H$ in \eqref{polexample} is defined by
\be\label{SubDisdef}\Delta^{(k)}_{H} =
c_{0}^{n-k-1}\sum_{\substack{I\subset\{1,\ldots,n\}\\
|I|=n-k}}\prod_{\substack{i<j\\(i,j)\in I^{2}}}(h_{i}-h_{j})^{2}\,
,\ee
where the sum in the right hand side of \eqref{SubDisdef} has
$\binom{n}{k}$ terms, running over all subsets
$I\subset\{1,\ldots,n\}$ of cardinality $n-k$.
\end{defn}
\begin{prop}\label{DiscHermProp} The $k^{\text th}$ subdiscriminant of
$H$ is a polynomial in the coefficients of $H$. Explicitly, if we
denote by $N_{j}=\sum_{i=1}^{n}h_{i}^{j}$ the $j^{\text{th}}$ Newton's
sum and by $\mathcal H^{(k)}=(N_{i+j-2})_{1\leq i,j\leq n-k}$ the
$k^{\text{th}}$ Hermite's matrix, then
\be\label{SDform} \Delta^{(k)}_{H}(c_{0},\ldots,c_{n}) =
c_{0}^{n-k-1}\det\mathcal H^{(k)}\, .\ee
\end{prop}
\begin{thm}\label{SDThm} The polynomial $H$ in \eqref{polexample} has
$p$ double roots (where a $q^{\text{th}}$ root is counted as $q-1$
double roots) if and only if the algebraic equations
\be\label{usesubdisc} \Delta^{(0)}_{H} = \cdots = \Delta^{(p-1)}_{H} =
0\ee
on its coefficients are satisfied.
\end{thm}
Prop.\ \ref{DiscHermProp} can be derived by noting that, if
\be\label{VdMdef}\mathcal 
V^{(k)} = (h_{j}^{i-1})_{\substack{1\leq i\leq n-k\\1\leq j\leq n}}\ee
is a truncated Van der Monde matrix, then $\mathcal H^{(k)}= \mathcal
V^{(k)}{}^{T}\mathcal V^{(k)}$. One then uses the Cauchy-Binet formula
for the determinant of the product of two matrices and the standard
result for the Van der Monde determinants to obtain \eqref{SDform}.
Th.\ \ref{SDThm} follows directly from the definition \eqref{DelH1}.

For a given rank $r$, the operator relations
\be\label{rankrrelations} \Delta_{P^{2}-4
q}^{(0)}=\cdots=\Delta_{P^{2}-4 q}^{(N-r-1)}=0\ee
are thus satisfied. For given $\{s_{+},s_{-}\}$, one has the 
relations
\be\label{spsmrelations}
\Delta_{P_{+}}^{(0)}=\cdots =
\Delta_{P_{+}}^{(s_{+}-1)}=0=
\Delta_{P_{-}}^{(0)}=\cdots =
\Delta_{P_{-}}^{(s_{-}-1)}\, .
\ee
These are not operator relations in the strict sense because
$P_{\pm}=P\mp 2 q^{1/2}$ and thus $q^{1/2}$ enters in the
coefficients, but any combination of the relations
\eqref{spsmrelations} that is invariant under $q^{1/2}\mapsto
-q^{-1/2}$ (or equivalently under the interchange of $P_{+}$ and
$P_{-}$) will be a proper operator relation.

\end{document}